\documentclass[12pt,english,round]{article}
\usepackage[T1]{fontenc}
\usepackage[latin9]{inputenc}
\usepackage{geometry}
\usepackage{color}
\usepackage{array}
\usepackage{float}
\usepackage{textcomp}
\usepackage{amsmath}
\usepackage{amssymb}
\usepackage{graphicx}
\usepackage{setspace}
\usepackage[authoryear]{natbib}
\usepackage{subscript}
\makeatletter

\providecommand{\tabularnewline}{\\}
\newcommand{\lyxdot}{.}

\makeatother

\usepackage{babel}
\begin{document}

\title{Climate Policy under Spatial Heat Transport: Cooperative and Noncooperative
Regional Outcomes \thanks{We thank participants at ASSA 2017 (Chicago), the 2017 Conference
on the Macro and Micro Economics of Climate Change (San Barbara),
the 2018 China Meeting of the Econometric Society (Shanghai), SEA
2018 (Washington DC), and INFORMS 2018 (Phoenix), for their helpful
comments. Cai would like to acknowledge support from the Hoover Institution
at Stanford University. This research is part of the Blue Waters sustained-petascale
computing project, which is supported by the National Science Foundation
(awards OCI-0725070 and ACI-1238993) and the State of Illinois. Blue
Waters is a joint effort of the University of Illinois at Urbana-Champaign
and its National Center for Supercomputing Applications. Earlier versions
of this paper include ``Climate Change Economics and Heat Transport
across the Globe: Spatial-DSICE'' and ``Climate Policy under Cooperation
and Competition between Regions with Spatial Heat Transport''. }}

\author{\vspace{1cm}
Yongyang Cai\thanks{The Ohio State University. cai.619@osu.edu} \quad{}William
Brock\thanks{University of Wisconsin and University of Missouri. wbrock@ssc.wisc.edu}\quad{}
Anastasios Xepapadeas\thanks{Athens University of Economics and Business and University of Bologna.
xepapad@aueb.gr}\quad{}Kenneth Judd\thanks{Hoover Institution. kennethjudd@mac.com}}
\maketitle
\begin{abstract}
We build a novel stochastic dynamic regional integrated assessment
model (IAM) of the climate and economic system including a number
of important climate science elements that are missing in most IAMs.
These elements are spatial heat transport from the Equator to the
Poles, sea level rise, permafrost thaw and tipping points. We study
optimal policies under cooperation and noncooperation between two
regions (the North and the Tropic-South) in the face of risks and
recursive utility. We introduce a new general computational algorithm
to find feedback Nash equilibrium. Our results suggest that when the
elements of climate science are ignored, important policy variables
such as the optimal regional carbon tax and adaptation could be seriously
biased. We also find the regional carbon tax is significantly smaller
in the feedback Nash equilibrium than in the social planner's problem
in each region, and the North has higher carbon taxes than the Tropic-South. 


\end{abstract}

\section{Introduction}

A major characteristic of leading integrated assessment models (IAMs)
such as RICE-2010 \citep{Nordhaus_RICE_2010} or DICE-2016 \citep{DICE2016}
is that the geophysical sector of the model determines the mean surface
temperature through the carbon cycle, which in turn determines the
damage function. Thus damages are related to the mean surface temperature
of the planet. 

A well-established fact in the science of climate change, however,
is that when the climate cools or warms, high latitude regions tend
to exaggerate the changes seen at lower latitudes (e.g., \citealp{Langen2007,IPCC2013}).
This effect is called polar amplification (PA) and indicates that,
under global warming, the temperature at the latitudes closer to the
Poles will increase faster than at latitudes nearer to the Equator.
PA is especially strong in the Arctic and is sometimes called \textquotedblleft Arctic
amplification\textquotedblright . For example, \citet{Bekryaev2010}
document a high-latitude (greater than 60 °N) warming rate of 1.36
degrees centigrade per century from 1875 to 2008. This trend is almost
twice that of the Northern Hemisphere trend of 0.79 degrees centigrade
per century.

Spatial heat and moisture transport, and the resulting PA, suggest
that a better representation of the climate science underlying IAMs
would be a geophysical sector structure which accounts for these phenomena.
This implies that, in the IAM output, the surface temperature anomaly
will be differentiated across spatial zones of the globe. The spatial
temperature differentiation is important for the economics of climate
change because it provides the impact of PA on the structure of the
economic damages. PA will accelerate the loss of Arctic sea ice, which
in turn has consequences for melting land ice that is associated with
a potential meltdown of the Greenland and West Antarctica ice sheets
which could cause serious global sea level rise (SLR). 

Another source of damage associated with PA relates to the thawing
of the permafrost, which is expected to bring about widespread changes
in ecosystems and damage to infrastructure, along with release of
greenhouse gases (GHGs) which exist in permafrost carbon stocks (see,
e.g., \citealp{IPCC2013,schuur_2015}). Furthermore, recent studies
suggest that Arctic amplification might increase the frequency of
extreme weather events \citep{Cohen2018}, although this remains a
controversial issue \citep{Overland2018}. 

The well-known \textquotedblleft burning embers\textquotedblright{}
diagram in \citet{Lenton2007} shows the ranking of tipping elements
by order of proximity to the present time. The \textquotedblleft nearest\textquotedblright{}
three potential tipping points are located in the high latitudes of
the Northern Hemisphere (Arctic summer sea ice loss, Greenland ice
sheet melt and boreal forest loss). Because of PA in the Arctic, each
degree increase in planetary yearly mean temperature leads to approximately
two degrees increase in the Arctic latitudes. Thus natural phenomena
occurring in high latitudes, due to spatial heat and moisture transport,
can cause economic damages in lower latitudes. These spatial impacts,
which could have important implications for climate change policies,
are not embodied in the current generation of IAMs. The RICE model
\citep{Nordhaus_RICE_2010} \textendash{} the regional version of
DICE \citep{DICE2016} \textendash{} still treats the climate system
by using the globally averaged measure of temperature and neglects
heat and moisture transport and especially PA. 

\citet{Hassler2012} extend \citet{Golosov2014} to multi-regions.
While their work is elegant, as is that of \citet{Golosov2014}, they
do not deal with poleward heat transport, multi-layer carbon cycles,
separation of atmospheric and oceanic layers, and regional tipping
points, as we do here. \citet{vanderPloeg_deZeeuw_2016} and \citet{JAAKKOLA2018}
develop interesting two-region models with tipping points that deal
explicitly with non-cooperative and cooperative institutional structures.
But they do not include geographic specification of regions, poleward
heat transport, recursive preferences, or the more realistic multi-layer
modeling of the carbon cycle and the temperature response to anthropogenic
forcing as we have. Thus neither they nor \citet{Golosov2014} and
\citet{Hassler2012} are able to study the effects of different values
of the intertemporal elasticity of substitution (IES) and risk aversion
parameters, or the effects of neglecting poleward heat transport on
regional carbon taxes, as we are able to do with our richer and more
realistic modeling of the interaction between climate dynamics and
economic dynamics. 

Another very recent IAM model of \citet{KrusellSmith2017} deals with
space at a much finer scale than the present paper and contrasts market
structures, including autarky and full borrowing and lending. However,
their model does not address issues related to heat and moisture transport,
SLR, permafrost thaw and the impacts of tipping points, as we do here.
Thus we feel that our work is complementary to that of \citet{KrusellSmith2017}
and not competitive. As mentioned above, one novel contribution of
the present paper is to develop an IAM which incorporates spatial
impacts associated with heat and moisture transport, along with treatment
of uncertainty and tipping points. As far as we know, no other IAM
treats these issues as we do here. 

The stochastic IAM developed in this paper is a complex two-regional
model with a more realistic \textendash{} relative to existing models
\textendash{} geophysical sector and its solution requires the use
of advanced numerical methods. Thus we adapt computational methods
related to the DSICE model of \citet{CaiJuddLontzek_DSICE}. The DSICE
framework is a stochastic generalization of DICE, which does not take
into account the heat and moisture transport dynamics of the climate
system. We adapt DSICE by disaggregating the globe into regions. However,
DSICE is only a social planner's model and its computational method
is for solving social planner's problems. Besides a social planner's
model, we also study feedback Nash equilibrium in this paper, and
we develop a new general computational method to solve the high-dimensional
dynamic stochastic game (with ten continuous state variables and one
discrete state variable) with recursive utility. 

Most IAMs are based on a social planner's problem assuming that all
countries are cooperative and the social planner can allocate resources
without border frictions, so they can just provide a polar solution
as there are very few unselfish sovereigns in the real world. It would
be interesting if we can find another polar solution under a noncooperative
Nash equiloibrium. To our best knowledge, \citet{vanderPloeg_deZeeuw_2016}
and \citet{JAAKKOLA2018} are the only work in finding optimal climate
policy under noncooperation. However, \citet{vanderPloeg_deZeeuw_2016}
just solve an open-loop Nash equilibrium whose optimal decisions depend
on time and the initial state only; \citet{JAAKKOLA2018} solve a
symmetric feedback Nash equilibrium (FBNE) under a low-dimensional
continuous-time model using the Hamilton-Jacobi-Bellman equation method,
but in most cases FBNE is not symmetric. This paper is the first to
numerically find optimal climate policy under both the social planner's
cooperative world and (asymmetrical) FBNE under a realistic discrete-time
dynamic stochastic model with more than ten continuous state variables. 

We call our model a Dynamic Integration of Regional Economy and Spatial
Climate under Uncertainty (DIRESCU).\footnote{\citet{Brock2017} considered a simple deterministic model and showed
that, by ignoring spatial heat and moisture transport and the resulting
PA, the regulator may overestimate or underestimate the tax on GHG
emissions. The structure of their economic model is, however, simplified
and this makes it difficult to discuss realistic policy options.} \textcolor{black}{In developing DIRESCU, we follow the two-region
approach of \citet{Langen2007} but change their regions as follows:
}region 1 is the region north of latitude 30°N to 90°N (called the
North), while region 2 is the region from latitude 90°S (the South
Pole) to 30°N (called the Tropic-South). Heat and moisture transport
take place northbound from the tropical belt of latitudes north of
the Equator which are included in the Tropic-South toward the North.\footnote{There is transport toward the South Pole from all latitudes south
of the Equator in the Tropic-South which we do not take into account
in order to ease the computational burden by reducing the number of
dimensions in our model and at the same time include the Southern
Hemisphere economies. Scientific evidence suggests that PA in Antarctica
is weaker than in the Arctic, because of weaker surface albedo feedback
and more efficient ocean uptake in the Southern Ocean, in combination
with Antarctic ozone depletion. Thus, for the time horizon of 100
years in which our solutions are focused, the majority of the effects
of heat transfer should be associated with heat transfer toward the
North Pole, and then we adopted the approximation of unifying the
Southern Hemisphere with the 0°-30°N belt and northbound heat transfer. } 

The interaction of the geophysical sector of DIRESCU with the economic
sector is reflected in the damage function. We introduce separate
damage functions in each region and allow for damages in the Tropic-South
to be caused by an increase in temperature (i.e., PA) in the North.
For example, the increased amplification of the temperature anomaly
in the high north latitudes increases the hazard rate of tipping events
in the high north latitudes toward earlier arrival times. Hence any
associated damages caused to lower latitudes by warming in the higher
north latitudes, e.g., increased melting of land ice leading to SLR
damages in the lower latitudes, will be increased by PA, even though
the high northern latitudes may benefit from additional warming. 

The rest of the economic module is based on a two-region differentiation
of DICE-2016 \citep{DICE2016}. We model the economic interactions
between the regions with an adjustment cost function, and we allow
for adaptation expenses in each region. \citet{KrusellSmith2017}
compare the two market structures of complete autarky and full international
borrowing and lending and find that the market structures do not have
a large impact on their results. While we can study autarky as \citet{KrusellSmith2017}
do by raising the cost of interaction to induce autarky, our formulation
of economic interactions does not include the case of full borrowing
and lending as in their model. We have ignored serious modeling of
market structure in order to focus on some elements of geophysics
that are ignored in other contributions, including that of \citet{KrusellSmith2017}.
We do this to provide new insights regarding the importance of spatial
heat and moisture transport phenomena in climate change policy. 

The present paper innovates relative to popular IAMs at the tractable
level of complexity in the literature (e.g., Nordhaus's DICE and RICE
models, the even more complex DSICE model and many others) in a number
of ways. In particular, (i) we incorporate an endogenous SLR module,
an endogenous permafrost melt module and, especially, we add the more
realistic geophysics of spatial heat and moisture transport from low
latitudes to high latitudes, while keeping the three-layer carbon
cycle of DICE and RICE, and expanding the two-layer temperature module
of DICE and RICE to a three-layer module; (ii) We introduce recursive
preferences and we consider a wide range of parameter values of risk
aversion and IES; (iii) We allow for adaptation to regional damage
from SLR and temperature increase;\footnote{The importance of relating damages from temperature increase to adaptation
has been emphasized by, for example, \citet{Barreca2016} who showed
remarkable reduction of damages to morbidity and mortality due to
heat stress in the U.S. after the introduction of technologies such
as air conditioning. Another example is \citet{burgess2014} who showed
large negative effects of extreme heat days in India, especially in
rural areas. Since lack of access to air conditioning is a difference
between India and the U.S., these results suggest that because many
areas in the Tropic-South are poorer than the North, we might expect
adaptation such as introduction of air conditioning to be slower in
the Tropic-South than in the wealthier parts of the North. } (iv) Our paper goes beyond the single-region DSICE model by developing
a new general computational method to numerically solve a dynamic
stochastic FBNE of two regions. 

We calibrate our many parameter values to match history as well as
to fit the representative concentration pathway (RCP) scenarios \citep{Meinshausen_RCP}.
The main results of this paper are summarized below.

First, stochastic processes of regional carbon taxes are derived and
various uncertainty fan charts are presented and compared with and
without heat and moisture transport as well as for a range of risk
aversion and IESs, under cooperation and noncooperation between regions.
Moreover, the North has much higher regional carbon taxes than the
Tropic-South. Our figures and tables provide a much more thorough
quantification of the multitude of types of uncertainties than the
received regional IAM literature at the DICE/RICE level of complexity. 

Second, the non-cooperative dynamic stochastic game between the regions
leads to much lower regional carbon taxes than the social planner's
model with economic interactions between the regions. If the economy
in each region is closed, then the social planner's model has much
larger regional carbon taxes and much larger surface temperature increases
than in the case with economic interactions. 

Third, neglecting heat and moisture transport as in RICE and other
regional IAMs that do not account for this added geophysics leads
to many biases, including inaccurate forecasting of the first time
of arrival of potential tipping points located in the high latitudes
of the Northern Hemisphere. The low (high) latitude regions would
be hotter (colder) if poleward heat transport were absent, hence damages
in the low latitude regions would be higher, since they are already
under heat stress and transporting some of that heat poleward helps
relieve this heat stress. For example, solutions without heat transport
will underestimate what actual heat-related damage there is in the
North, and overestimate the actual heat-related damage in the Tropic-South.
Without heat transport, the adaptation rates in the North will be
underestimated as its corresponding atmospheric temperature anomaly
is underestimated, and the adaptation rates in the Tropic-South will
be overestimated as its corresponding atmospheric temperature anomaly
is overestimated.

Fourth, endogenous SLR and adaptation are important new contributions
of our modeling. In this way we capture the projected increased diversion
into adapting to SLR (e.g., spending resources on SLR-protective infrastructure).
The projected earlier arrival of increased melting of land ice in
the Northern Hemisphere high latitudes due to our inclusion of heat
transport means a potential increase in SLR damages in coastal low
latitude areas relative to the projections when heat transport is
ignored. Ignoring SLR underestimates the regional carbon taxes significantly,
and ignoring adaptation overestimates the regional carbon taxes significantly. 

Fifth, the optimal regional carbon taxes for both regions tend to
be larger for larger IES values for climate tipping risks, in a cooperative
or non-cooperative world. This is consistent with empirical findings
in finance that greater IESs in Epstein\textendash Zin recursive preferences
\citep{Epstein-Zin-1989} imply that long-term risk matters and carbon
taxes are larger.\footnote{See \citet{BansalYaron2004} for financial risks, and \citet{Bansal2016}
for climate risks.} This result is also consistent with the findings in the DSICE model
\citet{CaiJuddLontzek_DSICE}.

\textcolor{black}{The paper is organized as follows. Section \ref{sec:Model}
builds }DIRESCU\textcolor{black}{. We calibrate our spatial climate
system and the economic system using DICE and RICE, as well as data
from other literature such as \citet{IPCC2013}. Section \ref{sec:Social-Planner-Nash}
discusses the social planner's problem and the regional feedback Nash
equilibrium under climate tipping risk and Epstein\textendash Zin
preferences to address the smoothness of consumption across time and
risk aversion. Section \ref{sec:Algorithm-FBNE} introduces our new
computational method to solve FBNE for a general dynamic game. Section
\ref{sec:res-sto} discusses the results of the DIRESCU model and
Section \ref{sec:Conclusion} concludes.}

\section{Model Setup\label{sec:Model}}

\begin{figure}
\begin{centering}
\includegraphics[width=0.5\textwidth,height=0.28\textheight]{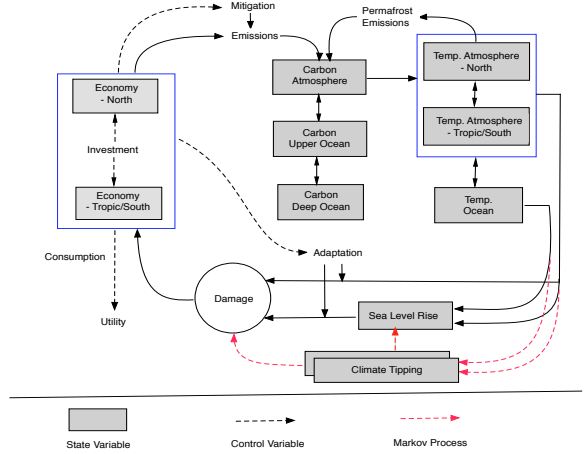}
\par\end{centering}
\caption{The DIRESCU model\label{fig:Spatial-DSICE-model}}
\end{figure}

The deterministic version of our DIRESCU model follows DICE-2016 \citep{DICE2016},
which maximizes social welfare with trade-offs between carbon dioxide
(CO$_{2}$) abatement, consumption and investment. Our model has been
augmented with an additional control, relative to DICE-2016, by including
adaptation to climate change following de Bruin et al. (2009). DIRESCU
has two regions: the first one (indexed with $i=1$) is the North
from latitude 30°N to 90°N and the second one (indexed with $i=2$)
is the Tropic-South from latitude 90°S to 30°N. We first model it
as a social planner problem with both economic and climate interaction
between two regions, SLR, permafrost thaw and climate tipping risks
(in the deterministic model, the risks are ignored). We then change
it to a dynamic stochastic game and solve its feedback Nash equilibrium.
The big picture of the model setup is depicted in Figure \ref{fig:Spatial-DSICE-model}
and its details are described below. 

\subsection{The Climate System}

The climate system contains four modules: the carbon cycle, the temperature
system, SLR and permafrost thaw. In our calibration of the climate
system, we use the four RCP scenarios (i.e., RCP2.6, RCP4.5, RCP6,
and RCP8.5) \citep{Meinshausen_RCP} and DICE/RICE optimal scenarios
(i.e., the optimal solutions of DICE-2016 and RICE-2010 with optimal
mitigation policy), where we define each scenario as a combination
of pathways of global emissions, atmospheric carbon concentration,
global radiative forcing, and globally averaged surface temperature
relative to 1900 levels. For the RCP scenarios, the pathways of atmospheric
carbon concentration, global radiative forcing, and globally averaged
surface temperature are generated from the software MAGICC \citep{Meinshausen_MAGICC},
using the corresponding pathways of global emissions as the input
to the software. The pathways of global emissions are provided by
MAGICC too. Appendix \ref{sec:Calibration-Climate} shows that our
calibrated system fits these scenarios, history, as well as other
data in the literature such as regional temperature projections reported
in \citet{IPCC2013}. 

\subsubsection{Carbon Cycle}

We follow DICE-2016 in using three-layer carbon concentrations: atmospheric
carbon, carbon in the upper ocean and carbon in the deep ocean. Let
$\mathbf{M}_{t}=(M_{t}^{\mathrm{AT}},M_{t}^{\mathrm{UO}},M_{t}^{\mathrm{DO}})^{\top}$
represent the carbon concentration in the atmosphere, the upper ocean
and the deep ocean. Then the three-layer carbon cycle system can be
represented as: 
\begin{equation}
\mbox{\textbf{M}}_{t+1}=\mathbf{\mathbf{\Phi}}_{\mathrm{M}}\mathbf{M}_{t}+\left(E_{t},0,0\right)^{\top},\label{eq:M-law}
\end{equation}
where $E_{t}$ is global carbon emissions (billions of metric tons)
and 
\[
\mathbf{\Phi}_{\mathrm{M}}=\left[\begin{array}{ccc}
1-\phi_{12} & \phi_{21}\\
\phi_{12} & 1-\phi_{21}-\phi_{32} & \phi_{32}\\
 & \phi_{23} & 1-\phi_{32}
\end{array}\right].
\]
The parameters $\phi_{ij}$ are calibrated against the four RCP scenarios
and the DICE-2016 optimal scenario. For every scenario, we can use
its pathway of emissions as the input $E_{t}$ to our carbon cycle
system (\ref{eq:M-law}). Our carbon cycle provides as output a pathway
of atmospheric carbon concentration for each scenario. We calibrate
a unique set of values for $\phi_{ij}$ so that the pathways of atmospheric
carbon concentration are close to the scenarios' pathways of atmospheric
carbon concentration, for all scenarios. 

\subsubsection{Temperature Subsystem\label{subsec:Temperature-subsystem}}

The global radiative forcing representing the impact of CO$_{2}$
concentrations on the surface temperature of the globe (watts per
square meter from 1900) is 
\begin{equation}
F_{t}=\eta\log_{2}\left(M_{t}^{\mathrm{AT}}/M_{*}^{\mathrm{AT}}\right)+F_{t}^{\mathrm{EX}},\label{eq:F-def}
\end{equation}
where $\eta=3.68$ as in DICE-2016 and $F_{t}^{\mathrm{EX}}$ is the
global exogenous radiative forcing. 

We use $\mathbf{T}_{t}=(T_{t,1}^{\mathrm{AT}},T_{t,2}^{\mathrm{AT}},T_{t}^{\mathrm{OC}})^{\top}$
to represent the temperature anomaly (relative to 1900 levels) in
the atmosphere (two regions) and the global ocean. Thus, the temperature
system is 
\begin{equation}
\mbox{\textbf{T}}_{t+1}=\mathbf{\Phi}_{\mathrm{T}}\mathbf{T}_{t}+\xi_{1}\left(F_{t},F_{t},0\right)^{\top},\label{eq:T-law}
\end{equation}
where we assume that the global radiative forcing has the same effect
on both regions, and 
\[
\mathbf{\Phi}_{\mathrm{T}}=\left[\begin{array}{ccc}
1-\xi_{2}-\xi_{4}-\xi_{6} & \xi_{4}+\xi_{5} & \xi_{2}\\
\xi_{4} & 1-\xi_{2}-\xi_{4}-\xi_{5}-\xi_{6} & \xi_{2}\\
\xi_{3} & \xi_{3} & 1-2\xi_{3}
\end{array}\right].
\]
In transition equation (\ref{eq:T-law}) and transition matrix $\mathbf{\Phi}_{\mathrm{T}}$,
the parameter $\xi_{1}$ is the temperature increase for each unit
of radiative forcing, $\xi_{2}$ and $\xi_{3}$ represent transport
between atmosphere and ocean, $\xi_{4}$ and $\xi_{5}$ are used to
capture spatial heat and moisture transport, and $\xi_{6}$ represents
the sensitivity of the outgoing long-wave radiation to atmospheric
temperature changes.

Similarly to the calibration of the carbon cycle, we calibrate $\xi_{1},...,\xi_{6}$
against the RCP scenarios, the DICE-2016 optimal scenario, and the
historical spatial temperatures in 1900-2015 from the Goddard Institute
for Space Studies \citeyearpar[GISTEMP Team,][]{GISS}. For each scenario,
we use its pathway of radiative forcing as the input $F_{t}$ to our
temperature system (\ref{eq:T-law}), which provides as output two
regional pathways of atmospheric temperatures in the regions. Then
we average the regional pathways to generate a globally averaged atmospheric
temperature pathway. We calibrate a unique set of values for $\xi_{1},...,\xi_{6}$
so that the generated globally averaged atmospheric temperature pathways
from our temperature system are close to the pathways of globally
averaged atmospheric temperature for all scenarios and, at the same
time, our regional temperatures in 2081-2100 are also close to the
projected regional temperatures in 2081-2100 provided in IPCC (2013). 

\subsubsection{Sea Level Rise}

Sea level rise is a serious problem caused by global warming. Table
4.1 of \citet{IPCC2013} shows that, if the whole Greenland ice sheet
melts, it will cause more than 7 meters (m) global mean SLR, and if
the whole Antarctic ice sheet melts, it will lead to about 58 m global
mean SLR. Moreover, once an ice sheet collapses, it is irreversible
for millennia even if forcing is reversed (\citealt{IPCC2013}, Table
12.4). \citet{Nerem2018} adduce evidence that SLR is accelerating
from the historical data. Table 13.5 of \citet{IPCC2013} shows SLR
in 2100 ranging from about 0.44 m for RCP2.6 to about 0.74 m for RCP8.5.
Figure 13.14 of \citet{IPCC2013} shows that the likely range of SLR
is from 1 to 3 m per degree Celsius of globally averaged surface temperature
increase if the warming is sustained for millennia. 

There are four main sources of SLR: thermal expansion, and melting
of glaciers and ice cap, the Greenland ice sheet, and the Antarctic
ice sheet. The west Antarctic ice sheet (WAIS) is vulnerable to ocean
warming as most of it is below sea level and extensively exposed to
the ocean. The contribution of a complete collapse of the marine WAIS
is estimated at 3.3 m of SLR \citep{Bamber2009}. Thermal expansion
is also due to ocean warming. The melting of glaciers and the Greenland
ice sheet is due to atmospheric warming in the North. Therefore, we
assume that the rate of SLR is dependent on north atmospheric temperature
$T_{t,1}^{\mathrm{TA}}$ and ocean temperature $T_{t}^{\mathrm{OC}}$,
and that a higher temperature implies a higher rate of SLR. We also
assume that SLR, $S_{t}$, is irreversible. Thus, we let 
\begin{equation}
S_{t+1}=S_{t}+\zeta_{1}^{\mathrm{SLR}}\left(T_{t,1}^{\mathrm{TA}}\right)^{\zeta_{2}^{\mathrm{SLR}}}+\zeta_{3}^{\mathrm{SLR}}T_{t}^{\mathrm{OC}},\label{eq:SLR-law}
\end{equation}
where $\zeta_{1}^{\mathrm{SLR}}$, $\zeta_{2}^{\mathrm{SLR}}$ and
$\zeta_{3}^{\mathrm{SLR}}$ are calibrated using the SLR data for
four RCP scenarios in Table 13.5 of \citet{IPCC2013} and Table 1
of \citet{Kopp2014}. 

\subsubsection{Permafrost Thaw}

With global warming, and in particular with PA, a large amount of
CO\textsubscript{2} and CH\textsubscript{4} could be emitted from
thawing permafrost in the Arctic and sub-Arctic regions, which contain
about 1,700 GtC (gigaton of carbon) in permafrost soils, while about
1,035 GtC are stored in the surface permafrost (0-3 m depth) and could
easily be emitted when they are thawed \citep{schuur_2015}. \citet{schuur_2015}
show that an average carbon release from the permafrost zone by 2100
across models is about 92 GtC with a standard deviation of 17 GtC
under RCP8.5. A higher atmospheric temperature in the North implies
a higher emission rate. Thus, we assume that carbon emission from
thawing permafrost is 
\begin{equation}
E_{t}^{\mathrm{Perm}}=\zeta_{1}^{\mathrm{Perm}}\left(1-\frac{1}{1+\zeta_{2}^{\mathrm{Perm}}T_{t,1}^{\mathrm{AT}}+\zeta_{3}^{\mathrm{Perm}}\left(T_{t,1}^{\mathrm{AT}}\right)^{2}}\right).\label{eq:E-perm}
\end{equation}
\citet{hope_NCC_2016} give a mean carbon emission path from thawing
permafrost for the A1B scenario in \citet{IPCC2007}, so we use its
annual time series\footnote{We thank Kevin Schaefer for providing the annual time series data.}
to calibrate $\zeta_{1}^{\mathrm{Perm}}$, $\zeta_{2}^{\mathrm{Perm}}$
and $\zeta_{3}^{\mathrm{Perm}}$. 

\subsection{Climate Tipping Points}

There are many uncertainties in the economic and climate system. For
example, DSICE \citep{CaiJuddLontzek_DSICE} discusses business cycle
shocks in productivity and climate risks, and also deals with parameter
uncertainty over the IES and risk aversion using uncertainty quantification.
\citet{LemoineTraeger2014} use a stylized model to study the impact
of climate tipping points. \citet{cai_NCC_2016} use DSICE to study
the impact of multiple interacting tipping points on the carbon tax
policy. In this paper we introduce uncertainty regarding the emergence
of endogenous tipping elements in the North into the spatial model
with heat and moisture transport.

We assume that there is a representative tipping element that will
take $D$ years to fully unfold its damage after it occurs. The final
damage level is $\overline{J}$ as a fraction of output, and the tipping
probability depends on the contemporaneous atmospheric temperature
in the North. Let $J_{t}$ represent the damage level (as a fraction
of output) of the tipping element, and let $\chi_{t}$ be the indicator
representing whether the tipping event has happened or not before
time $t$, so $\chi_{t}=0$ means that the tipping event has not happened,
and $\chi_{t}=1$ means that it has happened before time $t$. Thus,
the transition law of $J_{t}$ is 
\begin{equation}
J_{t+1}=\min(\overline{J},\,J_{t}+\Delta)\chi_{t},\label{eq:law-damLevel}
\end{equation}
where $\Delta=\overline{J}/D$ is the annual increment of damage level
after the tipping happens, and $\chi_{t}$ is a Markov chain with
the probability transition matrix 
\[
\left[\begin{array}{cc}
1-p_{t} & p_{t}\\
0 & 1
\end{array}\right],
\]
where $p_{t}$ is the tipping probability from state $\chi_{t}=0$
to $\chi_{t}=1$. We let $p_{t}=1-\exp\left(-\varrho\max\left(0,T_{t,1}^{\mathrm{AT}}-1\right)\right),$where
$\varrho$ is the hazard rate, so a higher atmospheric temperature
in the North implies a higher tipping probability. 

We use the Atlantic Meridional Overturning Circulation as the representative
tipping element, and employ its default setup as in \citet{cai_NCC_2016},
that is, $D=50$, $\overline{J}=0.15$ and $\varrho=0.00063$. To
introduce a general model, we let 
\begin{equation}
\chi_{t+1}=g(\chi_{t},\mathbf{T}_{t},\omega_{t})\label{eq:tipIndicator}
\end{equation}
denote the transition law for $\chi_{t}$. 

\subsection{The Economic System}

We calibrate our regional economic system to match RICE projections
over our regions. Appendix \ref{sec:Calibration-Econ} shows that
our calibrated system fits RICE projections. 

\subsubsection{Production}

The gross output at time $t$ in each region is determined by a Cobb-Douglas
production function,

\begin{equation}
\mathcal{Y}_{t,i}\equiv A_{t,i}K_{t,i}^{\alpha}L_{t,i}^{1-\alpha},\label{eq:grossY-def}
\end{equation}
with $\alpha=0.3$ and $L_{t,i}$ the exogenous population at time
$t$ and region $i$ aggregated from RICE.\footnote{For region $i$ and time $t$, we sum up population over the RICE
subregions located in region $i$ (if one RICE subregion is located
across our border line 30°N, then we give a rough estimate with the
ratio of land of the subregion located in the region $i$).} 

We use region-specific total factor productivity (TFP) $A_{t,i}$.
\citet{Sachs2001} stresses ecological specific technical progress
and lists five reasons why TFPs in low latitude zones tend to be smaller
than temperate latitude zones. Of course there are exceptions, as
Sachs points out (e.g., Hong Kong and Singapore and, now, lower latitude
parts of China and parallel parts of \textquotedblleft Asian Tigers\textquotedblright ).
However, theory suggests that the economies that are \textquotedblleft behind\textquotedblright{}
should grow faster than the leaders because the leaders have already
done the \textquotedblleft heavy lifting\textquotedblright{} of the
TFP R\&D which the followers could presumably copy. For example, \citet{Sachs2002}
discuss the transition from \textquotedblleft adopter\textquotedblright{}
to \textquotedblleft innovator\textquotedblright{} for countries. 

We let $A_{t,i}=A_{0,i}\exp\left(\alpha_{i}^{\mathrm{TFP}}\left(1-\exp\left(-d_{i}^{\mathrm{TFP}}t\right)\right)/d_{i}^{\mathrm{TFP}}\right),$
where $A_{0,i}$, $\alpha_{i}^{\mathrm{TFP}}$ and $d_{i}^{\mathrm{TFP}}$
are calibrated to match the TFP path in region $i$ which is computed
from RICE by aggregating across the RICE subregions in region $i$.\footnote{We first estimate $K_{t,i}$, $L_{t,i}$ and $\mathcal{Y}_{t,i}$
by summing over those in RICE subregions located in our region $i$
for each time $t$, and then compute the TFP paths $A_{t,i}=\mathcal{Y}_{t,i}/(K_{t,i}^{\alpha}L_{t,i}^{1-\alpha})$
for region $i$.} 

\subsubsection{Damages}

In the deterministic case of DIRESCU, we consider two types of damages:
damages to output from SLR and damages to output directly from temperature
increase.

We follow RICE to let 
\begin{equation}
D_{t,i}^{\mathrm{SLR}}=\pi_{1,i}S_{t}+\pi_{2,i}S_{t}^{2}\label{eq:SLR-dam}
\end{equation}
reflect the damage from SLR, $S_{t}$, as a fraction of output. We
calibrate $\pi_{1,i}$ and $\pi_{2,i}$ to match the damage from SLR
which is computed from RICE.\footnote{We estimate $\mathcal{Y}_{t,i}$ and $\mathcal{D}_{t,i}^{\mathrm{SLR}}=D_{t,i}^{\mathrm{SLR}}\mathcal{Y}_{t,i}$
by summing over those in RICE regions located in our region $i$ for
each time $t$, and then compute $D_{t,i}^{\mathrm{SLR}}=\mathcal{D}_{t,i}^{\mathrm{SLR}}/\mathcal{Y}_{t,i}$
for region $i$. With the data on the SLR path in RICE and $D_{t,i}^{\mathrm{SLR}}$,
we then calibrate $\pi_{1,i}$ and $\pi_{2,i}$ so that equation (\ref{eq:SLR-dam})
holds approximately.} 

We follow DICE and RICE to assume a quadratic damage function for
temperature increase 
\begin{equation}
D_{t,i}^{\mathrm{T}}=\pi_{3,i}T_{t,i}^{\mathrm{AT}}+\pi_{4,i}(T_{t,i}^{\mathrm{AT}})^{2},\label{eq:Dam_Temp_Quad}
\end{equation}
where $\pi_{3,i}$ and $\pi_{4,i}$ are calibrated to fit the aggregated
projected damage from surface temperature change over RICE subregions.\footnote{We use the radiative forcing path in RICE to estimate $T_{t,i}^{\mathrm{AT}}$
using our calibrated climate equation (\ref{eq:T-law}). We also estimate
$\mathcal{Y}_{t,i}$ and $\mathcal{D}_{t,i}^{\mathrm{T}}=D_{t,i}^{\mathrm{T}}\mathcal{Y}_{t,i}$
by summing over those in RICE regions located in our region $i$ for
each time $t$, and then compute $D_{t,i}^{\mathrm{T}}=\mathcal{D}_{t,i}^{\mathrm{T}}/\mathcal{Y}_{t,i}$
for region $i$. With the data on $T_{t,i}^{\mathrm{AT}}$ and $D_{t,i}^{\mathrm{T}}$,
we then calibrate $\pi_{1,i}$ and $\pi_{2,i}$ so that equation (\ref{eq:Dam_Temp_Quad})
holds approximately.}

With the quadratic damage function (\ref{eq:Dam_Temp_Quad}) for the
Tropic-South, this region has damage of only 9\% of its output if
its regional surface temperature increase is the same as the global
mean surface temperature in 2100 under RCP8.5, i.e., 4.7°C, as RICE
does not take into account catastrophic damages. However, \citet{burke_nature_2015}
show that damages from high temperature increase in low- and mid-latitude
regions are much higher, reducing projected global output by 23\%
in 2100 under RCP8.5, with the poorest 40\% of countries (most being
in our Tropic-South) having 75\% reduction relative to a world without
climate change. \citet{Dell2012} show that there are large and negative
effects of higher temperatures on growth in poor countries, with about
1.3\% economic growth reduction for a 1°C increase. \citet{burke_nature_2015}
show that climate change may lead to negative economic growth for
some countries. Our estimate of damages follows DICE and RICE, that
is, damages are proportional to instantaneous output, not to growth
of TFP, so they may be underestimated for the Tropic-South and overestimated
for the North. However, our estimate of damages to the high-latitude
regions may also be underestimated, as we ignored potential damages
from inequality between the regions \citep{Hsiang2017}.

\subsubsection{Emissions, Adaptation, and Mitigation}

Global carbon emissions at time $t$ are defined as
\[
E_{t}\equiv\sum_{i=1}^{2}E_{t,i}^{\mathrm{Ind}}+E_{t}^{\mathrm{Perm}}+E_{t}^{\mathrm{Land}},
\]
where $E_{t}^{\mathrm{Land}}$ is exogenous global carbon emissions
from biological processes, $E_{t}^{\mathrm{Perm}}$ is emissions from
permafrost thawing estimated by equation (\ref{eq:E-perm}), and $E_{t,i}^{\mathrm{Ind}}=\sigma_{t,i}(1-\mu_{t,i})\mathcal{Y}_{t,i}$
is industrial emissions, where $\mu_{t,i}$ is an emission control
rate and $\sigma_{t,i}$ is the carbon intensity in region $i$. We
let
\begin{equation}
\sigma_{t,i}=\sigma_{0,i}\exp\left(-\alpha_{i}^{\sigma}\left(1-\exp\left(-d_{i}^{\sigma}t\right)\right)/d_{i}^{\sigma}\right),\label{eq:carbon-intensity}
\end{equation}
where $\sigma_{0,i}$, $\alpha_{i}^{\sigma}$ and $d_{i}^{\sigma}$
are calibrated to match the carbon intensity paths in region $i$
which are computed from RICE by aggregating across the RICE subregions
in region $i$.\footnote{We use the business-as-usual (BAU) results (i.e., with $\mu_{t,i}\equiv0$)
of RICE to estimate the carbon intensity paths. We first estimate
$E_{t,i}^{\mathrm{Ind}}$ and $\mathcal{Y}_{t,i}$ under BAU by summing
over those in RICE subregions located in our region $i$ for each
time $t$, and then compute the carbon intensity paths $\sigma_{t,i}=E_{t,i}^{\mathrm{Ind}}/\mathcal{Y}_{t,i}$
for region $i$.} 

We include an adaptation choice variable $P_{t,i}$ for each region
in our model as in \citet{deBruin2009}. Adaptation reduces damages
to output from SLR and temperature increase. The output net of all
damages including SLR, temperature anomaly and tipping becomes
\begin{equation}
Y_{t,i}\equiv(1-J_{t})\Omega_{t,i}\mathcal{Y}_{t,i}\equiv\frac{(1-J_{t})\mathcal{Y}_{t,i}}{1+(1-P_{t,i})(D_{t,i}^{\mathrm{SLR}}+D_{t,i}^{\mathrm{T}})}.\label{eq:Y-def-sto}
\end{equation}
where $P_{t,i}\in[0,1]$ is the adaptation rate, and $\Omega_{t,i}$
is the damage factor after adaptation. We follow \citet{deBruin2009}
to assume that adaptation expenditure is $\Upsilon_{t,i}\equiv\eta_{1}P_{t,i}^{\eta_{2}}Y_{t,i},$
with $\eta_{1}=0.115$ and $\eta_{2}=3.6$. 

We follow DICE to assume that mitigation expenditure is $\Psi_{t,i}\equiv\theta_{1,t,i}\mu_{t,i}^{\theta_{2}}Y_{t,i},$
where $\theta_{1,t,i}$ is the abatement cost as a fraction of output
in region $i$ at time $t$. We use the DICE/RICE form to define $\theta_{1,t,i}=b_{0,i}\exp\left(-\alpha_{i}^{b}t\right)\sigma_{t,i}/\theta_{2},$
where $\alpha_{i}^{b}$ and $\theta_{2}$ are parameters given by
RICE, and $b_{0,i}$ is the initial backstop price in region $i$. 

Let $\widehat{Y}_{t,i}$ denote the output net of climate damage,
mitigation expenditure and adaptation cost, that is, $\widehat{Y}_{t,i}=Y_{t,i}-\Psi_{t,i}-\Upsilon_{t,i}.$

\subsubsection{Economic Interactions between Regions}

In the economic system, each region's stock of capital is the state
variable $K_{t,i}$. Its law of motion is: 
\begin{equation}
K_{t+1,i}=(1-\delta)K_{t,i}+I_{t,i},\label{eq:k-law}
\end{equation}
where $\delta=0.1$ is the depreciation rate and $I_{t,i}$ is investment
in region $i$. We model the economic interactions between two regions
with the following adjustment cost function: 
\begin{equation}
\Gamma_{t,i}\equiv\frac{B}{2}\widehat{Y}_{t,i}\left(\frac{I_{t,i}+c_{t,i}L_{t,i}}{\widehat{Y}_{t,i}}-1\right)^{2},\label{eq:adjust-cost}
\end{equation}
where $B$ is the intensity of the friction, and $c_{t,i}$ is per
capita consumption in region $i$. The open economy situation corresponds
to $B=0$, while a large $B$ approximates the closed economy with
$\widehat{Y}_{t,i}=I_{t,i}+c_{t,i}L_{t,i}$ (note that $\Gamma_{t,i}=0$
could be caused by either the open economy or the closed economy,
so we use $B=0$ and large $B$ to distinguish the two cases). \citet{Anderson2003}
discuss border barriers and how costly they are. Similar adjustment
cost functions have been used in \citet{Goulder2016}. The economic
interaction cost also includes the cost of avoiding carbon leakage
between two regions. \citet{Eaton-Kortum-2002} find that if all countries
(in their data set) collectively remove tariffs, then most countries
will gain around 1\% of output with mobile labor, and less than 0.5\%
with immobile labor. Since we assume mobile labor within each region
but immobile between two regions, we estimate the economic interaction
cost to be roughly 0.5\% of output for each region and use this to
calibrate $B$. 

The market clearing condition with adjustment costs becomes

\begin{equation}
\sum_{i=1}^{2}\left(I_{t,i}+c_{t,i}L_{t,i}+\Gamma_{t,i}\right)=\sum_{i=1}^{2}\widehat{Y}_{t,i}.\label{eq:budget}
\end{equation}

\section{Social Planner versus Nash Equilibrium\label{sec:Social-Planner-Nash}}

We use the Epstein\textendash Zin preference \citep{Epstein-Zin-1989}
to isolate the IES and risk aversion for our stochastic models under
cooperation (a social planner's model) or noncooperation (a dynamic
stochastic game) between the regions. 

\subsection{Social Planner's Model}

With time separable utilities, a social planner maximizes the sum
of discounted expected regional utilities subject to economic and
climatic constraints for time separable utilities. Using the recursive
utility, we use a transformation similar to that in \citet{CaiJuddLontzek_DSICE}
and then solve the Bellman equation:
\begin{equation}
V_{t}^{\mathrm{SP}}(\mathbf{x}_{t})=\max_{\mathbf{a}_{t}}\left\{ \sum_{i=1}^{2}u(c_{t,i})L_{t,i}+\frac{\beta}{1-\frac{1}{\psi}}\left[\mathbb{E}_{t}\left(\left(\left(1-\frac{1}{\psi}\right)V_{t+1}^{\mathrm{SP}}(\mathbf{x}_{t+1})\right)^{\varTheta}\right)\right]^{1/\varTheta}\right\} ,\label{eq:BellmanEq}
\end{equation}
subject to (\ref{eq:M-law}), (\ref{eq:T-law})-(\ref{eq:SLR-law}),
(\ref{eq:law-damLevel})-(\ref{eq:tipIndicator}), (\ref{eq:Y-def-sto})-(\ref{eq:k-law}),
and (\ref{eq:budget}), where 
\begin{equation}
\mathbf{x}_{t}=(K_{t,1},K_{t,2},M_{t}^{\mathrm{AT}},M_{t}^{\mathrm{UO}},M_{t}^{\mathrm{DO}},T_{t,1}^{\mathrm{AT}},T_{t,2}^{\mathrm{AT}},T_{t}^{\mathrm{OC}},S_{t},J_{t},\chi_{t})\label{eq:statevar}
\end{equation}
is the vector of eleven state variables (the first ten variables are
continuous), $\mathbf{a}_{t}=\left(I_{t,1},I_{t,2},c_{t,1},c_{t,2},\mu_{t,1},\mu_{t,2},P_{t,1},P_{t,2}\right)$
is the vector of decision variables (all are continuous), $\mathbb{E}_{t}$
is the expectation operator conditional on the time-$t$ information,
$\beta$ is the discount factor, $u$ is a per capita utility function:
$u(c)=c^{(1-1/\psi)}/(1-1/\psi),$with $\psi$ the IES, and $\varTheta=(1-\gamma)/(1-1/\psi)$
with $\gamma$ the risk aversion parameter. If $\psi\gamma=1$, then
the recursive utility becomes time separable, and (\ref{eq:BellmanEq})
becomes a standard Bellman equation. We use annual time steps, where
the initial year ($t=0$) is 2015, and the terminal time ($t=T=300$)
is the year 2315. The terminal value function $V_{T}^{\mathrm{SP}}(\mathbf{x}_{T})$
is computed as shown in Appendix \ref{sec:Terminal-Value-Function}.
We use $\psi=1.5$ and $\gamma=3.066$ as in \citet{Pindyck_Wang_2013}
for our benchmark stochastic case. We use the parallel dynamic programming
method \citep{Caietal2015_parallel} with simplicial complete Chebyshev
polynomial approximation \citep{CaiJuddLontzek2018_Comp} and time-varying
approximation domains \citep{cai_ARRE2019} on the Blue Waters supercomputer
to solve the social planner's problem. The computational method has
been applied successfully to solve DSICE \citep{cai_NCC_2016,CaiJuddLontzek_DSICE}. 

\subsection{Feedback Nash Equilibrium}

A non-cooperative world can be described as a situation in which each
region acts a separate social planner and chooses emissions paths
to maximize its own welfare subject to the relevant economic and climatic
constraints and by taking emission of the other region as given. For
this dynamic game two solution concepts are studied in the literature,
the open loop Nash equilibrium and the feedback Nash equilibrium (FBNE).
As is well known, the open loop Nash equilibrium does not possess
the Markov perfect property and is not robust against unexpected changes
in the state of the system. Therefore, the FBNE is considered to be
a more satisfactory solution concept.

We derive the FBNE in a dynamic programming framework (e.g. \citealt{BasarOlsder1995}),
so that the controls of each region depend on the states, and the
solution is a Markov perfect non-cooperative regional equilibrium
by construction. As shown in Appendix \ref{sec:No-Transfer-of} in
our regional FBNE there is no transfer of capital between the regions,
so the market clearing condition (\ref{eq:budget}) is changed to
\begin{eqnarray}
I_{t,i}+c_{t,i}L_{t,i} & = & \widehat{Y}_{t,i}\label{eq:budget-compequi}
\end{eqnarray}
for $i=1,2$, and then adjustment costs $\Gamma_{t,i}$ are zero,
so we can rewrite the transition laws of capital as 
\begin{equation}
K_{t+1,i}=(1-\delta)K_{t,i}+\widehat{Y}_{t,i}-c_{t,i}L_{t,i}\label{eq:k-law-1}
\end{equation}
after the substitution of $I_{t.i}$.

The FBNE is solved for a nonlinear dynamic stochastic programming
problem, with recursive preferences, without symmetry, and regional
feedback strategies which depend on a vector of eleven state variables
(ten of which are continuous). We solve the following system of two
Bellman equations: 

\begin{equation}
V_{t,i}^{\mathrm{FBNE}}(\mathbf{x}_{t})=\max_{c_{t,i},P_{t,i},\mu_{t,i}}\left\{ u(c_{t,i})L_{t,i}+\beta\mathcal{G}_{t,i}(\mathbf{x}_{t+1})\right\} ,\label{eq:BellmanCE}
\end{equation}
subject to 
\begin{equation}
\mathbf{x}_{t+1}=\mathbf{f}_{t}(\mathbf{x}_{t},\mathbf{a}_{t}^{\mathrm{FBNE}},\omega_{t})\label{eq:lawCE}
\end{equation}
for $i=1,2$, where $\mathbf{x}_{t}$ is the vector of state variables
defined in (\ref{eq:statevar}), $\mathbf{a}_{t}^{\mathrm{FBNE}}=\left(c_{t,1},c_{t,2},\mu_{t,1},\mu_{t,2},P_{t,1},P_{t,2}\right)$
is the vector of decision variables for the FBNE, 
\begin{equation}
\mathcal{G}_{t,i}(\mathbf{x}_{t+1})\equiv\frac{1}{1-\frac{1}{\psi}}\left[\mathbb{E}_{t}\left(\left(\left(1-\frac{1}{\psi}\right)V_{t+1,i}^{\mathrm{FBNE}}(\mathbf{x}_{t+1})\right)^{\varTheta}\right)\right]^{1/\varTheta},\label{eq:G}
\end{equation}
and $\mathbf{f}_{t}$ represents the vector of the transition laws
of state variables: (\ref{eq:M-law}), (\ref{eq:T-law}), (\ref{eq:SLR-law}),
(\ref{eq:law-damLevel}), (\ref{eq:tipIndicator}), and (\ref{eq:k-law-1})
for $i=1,2$. 

For each $i$, the maximization problem in (\ref{eq:BellmanCE}) has
only three decisions: $c_{t,i},P_{t,i},\mu_{t,i}$. Thus, for an interior
solution,\footnote{If there is a binding constraint, then we just need to add its associated
complementarity condition. } we have the system of first-order conditions (FOCs) of (\ref{eq:BellmanCE})
over $c_{t,i},P_{t,i},\mu_{t,i}$: 
\begin{equation}
\mathbf{FOC}_{t}(\mathbf{x}_{t},\mathbf{a}_{t}^{\mathrm{FBNE}},\mathbf{x}_{t+1})=0\label{eq:FOC}
\end{equation}
which contains six equations provided in Appendix \ref{sec:First-Order-Conditions}.
After we substitute $\mathbf{x}_{t+1}$ by $\mathbf{f}_{t}(\mathbf{x}_{t},\mathbf{a}_{t}^{\mathrm{FBNE}},\omega_{t})$,
for each current-period state $\mathbf{x}_{t}$, the system (\ref{eq:FOC})
has six equations and six unknowns $\mathbf{a}_{t}^{\mathrm{FBNE}}$,
so we can solve it to find a solution. We then use the solution to
compute the FBNE value functions $V_{t,i}^{\mathrm{FBNE}}(\mathbf{x}_{t})=u(c_{t,i})L_{t,i}+\beta\mathcal{G}_{t,i}(\mathbf{x}_{t+1}),$
for $i=1,2$. 

\subsection{Social Cost of Carbon and Carbon Tax}

In \citet{DICE2016}, the regional social cost of carbon (SCC) is
defined to be the present value of future damages in a region caused
by one extra ton of global carbon emissions in the current period.
While the global SCC is equal to the optimal global carbon tax (see
\citet{CaiJuddLontzek_DSICE}), we cannot derive that the optimal
regional carbon tax is equal to the regional SCC, as the global SCC
is the sum of regional SCCs over all regions but the global carbon
tax cannot be the sum of regional carbon taxes. 

In a cooperative world, we follow \citet{vanderPloeg_deZeeuw_2016}
and \citet{CaiJuddLontzek_DSICE} to define the regional cooperative
SCC:
\[
\mathrm{\tau}_{t,i}^{\mathrm{SP}}=-1000\left(\frac{\partial V_{t}^{\mathrm{SP}}}{\partial M_{t}^{\mathrm{AT}}}\right)\left/\left(\frac{\partial V_{t}^{\mathrm{SP}}}{\partial K_{t,i}}\right)\right.
\]
which is equal to the optimal regional carbon tax in the cooperative
world if the emission control rate does not hit its bound \citep{CaiJuddLontzek_DSICE}.
If the economy is open, that is, the adjustment cost of border friction
is zero (i.e., $B=0$), then the regional carbon tax is the same across
regions as the marginal return of capital will be the same across
regions. However, if there is nonzero friction cost between regions
or even the economy is closed, then the regional carbon tax is different
across regions.

In a non-cooperative world, we follow \citet{vanderPloeg_deZeeuw_2016}
and \citet{CaiJuddLontzek_DSICE} to define the regional non-cooperative
SCC:
\[
\mathrm{\tau}_{t,i}^{\mathrm{FBNE}}=-1000\left(\frac{\partial V_{t,i}^{\mathrm{FBNE}}}{\partial M_{t}^{\mathrm{AT}}}\right)\left/\left(\frac{\partial V_{t,i}^{\mathrm{\mathrm{FBNE}}}}{\partial K_{t,i}}\right)\right.
\]
which is equal to the optimal regional carbon tax in FBNE if the emission
control rate is not binding at its bound \citep{CaiJuddLontzek_DSICE}.
Since the economy is closed in the FBNE, the regional carbon tax is
different across regions. 

\section{Computational Method for Solving FBNE\label{sec:Algorithm-FBNE}}

In the literature for solving discrete time FBNE with continuous state
and control variables, there are two popular methods. One is the projection
method \citep{Judd1992,Judd1998}, which is first applied by \citet{RuiMiranda1996}
to solve nonlinear dynamic games. Another is the discretization method,
that is, we discretize state and control variables and then implement
the Pakes-McGuire method \citep{PakesMcGuire1994} or its variants
(e.g., \citealt{Cai_etal_2018_RSUE}). But the projection method is
only for solving infinite-horizon stationary problems, and the discretization
method can only work for low-dimensional problems due to the curse-of-dimensionality. 

In this section we introduce a new time backward iteration method
to solve FBNE in general cases including finite-horizon nonstationary
problems with multiple players $i\in\mathbb{I}$ where $\mathbb{I}$
represent the set of players. In this study, we have two players (one
player per region). The value functions for our problems are continuous
for each discrete state and each player, and almost everywhere differentiable
in the continuous state variables. For each player $i$, we approximate
its value function $V_{i}^{\mathrm{FBNE}}(\mathbf{x})$ using a functional
form $\hat{V}(\mathbf{x};{\bf b}_{i})$ with a finite number of parameters,
${\bf b}_{i}$, where $\mathbf{x}$ is the vector of state variables.
In our study, $\mathbf{x}$ represents the eleven-dimensional vector
defined in (\ref{eq:statevar}). The functional form $\hat{V}$ may
be a linear combination of polynomials, a rational function, or a
neural net. \textcolor{black}{Detailed discussion of approximation
methods can be found in \citet{Judd1998}, \citet{MirandaFackler2002},
and \citet{cai_ARRE2019}.} In this study, we choose simplicial complete
Chebyshev polynomial approximation \citep{CaiJuddLontzek2018_Comp}.
After the functional form is set, at each time $t$, we find $\{{\bf b}_{t,i}:i\in\mathbb{I}\}$
such that $\{\hat{V}_{t,i}(\mathbf{x};{\bf b}_{t,i}):i\in\mathbb{I}\}$
approximately solve the following general Bellman equations simulataneously:
\begin{eqnarray}
V_{t,i}^{\mathrm{FBNE}}(\mathbf{x}) & = & \max_{\mathbf{a}\in{\cal D}(x,t)}\text{ \ }u_{t,i}(\mathbf{x},\mathbf{a})+\beta\mathcal{G}_{t,i}\left(\mathbf{x}_{+}\right),\label{eq:Bellman_FBNE_general}\\
 &  & \text{{\rm \ s.t. \ \ }}\mathbf{x}_{+}=\mathbf{f}_{t}(\mathbf{x},\mathbf{a},\omega),\nonumber 
\end{eqnarray}
for all $i\in\mathbb{I}$ such that $\hat{V}_{t,i}(\mathbf{x};{\bf b}_{t,i})\approx V_{t,i}^{\mathrm{FBNE}}(\mathbf{x})$.
Here $V_{t,i}^{\mathrm{FBNE}}(\mathbf{x})$ is the value function
of player $i$ at time $t\leq T$ (the terminal value function $V_{T,i}^{\mathrm{FBNE}}(\mathbf{x})$
is given), $\mathbf{x}_{+}$ is the next-stage state, ${\cal D}(\mathbf{x},t)$
is a feasible set of $\mathbf{a}$, $\omega$ is a random variable
vector, $\mathbf{f}_{t}$ is the vector of the transition laws of
$\mathbf{x}$, $\beta$ is a discount factor and $u_{t,i}(\mathbf{x},\mathbf{a})$
is the utility function of player $i$ (in this study, $u_{t,i}(\mathbf{x},\mathbf{a})=u(c_{t,i})L_{t,i}$),
and $\mathcal{G}_{t,i}$ is a function generated from $V_{t+1,i}$,
which can be written as $\mathcal{G}_{t,i}\equiv\mathcal{H}_{t}\left(V_{t+1,i}^{\mathrm{FBNE}}\right)$
where $\mathcal{H}_{t}$ is a functional operator. A typical functional
operator is $\mathbb{E}_{t}$, the expectation operator conditional
on time-$t$ information. In this study, $\mathcal{H}_{t}$ is defined
in a way such that $\mathcal{G}_{t,i}$ has the form defined in (\ref{eq:G}).

To solve the system of maximization problems (\ref{eq:Bellman_FBNE_general}),
we transform them to solve a system of FOCs (for an interior solution):
\begin{equation}
\nabla_{\mathbf{a}}\left(u_{t,i}(\mathbf{x},\mathbf{a})+\beta\mathcal{G}_{t,i}\left(\mathbf{f}_{t}(\mathbf{x},\mathbf{a},\omega)\right)\right)=0,\ i\in\mathbb{I},\label{eq:FOC_general}
\end{equation}
where $\nabla_{\mathbf{a}}$ is the gradient vector over $\mathbf{a}$.
In our study, it corresponds to the system (\ref{eq:FOC}) after the
next-period state is substituted by the transition functions according
to (\ref{eq:lawCE}). However, it is often challenging to find a solution
of the system (\ref{eq:FOC_general}), in particular, when it includes
complementarity conditions for a solution binding at an inequality
constraint, or when $\mathbf{x}$ is near to or on the border of approximation
domains. Thus, we instead solve the following minimization model 
\begin{eqnarray}
\min &  & \sum_{i\in\mathbb{I}}\left\Vert \mathbf{\epsilon}_{i}\right\Vert ,\label{eq:minFeas-1}\\
\mathrm{s.t.} &  & \nabla_{\mathbf{a}}\left(u_{t,i}(\mathbf{x},\mathbf{a})+\beta\mathcal{G}_{t,i}\left(\mathbf{f}_{t}(\mathbf{x},\mathbf{a},\omega)\right)\right)=\mathbf{\epsilon}_{i},\ i\in\mathbb{I},\nonumber \\
 &  & \mathbf{a}\in\mathcal{D}(\mathbf{x},t),\nonumber \\
 &  & \left\Vert \mathbf{\epsilon}_{i}\right\Vert \leq\bar{\epsilon},\ i\in\mathbb{I},\nonumber 
\end{eqnarray}
where $\mathbf{\epsilon}_{i}$ is a vector, $\bar{\epsilon}$ is a
given small positive number for guaranteeing that the left hand side
of (\ref{eq:FOC_general}) is close to zero, and $\left\Vert \cdot\right\Vert $
is a norm. If $\left\Vert \cdot\right\Vert $ is the $\mathcal{L}^{1}$
norm, then there are kinks which create challenges for optimization.
Appendix \ref{sec:L1-fitting} describes computational techniques
to overcome the kink problem.

The following is our new algorithm with time backward iteration for
solving FBNE of finite horizon problems.
\begin{description}
\item [{Algorithm}] 1. \textit{Time Backward Iteration for Solving FBNE}
\begin{doublespace}
\item [{\textit{Initialization}\textmd{\textit{.}}}] \textit{Choose the
time-varying approximation nodes, $\mathbb{X}_{t}=\{\mathbf{x}_{t,j}:$
$1\leq j\leq m_{t}\}$ for every $t<T$, and choose a functional form
for $\hat{V}(\mathbf{x};{\bf b})$. Let $\hat{V}_{T,i}(\mathbf{x};{\bf b}_{T,i})\equiv V_{T,i}^{\mathrm{FBNE}}(\mathbf{x})$.
Then for $t=T-1,T-2,\ldots,0$, iterate through steps 1 and 2. }
\item [{\textit{Step}}] \textbf{\textit{1.}}\textit{ Minimization step.
For each approximation node} \textit{$\mathbf{x}_{t,j}$, solve (\ref{eq:minFeas-1})
to find the optimal solution $\mathbf{a}^{*}$ and compute $v_{i,j}=u_{t,i}(\mathbf{x}_{t,j},\mathbf{a}^{*})+\beta\mathcal{G}_{t,i}\left(\mathbf{f}_{t}(\mathbf{x}_{t,j},\mathbf{a}^{*},\omega)\right)$
for each $i\in\mathbb{I}$. Here $\mathcal{G}_{t,i}\equiv\mathcal{H}_{t}\left(\hat{V}_{t+1,i}\right)$. }
\item [{\textit{Step}}] \textbf{\textit{2.}}\textit{ Fitting step. Using
an appropriate approximation method, compute the ${\bf b}_{t,i}$
such that $\hat{V}_{t,i}(\mathbf{x};{\bf b}_{t,i})$ approximates
$\{(\mathbf{x}_{t,j},v_{i,j}):1\leq j\leq m_{t}\}$, i.e., $\hat{V}_{t,i}(\mathbf{x}_{t,j};{\bf b}_{t,i})\approx v_{i,j}$,
for every $i\in\mathbb{I}$ and $1\leq j\leq m_{t}.$ }
\end{doublespace}
\end{description}
Note that the minimization step can be easily parallelized over \textit{$i\in\mathbb{I}$
and $1\leq j\leq m_{t}.$} Thus, we can also use the Master-Worker
framework of the parallel dynamic programming in \citet{Caietal2015_parallel}
to implement parallism on Algorithm 1, and can reach the similar parallelism
efficiency. In this study, the terminal value functions $V_{T,i}^{\mathrm{FBNE}}(\mathbf{x}_{T})$
are computed as shown in Appendix \ref{sec:Terminal-Value-Function}.
Since the state space is high-dimensional, we require a large number
of approximation nodes $\{\mathbf{x}_{t,j}\}$, so we implement parallelism
across $\mathbf{x}_{t,j}$ and $i\in\mathbb{I}$ on the Blue Waters
supercomputer to find FBNE. 

While Algorithm 1 solves finite-horizon problems, it is simple to
adapt it to solve infinite-horizon stationary problems, in which we
use time-invariant approximation domains and nodes. We can just replace
the $T$ iterations by the unlimited number of iterations with the
following stopping criterion: the difference between two consecutive
value functions for every player is small enough. See \citet{cai_ARRE2019}
for a discussion about stopping criteria. 

To validate Algorithm 1, we first implement it (after being adapted
for infinite-horizon stationary problems) to solve a linear-quadratic
differential game given in \citet{Engwerda2009}: its example 4.6.
The game has three FBNEs. We first discretize time to change it to
be a discrete time problem and then implement our algorithm. We find
that it converges to one of three FBNEs. We also employ Algorithm
1 (after being adapted for infinite-horizon stationary problems) to
solve the dynamic stochastic games in international commodity markets
in \citet{RuiMiranda1996}, and find that it provides solutions close
to those in \citet{RuiMiranda1996}. Thus, we see that our time backward
iteration can find FBNE, at least one if it exists. In our study,
it might have only a unique FBNE for each case although we cannot
prove it in theory. 

\section{Results \label{sec:res-sto} }

We solve the Bellman equations (\ref{eq:BellmanEq}) and (\ref{eq:BellmanCE})
with $\psi=1.5$ and $\gamma=3.066$ as our default utility preference
values. After we solve the Bellman equations, we use the optimal policy
functions to generate 10,000 simulation paths forward. That is, each
simulation path starts at the observed initial states, we simulate
one sample of the shock for the tipping point at time $t$, and then
with the realized sample and the optimal control policy at $t$, we
obtain the optimal states at $t+1$. 

\begin{figure}
\begin{centering}
\begin{tabular}{cc}
\includegraphics[width=0.45\textwidth,height=0.21\textheight]{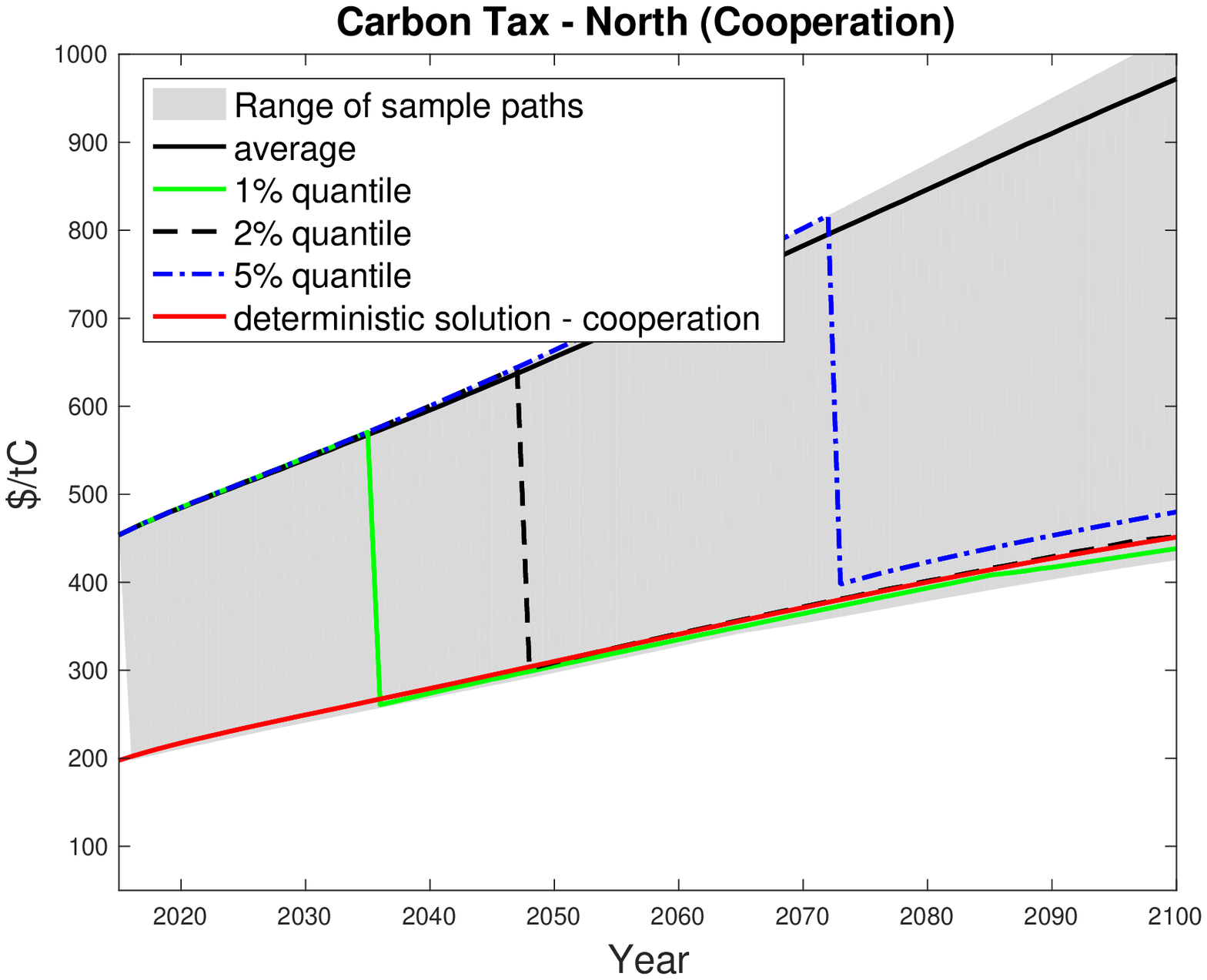} & \includegraphics[width=0.45\textwidth,height=0.21\textheight]{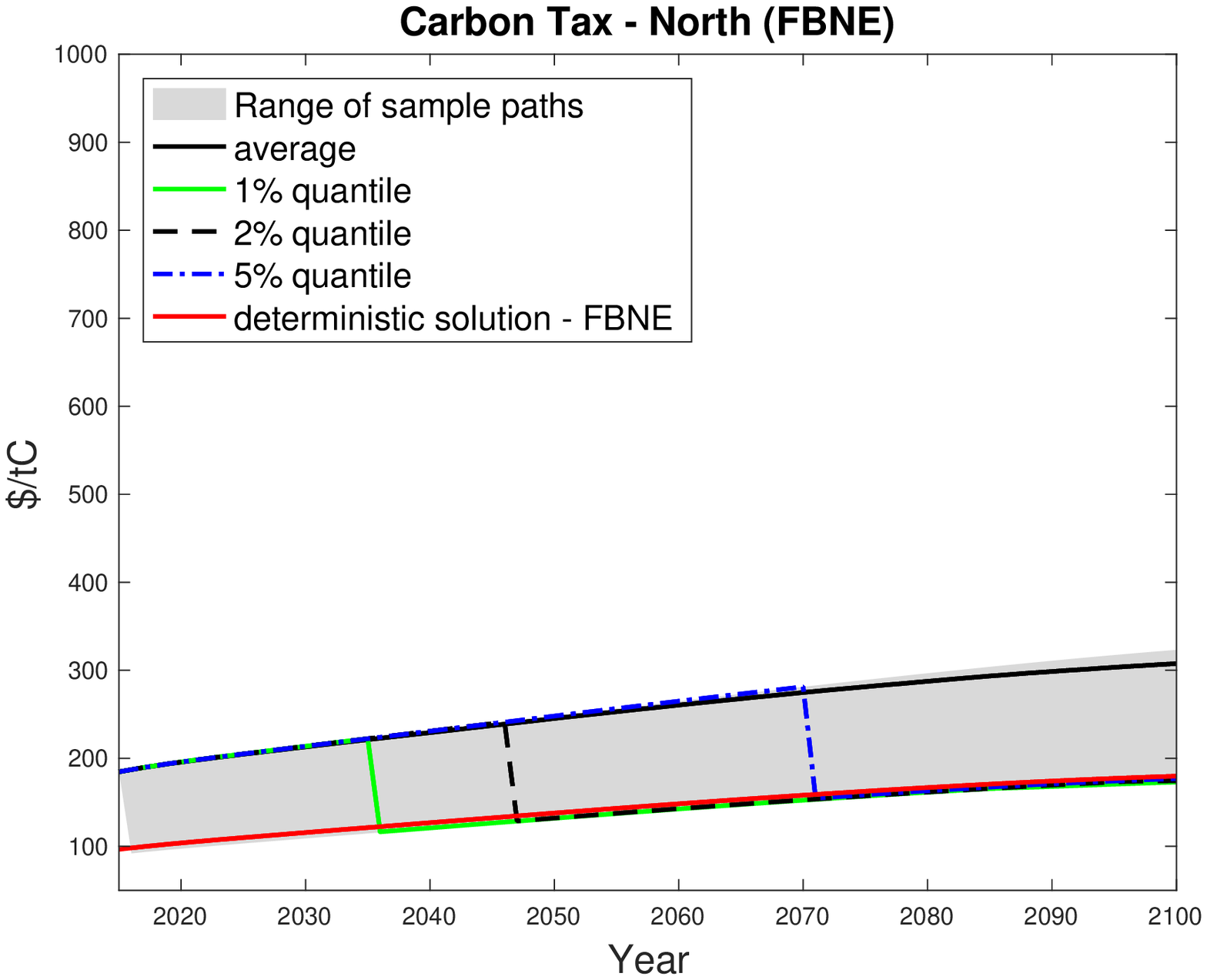}\tabularnewline
\includegraphics[width=0.45\textwidth,height=0.21\textheight]{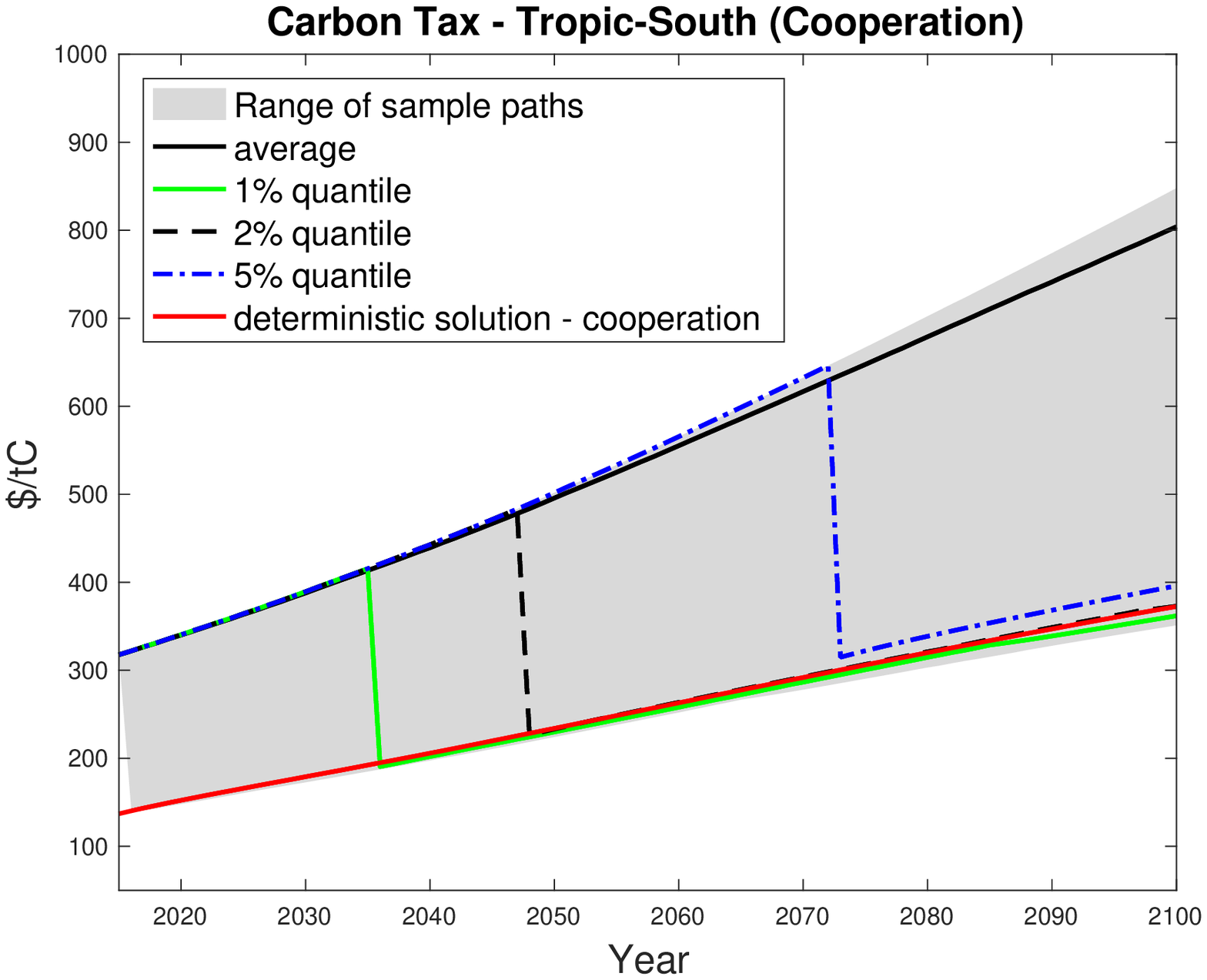} & \includegraphics[width=0.45\textwidth,height=0.21\textheight]{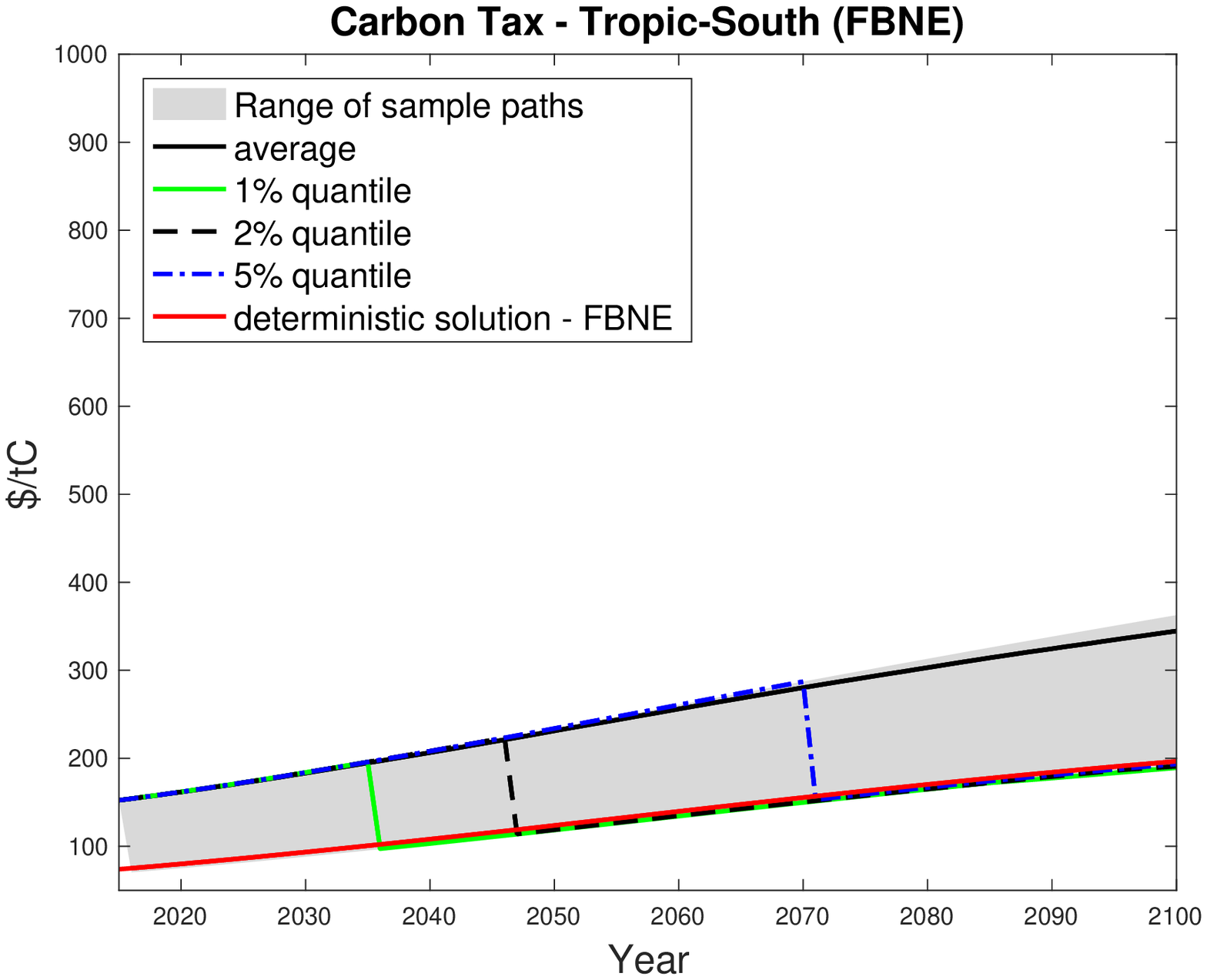}\tabularnewline
\end{tabular}
\par\end{centering}
\caption{Regional Carbon Tax\label{fig:tax-sto}}
\end{figure}

The two left panels of Figure \ref{fig:tax-sto} show the distributions
of the simulated regional carbon taxes for both regions under the
social planner's model, and the two right panels of Figure \ref{fig:tax-sto}
are for the feedback Nash equilibrium. All the panels in Figure \ref{fig:tax-sto},
as well as in the following figures, include a line representing the
deterministic case derived with the same IES used in the stochastic
model. The shaded area represents the range of the 10,000 sample paths,
along with the average, 1\%, 2\% and 5\% quantile paths (that is,
at each point in time, we compute the average and these quantiles
of 10,000 values). The initial regional carbon tax increases significantly
from the deterministic case to the stochastic case under either cooperation
or non-cooperation, and the FBNE solution has significantly less regional
carbon taxes than the cooperative solution. In the social planner's
problem, the initial regional carbon tax for the stochastic case is
\$454/tC for the North and \$318/tC for the Tropic-South, about 2.3
times that of the corresponding deterministic case (with $\text{\ensuremath{\psi}}=1.5$)
with cooperation in each region. In the FBNE, the initial regional
carbon tax for the stochastic case is \$185/tC for the North and \$152/tC
for the Tropic-South, more than double of that of the corresponding
deterministic case of FBNE in each region, but less than a half of
the initial regional carbon tax of the social planner's stochastic
case in each region. 

\begin{figure}
\begin{centering}
\begin{tabular}{cc}
\includegraphics[width=0.45\textwidth,height=0.21\textheight]{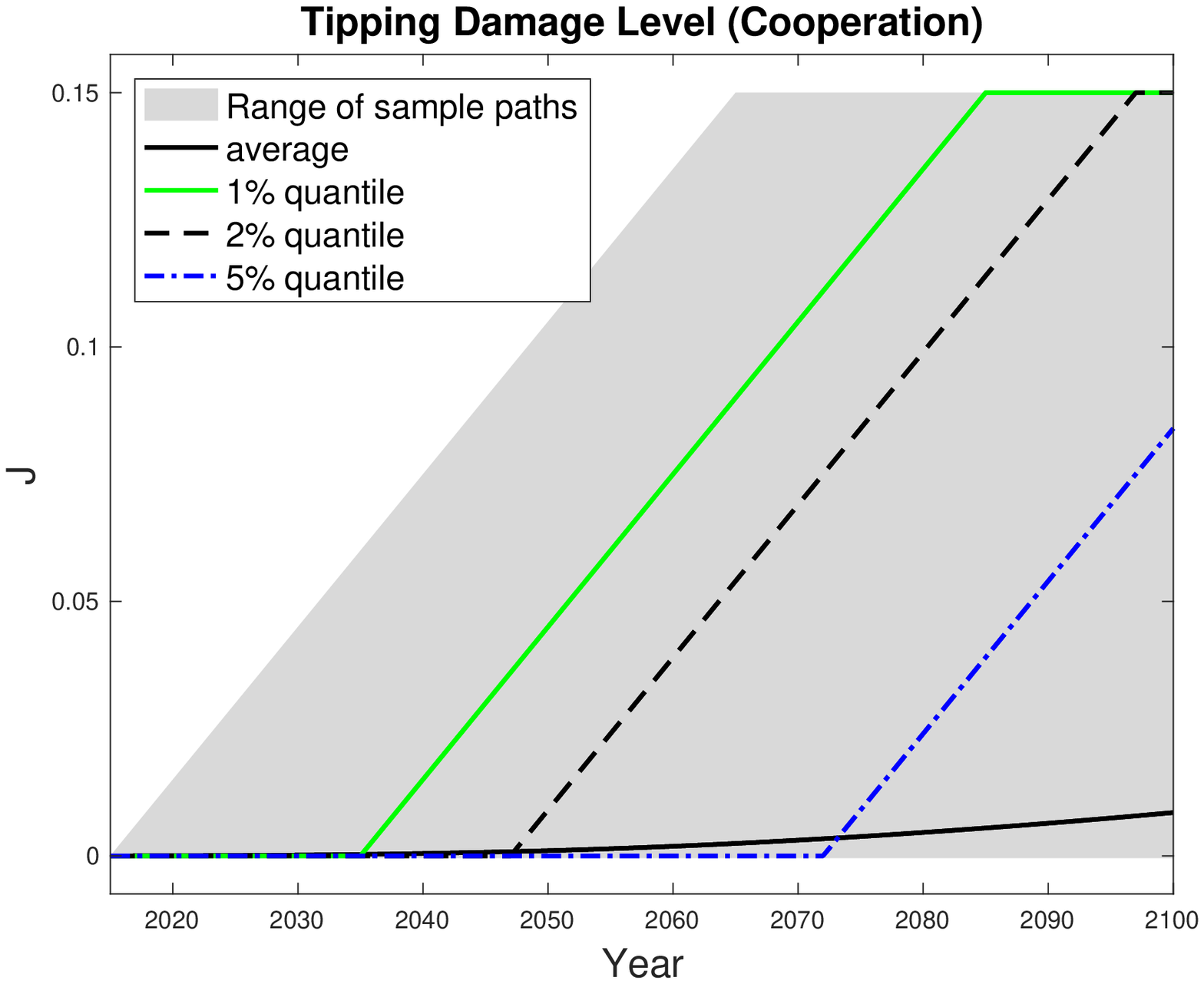} & \includegraphics[width=0.45\textwidth,height=0.21\textheight]{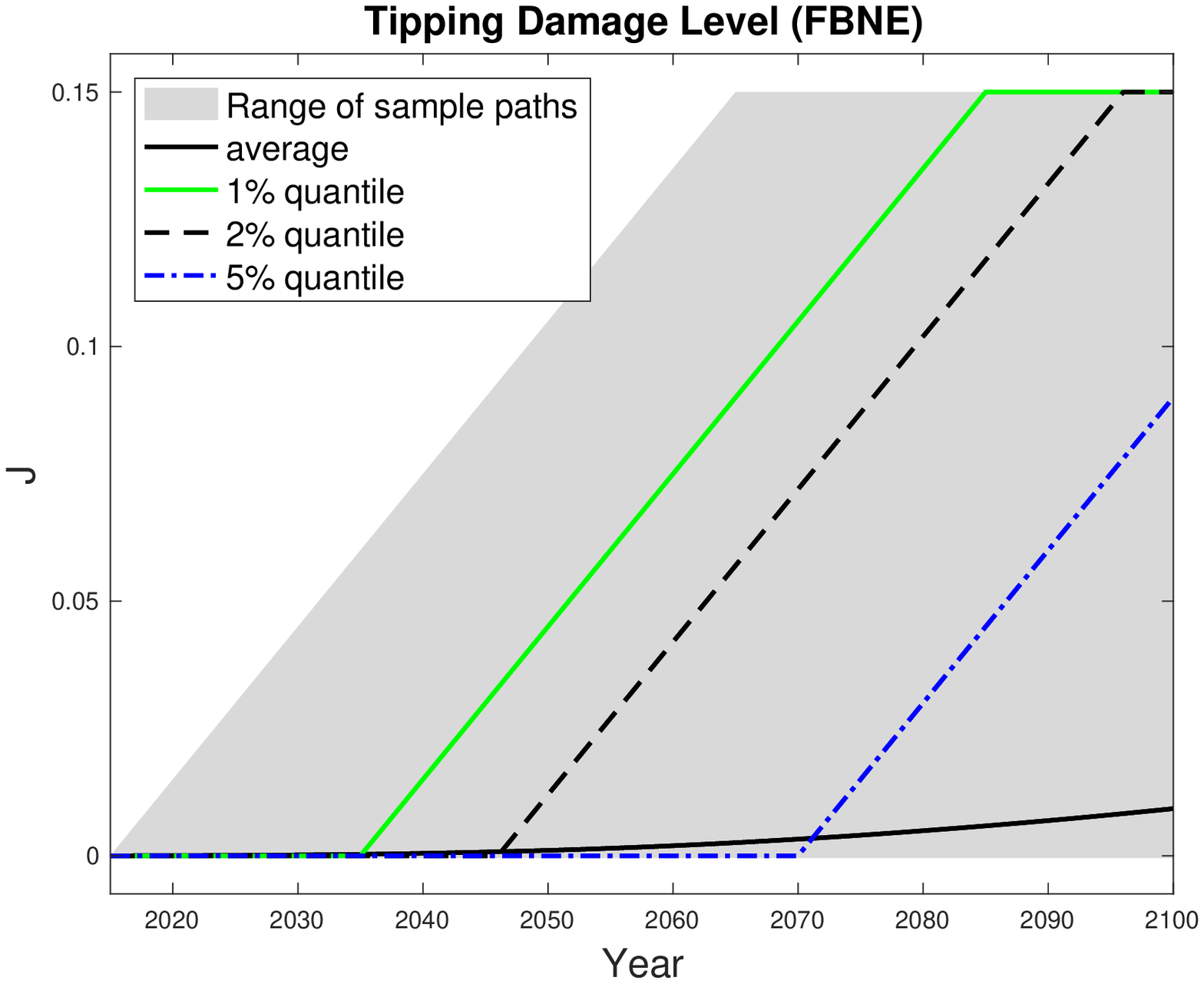}\tabularnewline
\includegraphics[width=0.45\textwidth,height=0.21\textheight]{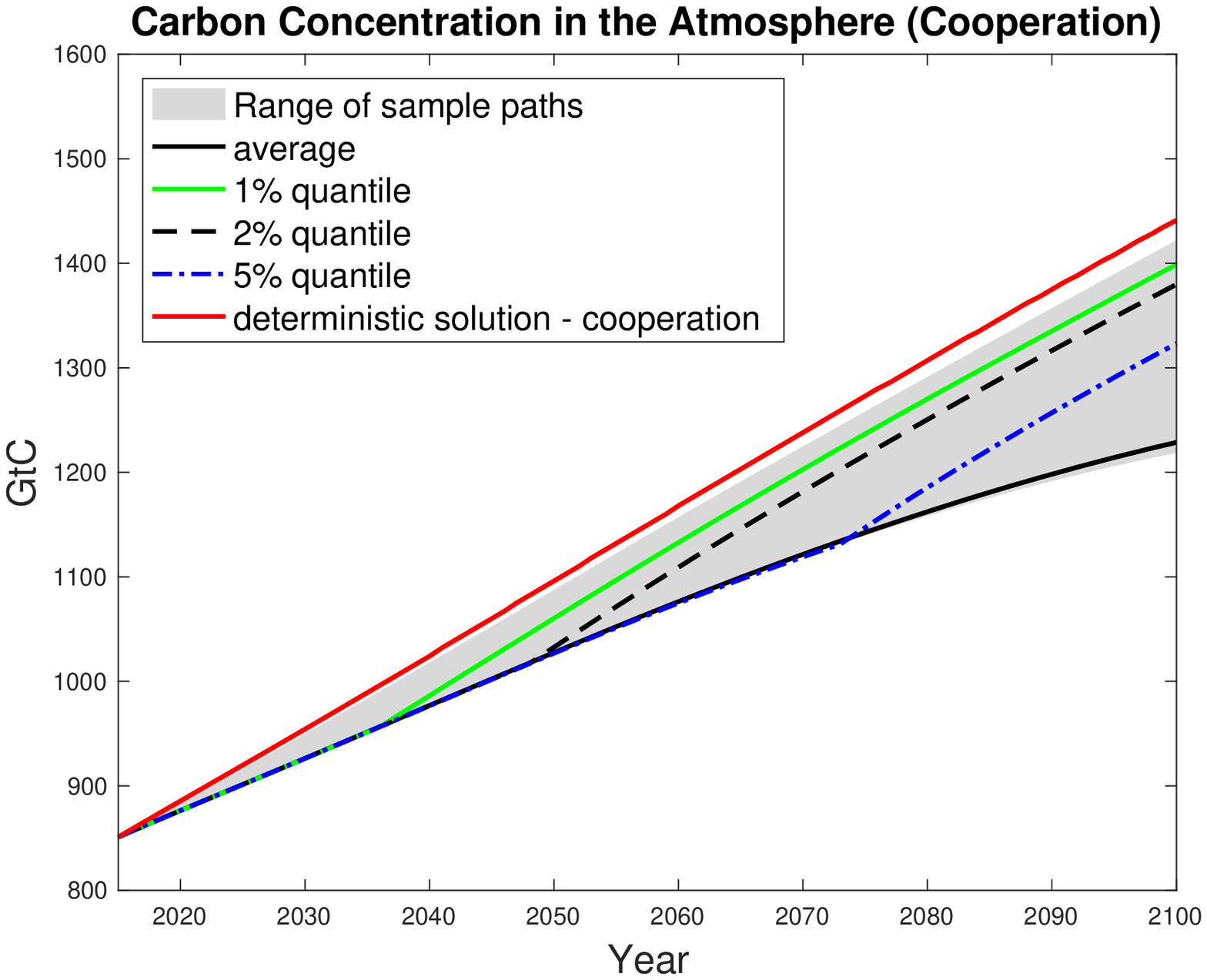} & \includegraphics[width=0.45\textwidth,height=0.21\textheight]{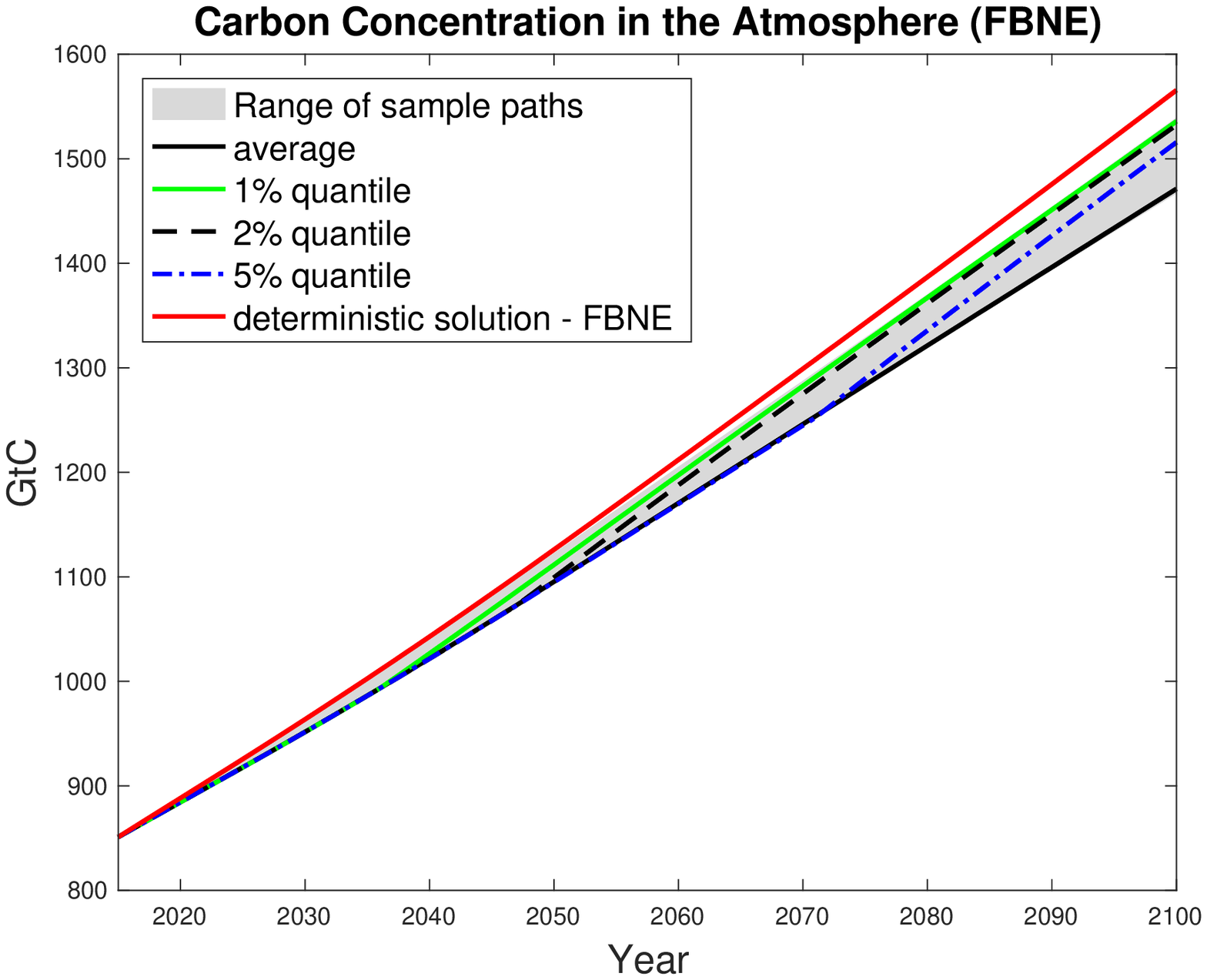}\tabularnewline
\end{tabular}
\par\end{centering}
\caption{Tipping damage levels and atmospheric carbon concentration \label{fig:tiplevel-CA}}
\end{figure}

Figure \ref{fig:tiplevel-CA} shows the distributions of the optimal
simulation paths for tipping damage levels $J_{t}$ and atmospheric
carbon concentration under cooperation (the left panels) and non-cooperation
(the right panels). For the social planner's problem, the cumulative
probability that the tipping event will occur before 2100 is only
9.6\%, while the 1\%, 2\% and 5\% cumulative probability of tipping
occurs in years 2035, 2047, and 2072 respectively. For the FBNE, the
cumulative probability that the tipping event will occur before 2100
increases to 10.7\%, and the 2\% or 5\% cumulative probability of
tipping occurs at an earlier year. Once the tipping event happens,
the regional carbon tax immediately falls significantly, but damages
unfold over a 50-year period, as shown in the top panels for the tipping
damage level $J_{t}$, under both cooperation and non-cooperation.
This happens because the high carbon tax before tipping is intended
to prevent or delay the tipping point as its occurrence depends on
the contemporaneous temperature. However, after the tipping event
happens, this incentive disappears as the damage will unfold in a
deterministic way. This result is consistent with the finding in DSICE.
The bottom panels show that, with the stricter mitigation policy,
the stochastic model has smaller carbon concentration in the atmosphere
and in 2100 it is on average 200 GtC less than the corresponding deterministic
simulation (with $\text{\ensuremath{\psi}}=1.5$) for the cooperative
case, and 100 GtC less than the corresponding deterministic FNBE simulation
for the noncooperative case. If the tipping event occurs, then the
carbon concentration has a higher rate of increase as the corresponding
mitigation policy is less strict. 

\begin{figure}
\begin{centering}
\begin{tabular}{cc}
\includegraphics[width=0.45\textwidth,height=0.21\textheight]{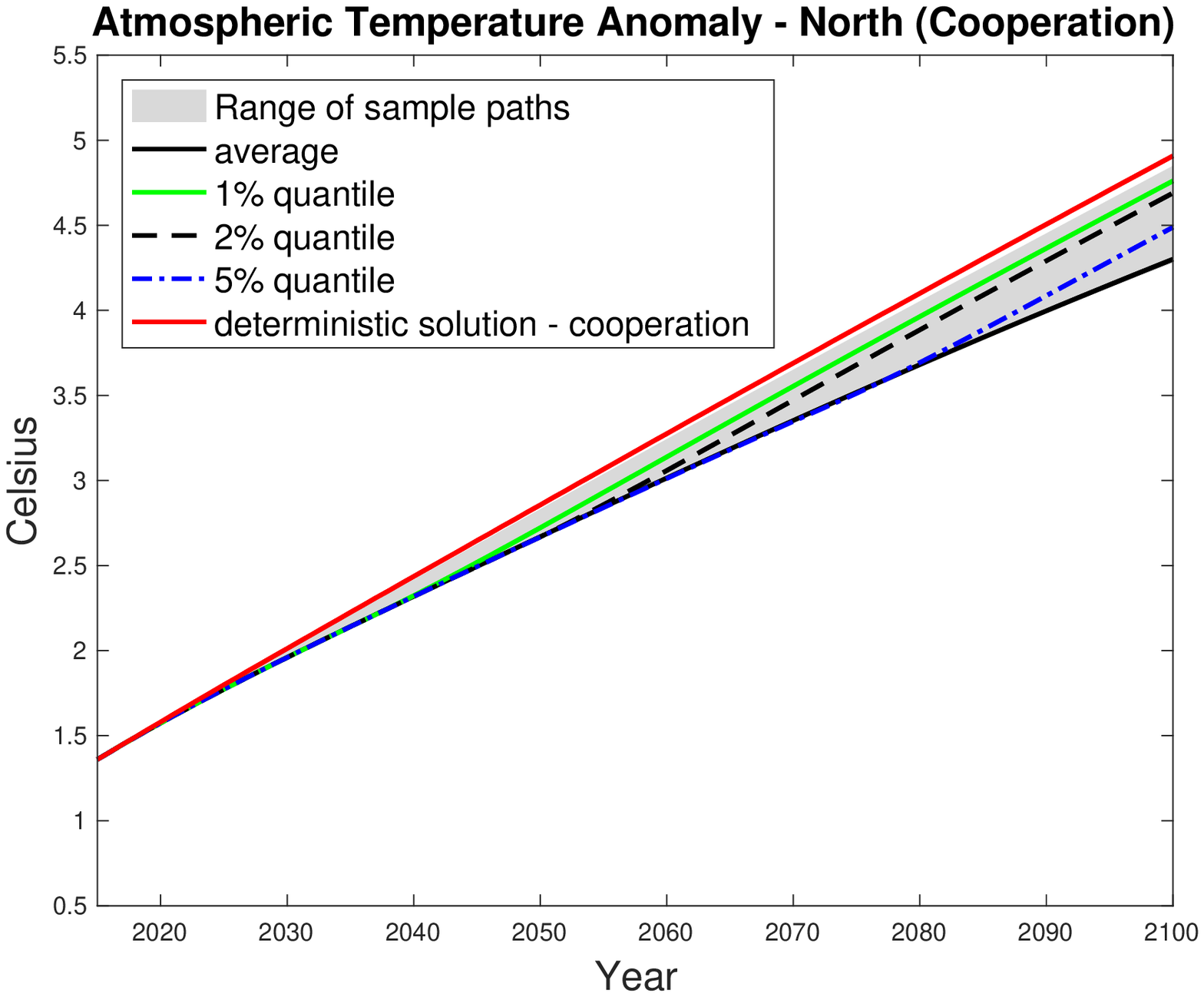} & \includegraphics[width=0.45\textwidth,height=0.21\textheight]{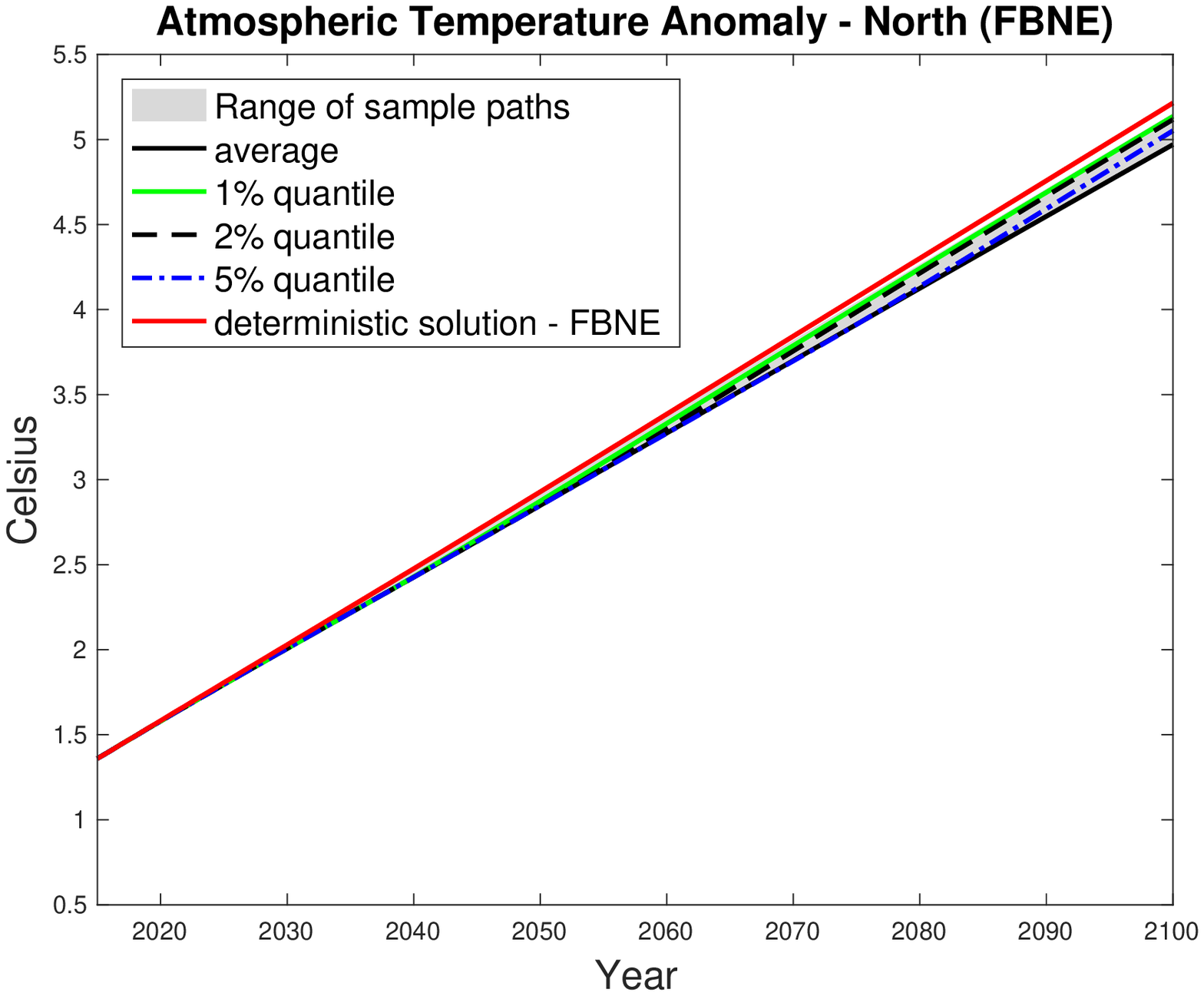}\tabularnewline
\includegraphics[width=0.45\textwidth,height=0.21\textheight]{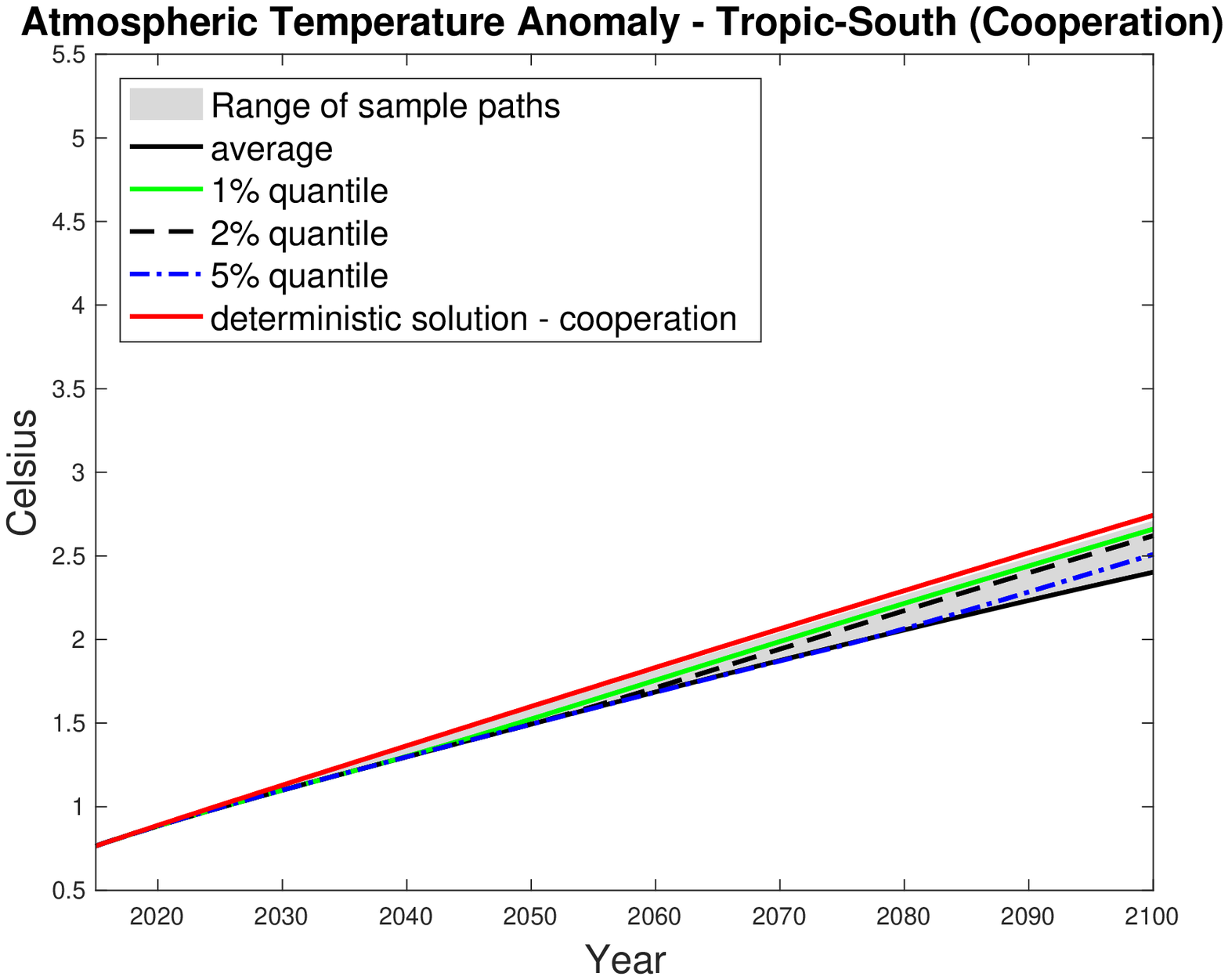} & \includegraphics[width=0.45\textwidth,height=0.21\textheight]{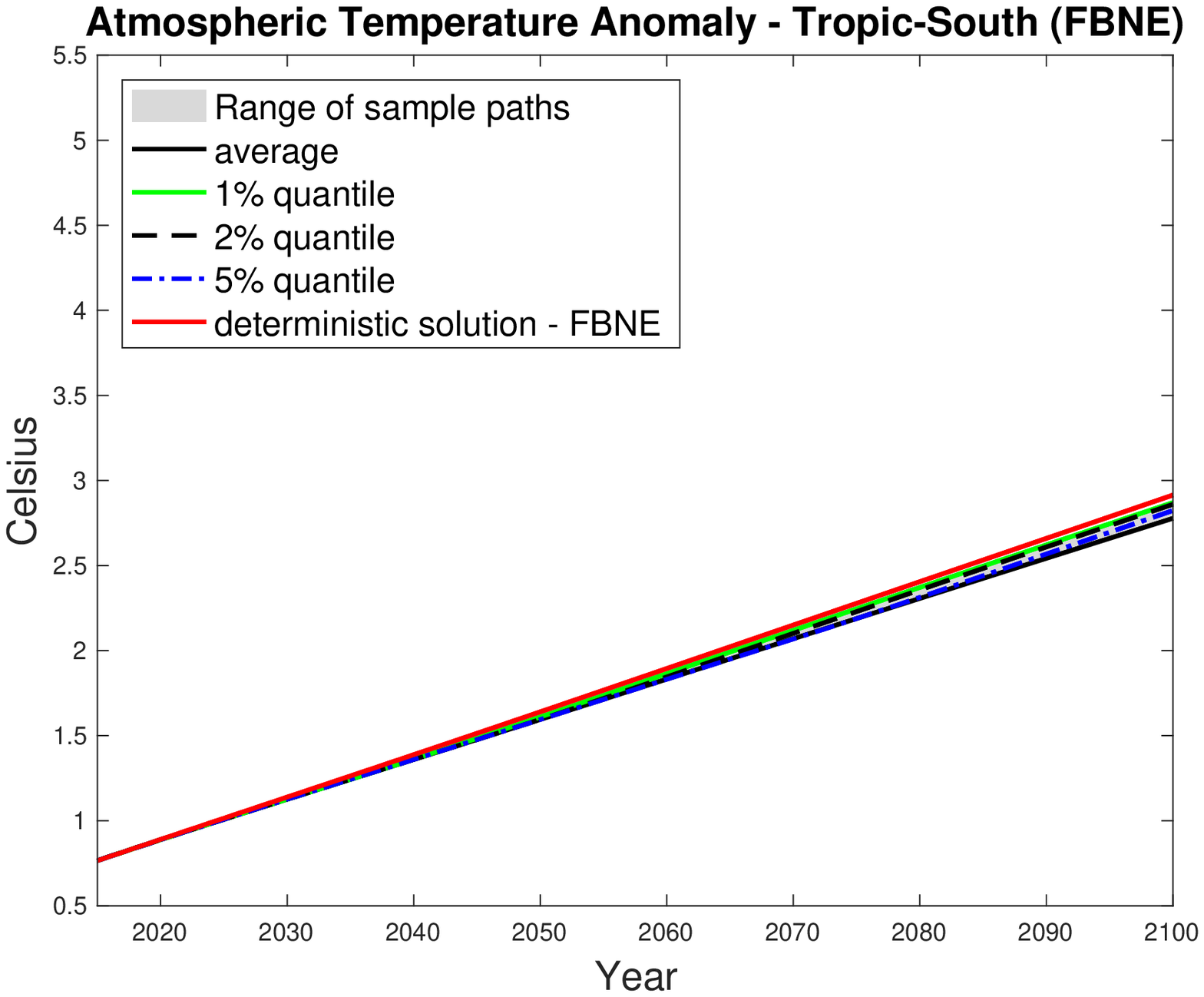}\tabularnewline
\end{tabular}
\par\end{centering}
\caption{Surface Temperature\label{fig:TA}}
\end{figure}

Figure \ref{fig:TA} shows the distributions of the optimal simulation
paths for atmospheric temperatures in both regions for the cooperative
(the left panels) and non-cooperative (the right panels) cases. We
see that the North has a much higher temperature anomaly than the
Tropic-South. Moreover, the atmospheric temperature anomaly in the
North is more than 2\textcelsius{} higher than in the Tropic-South
in 2100. The non-cooperative case has higher temperatures than the
cooperative case in both regions, but narrower ranges in 2100. 

\begin{figure}
\begin{centering}
\begin{tabular}{cc}
\includegraphics[width=0.45\textwidth,height=0.21\textheight]{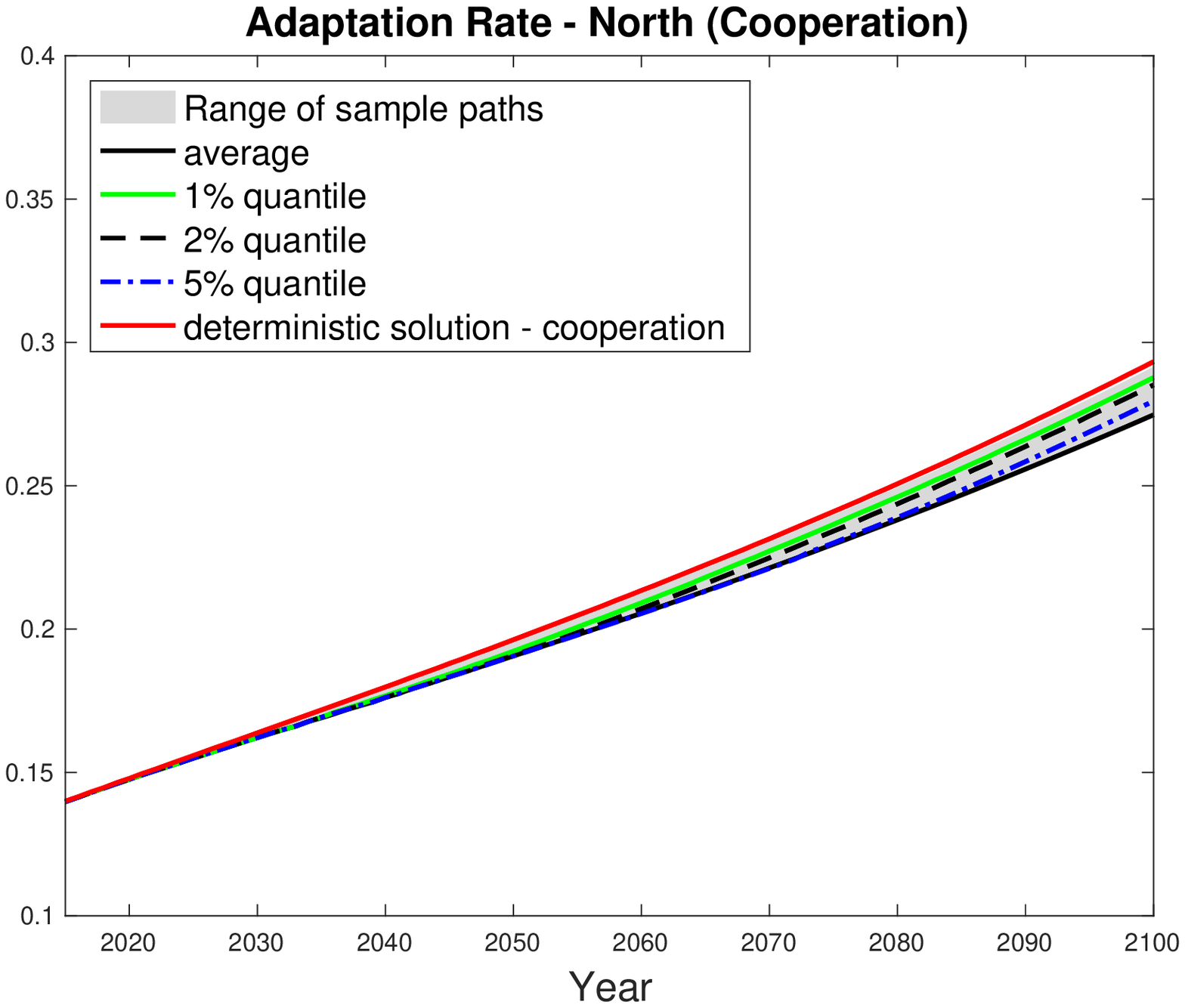} & \includegraphics[width=0.45\textwidth,height=0.21\textheight]{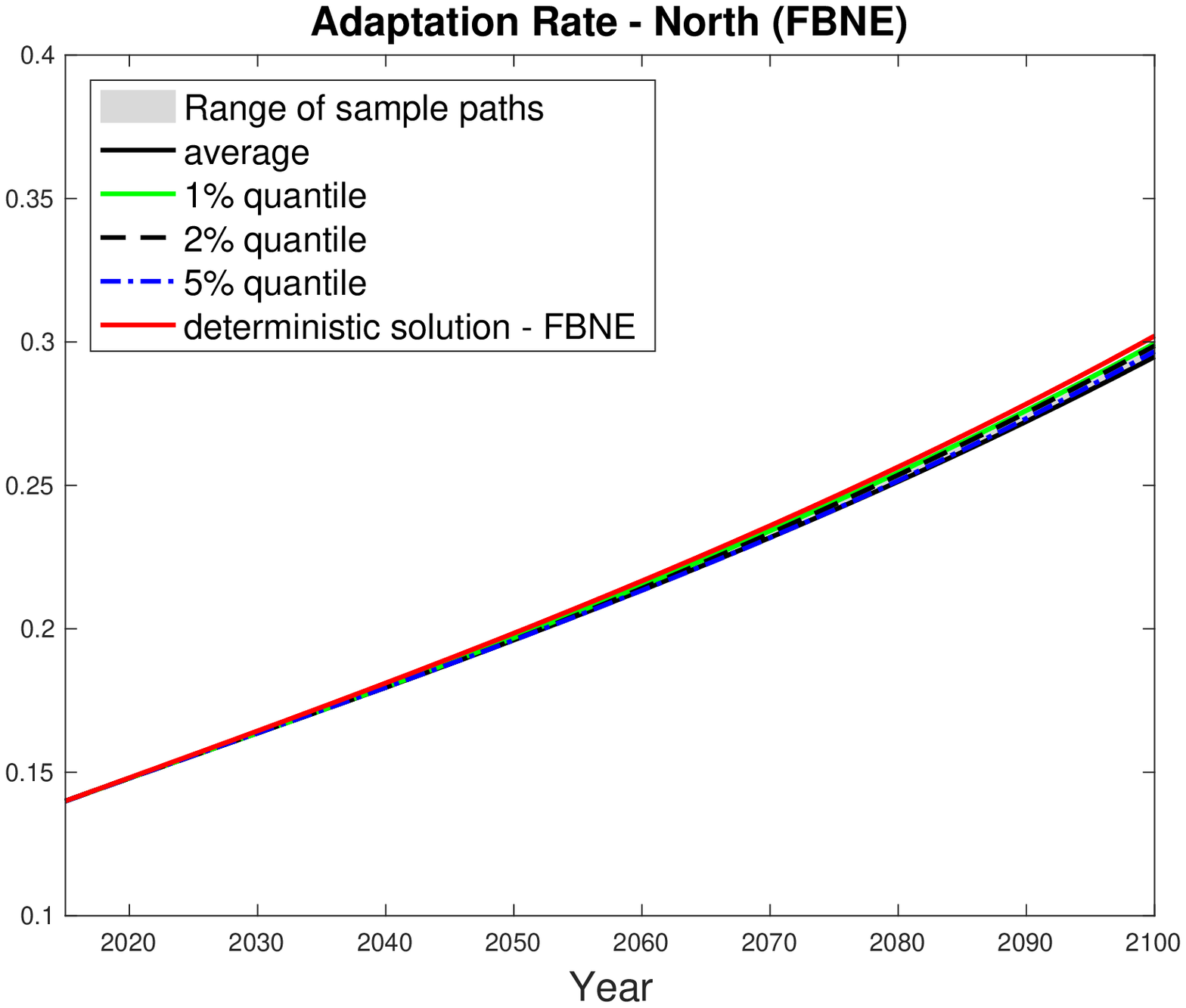}\tabularnewline
\includegraphics[width=0.45\textwidth,height=0.21\textheight]{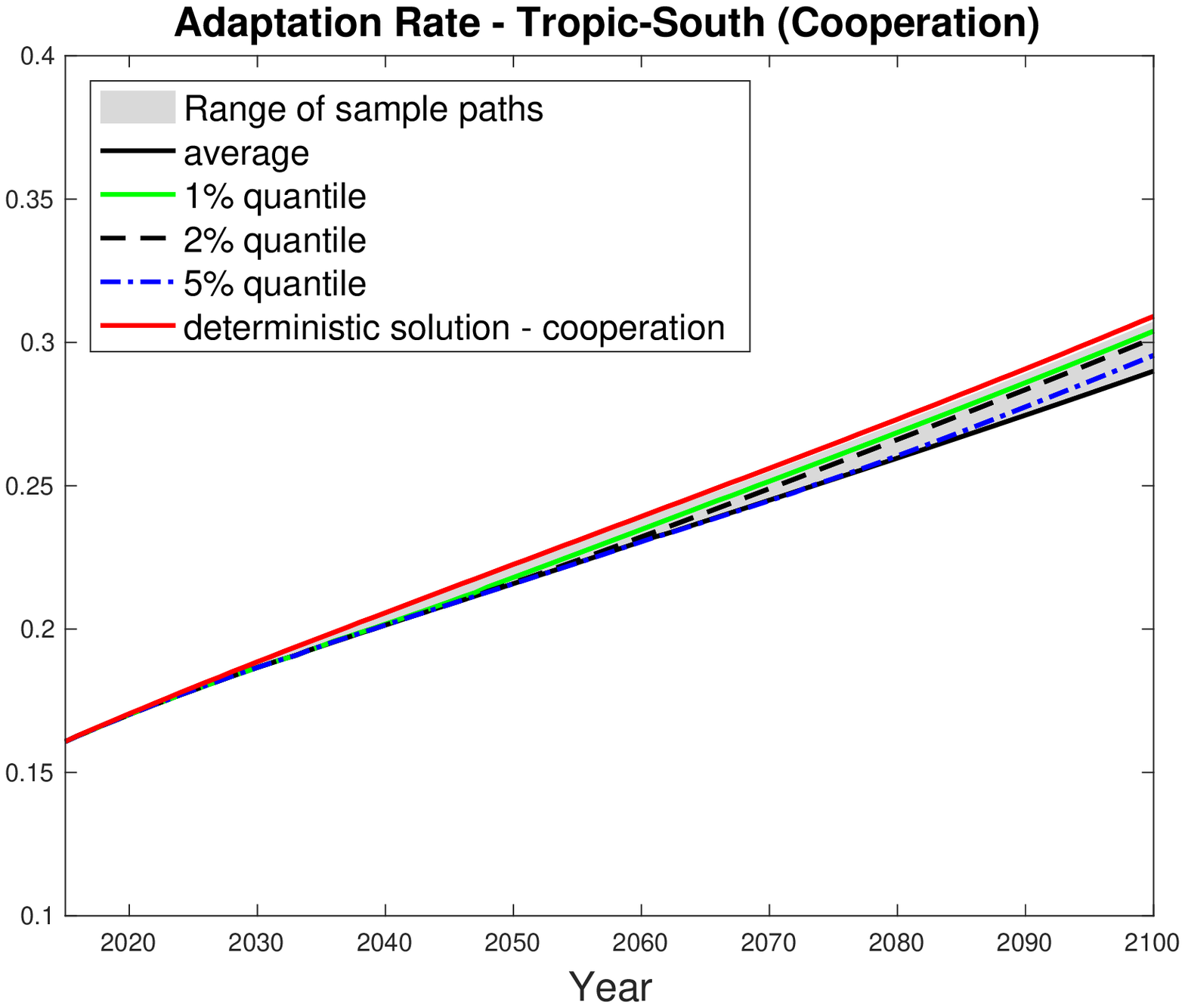} & \includegraphics[width=0.45\textwidth,height=0.21\textheight]{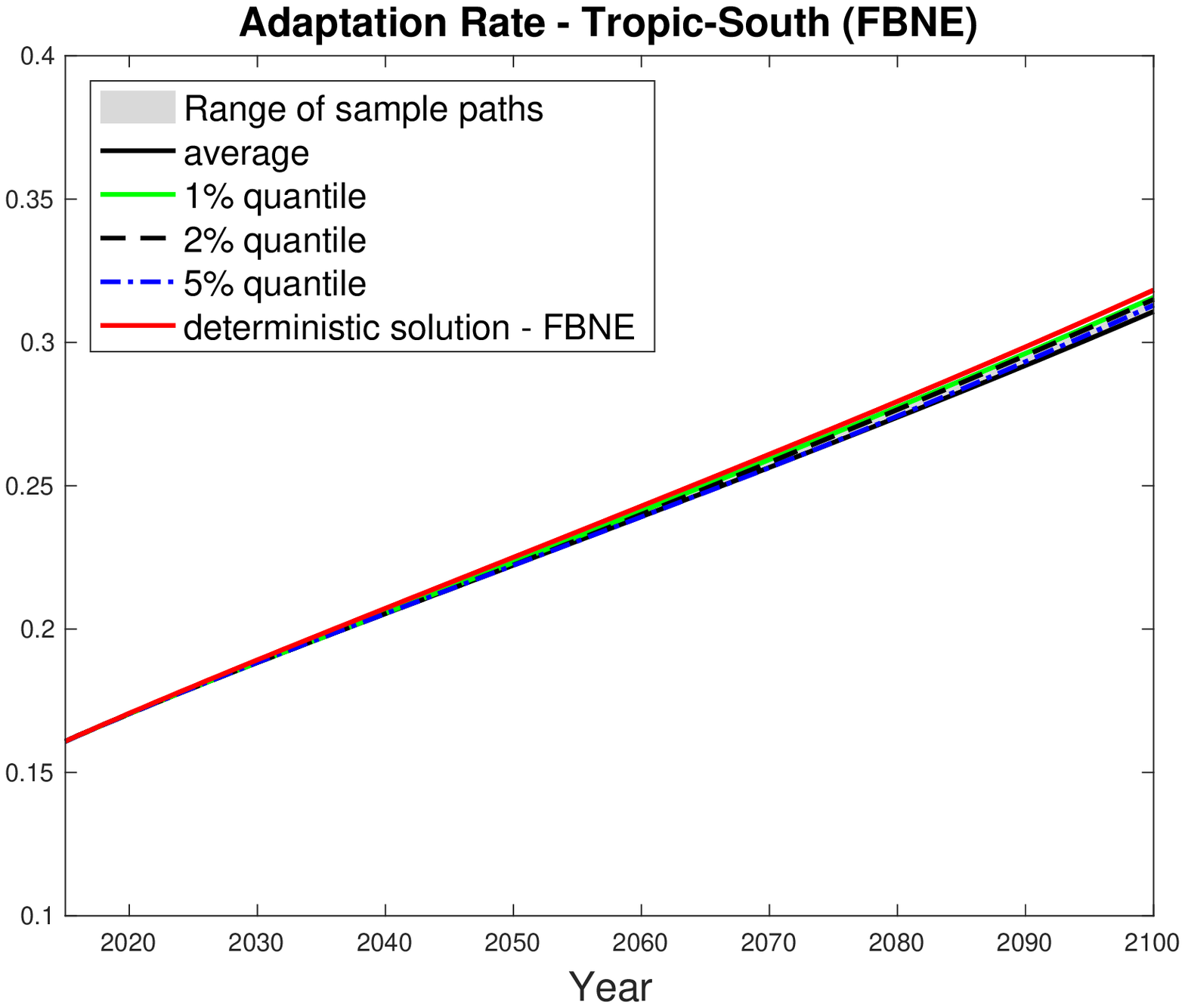}\tabularnewline
\end{tabular}
\par\end{centering}
\caption{Optimal Adaptation \label{fig:adapt}}
\end{figure}

Figure \ref{fig:adapt} shows the optimal adaptation rates in both
regions for the cooperative (the left panels) and non-cooperative
(the right panels) cases. We see that the North has lower adaptation
rates than the Tropic-South, and that the stochastic results have
lower adaptation rates than the corresponding deterministic case,
since with the stricter mitigation policy and the resulting lower
temperatures in the stochastic case, there is less need to adapt.
The non-cooperative case has higher adaptation rates than the cooperative
case in both regions, but narrower ranges in 2100. This adaptation
policy has a direction between the regions opposite to the carbon
tax policy, because the optimal regional adaptation rate is a solution
depending only on the regional damage function, the regional mitigation
cost function, and the regional adaptation cost function. This can
be seen by the FBNE's first-order condition (\ref{eq:FOC_P}) and
(\ref{eq:dY_dP}), which also hold for the social planner's problem.
That is, adaptation has no direct impact on the common good, carbon
concentration. Thus, the cooperation and non-cooperation have no direct
impact on adaptation. But since the FBNE has lower carbon tax, which
leads to higher temperature and then higher damage, the optimal adaptation
is then higher. 

\subsection{Bias from Ignoring Heat Transport and PA\label{subsec:Bias-sto}}

\begin{figure}
\begin{centering}
\begin{tabular}{cc}
\includegraphics[width=0.45\textwidth,height=0.21\textheight]{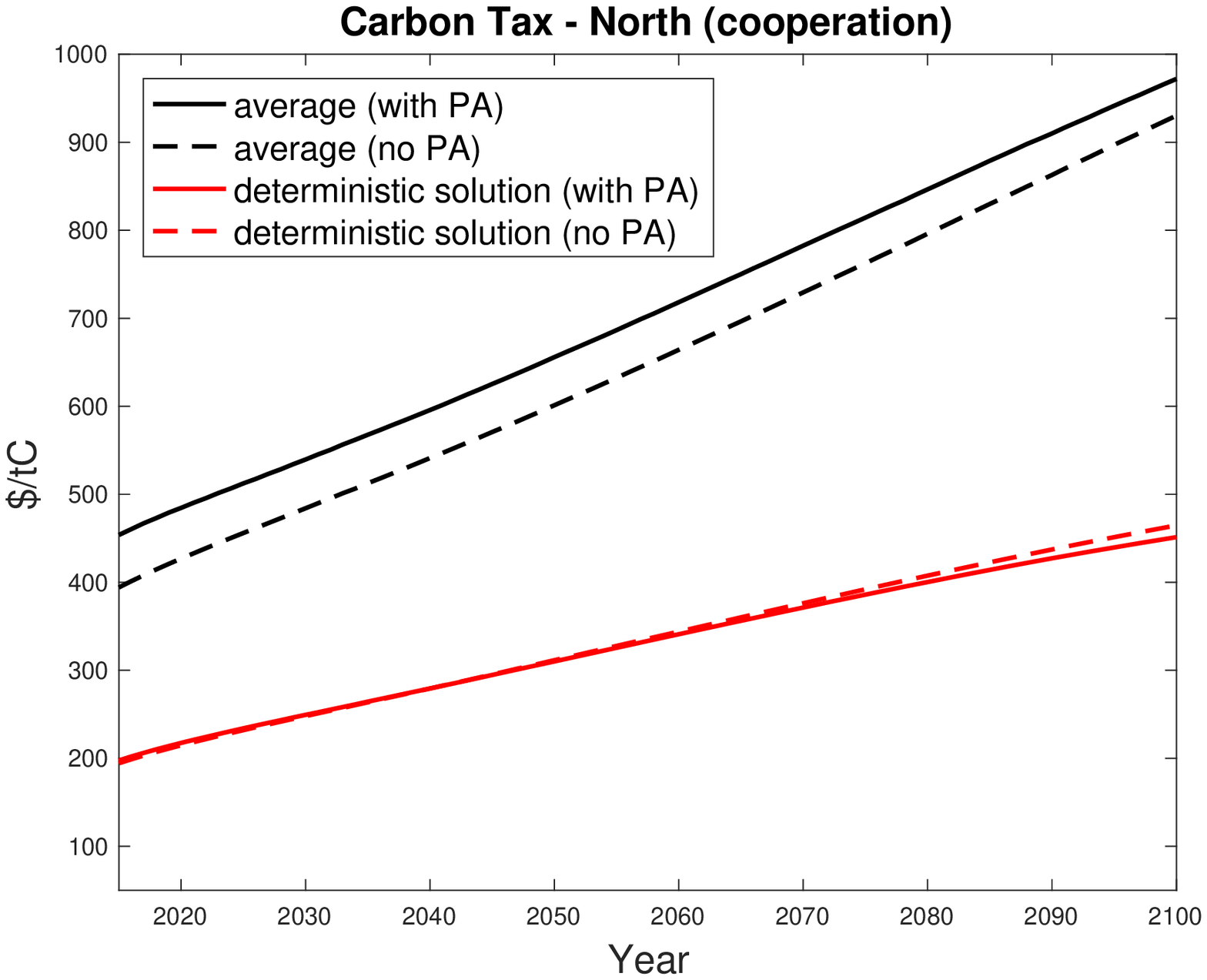} & \includegraphics[width=0.45\textwidth,height=0.21\textheight]{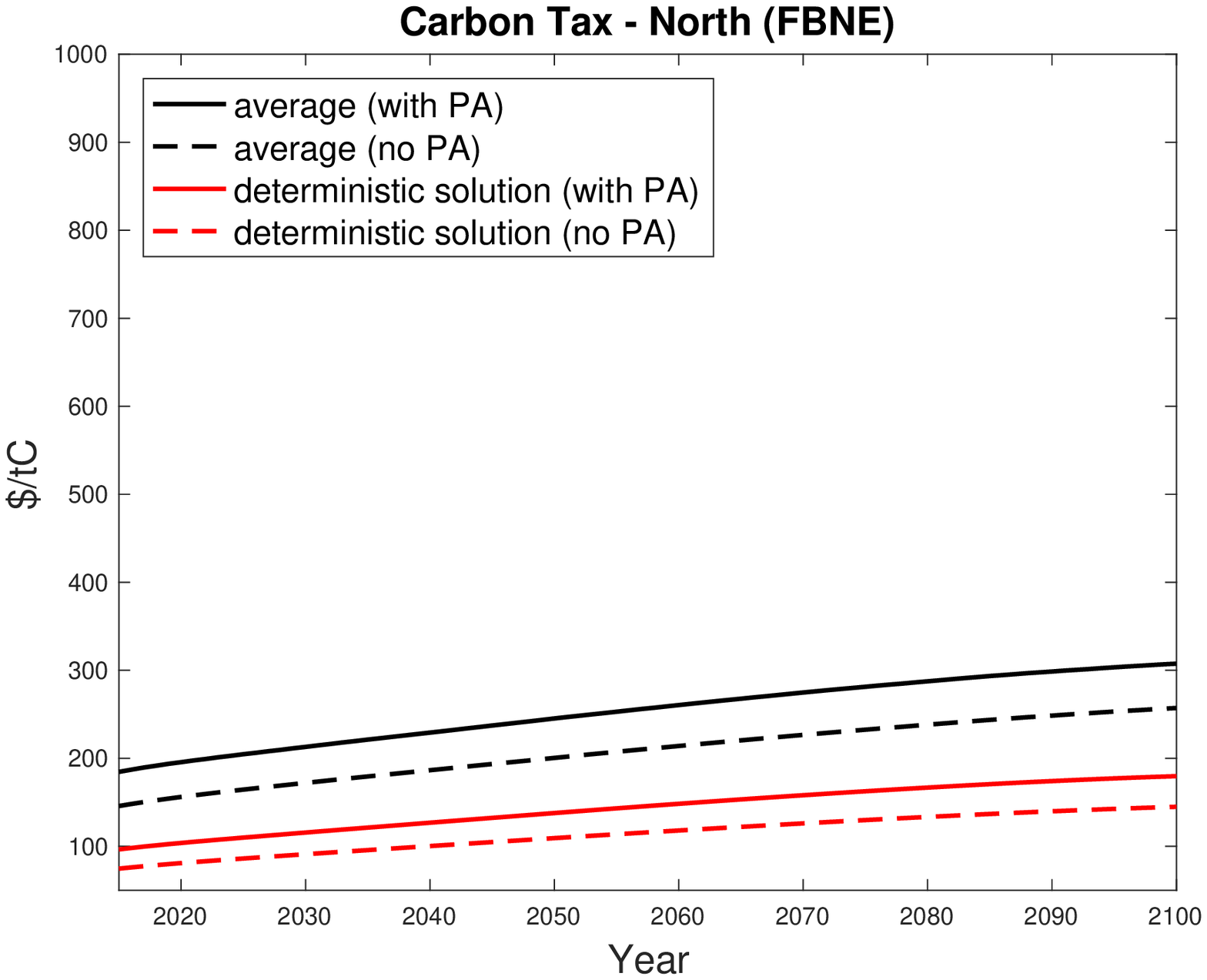}\tabularnewline
\includegraphics[width=0.45\textwidth,height=0.21\textheight]{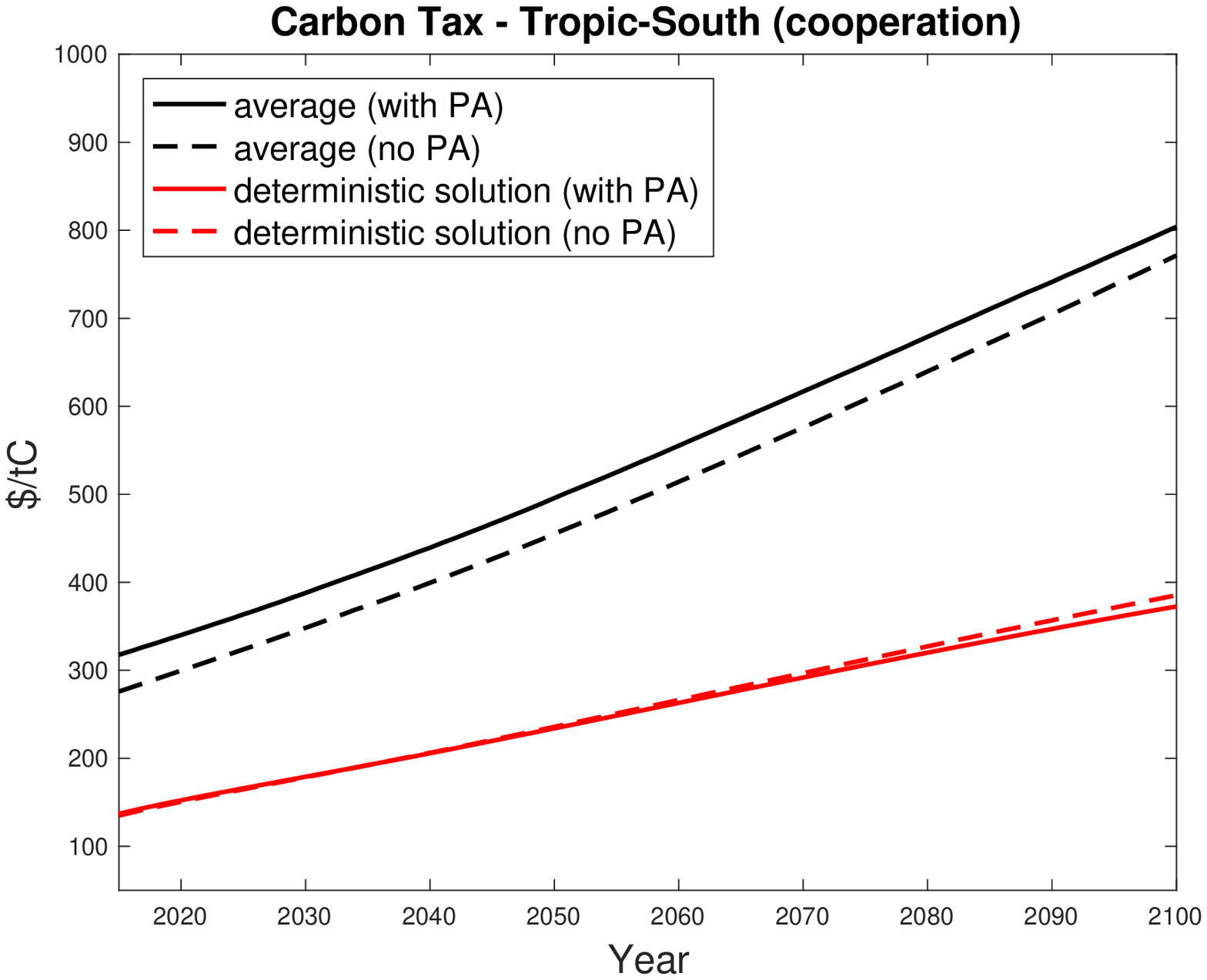} & \includegraphics[width=0.45\textwidth,height=0.21\textheight]{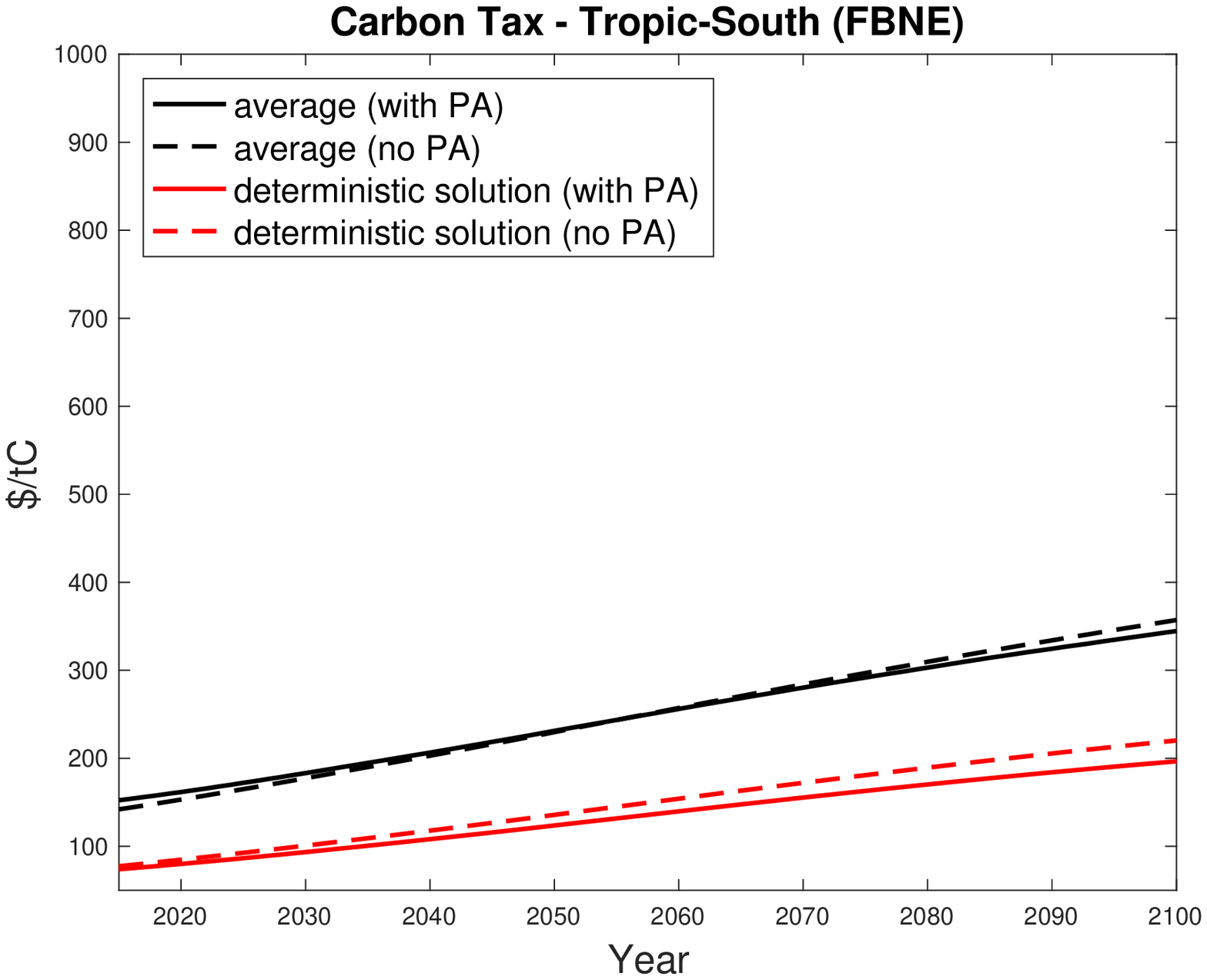}\tabularnewline
\end{tabular}
\par\end{centering}
\caption{Bias of Regional Carbon Taxes from Ignoring Heat Transport and PA
\label{fig:tax-bias-noPA}}
\end{figure}

We examine the bias from ignoring heat and moisture transport and
PA (i.e., $\xi_{4}=\xi_{5}=0$), shown in Figure \ref{fig:tax-bias-noPA},
for the regional carbon tax under cooperation (the left panels) and
non-cooperation (the right panels). In each panel of the figure, the
black lines represent the average paths and the red lines represent
the solution paths of the deterministic cases. The solid lines show
the case with heat and moisture transport and PA, while the dashed
lines show the case without it. Figure \ref{fig:tax-bias-noPA} shows
that ignoring heat transport and PA underestimates the average regional
carbon taxes for both regions under cooperation and non-cooperation
in the initial periods for the stochastic results. For example, the
initial regional carbon tax without PA is about 13\% less than in
the case with PA for both regions in the social planner's model. This
is because ignoring heat transport leads to a lower temperature in
the North and then underestimates the tipping probability which depends
on the atmospheric temperature in the North. This lower tipping probability
means a less risky tipping element which leads to smaller regional
carbon taxes. Note that in the deterministic cases, ignoring heat
transport has little impact on the regional carbon tax for the social
planner's problem, but in the FBNE, the impact of ignoring heat and
moisture transport and PA is in the opposite direction for the regions:
it underestimates the regional carbon tax in the North, but a bit
overestimates in the Tropic-South. 

\begin{figure}
\begin{centering}
\begin{tabular}{cc}
\includegraphics[width=0.45\textwidth,height=0.21\textheight]{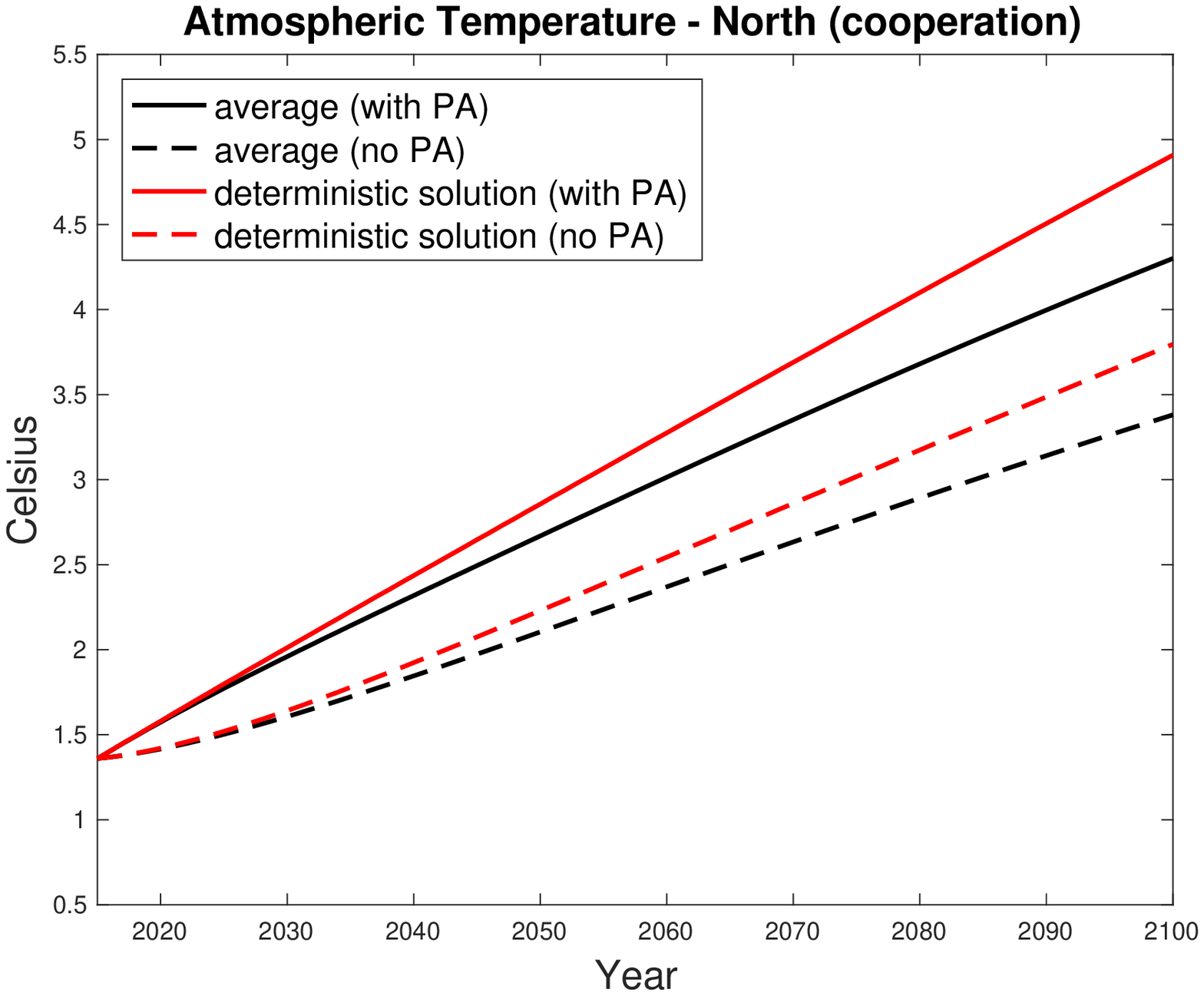} & \includegraphics[width=0.45\textwidth,height=0.21\textheight]{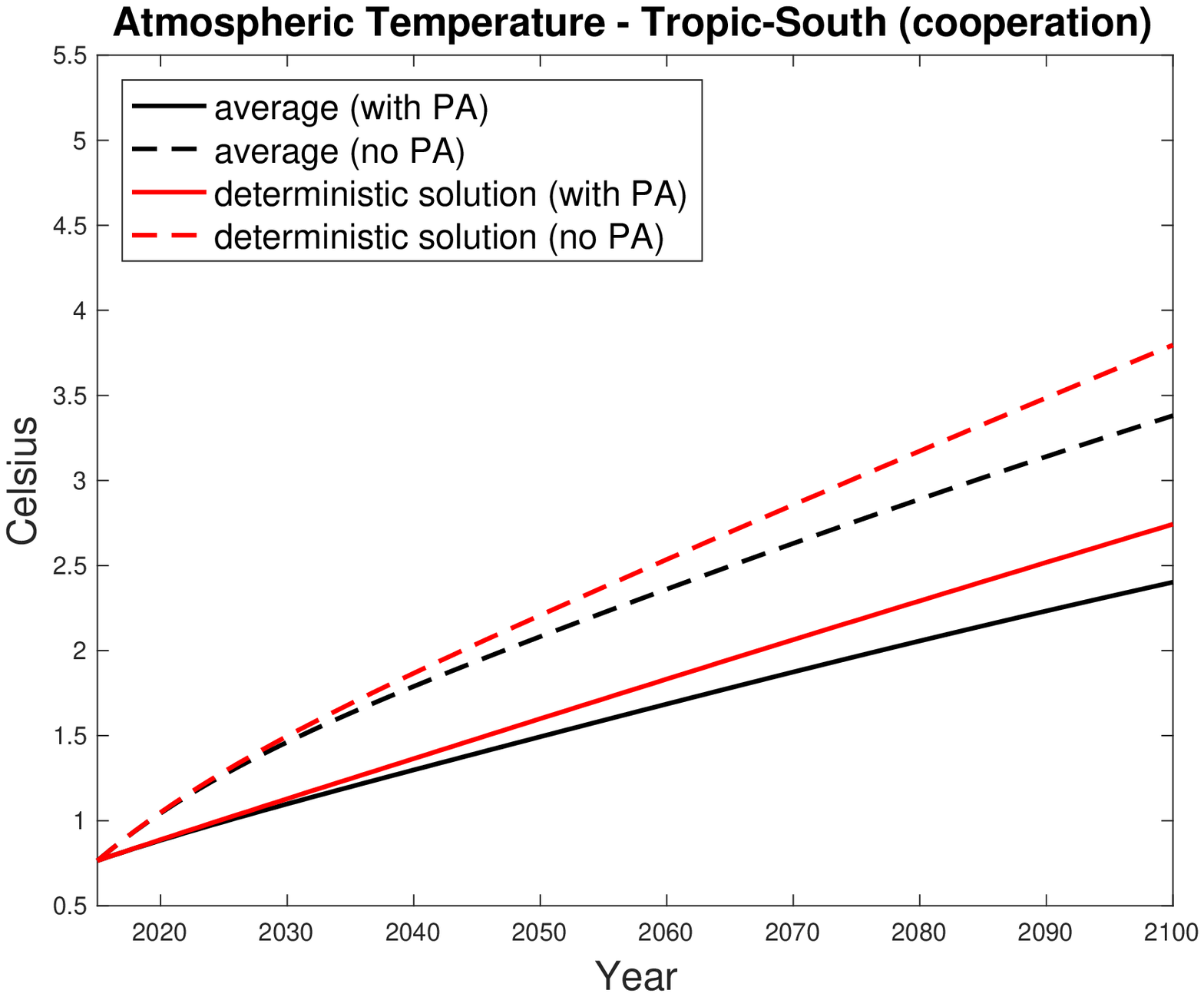}\tabularnewline
\includegraphics[width=0.45\textwidth,height=0.21\textheight]{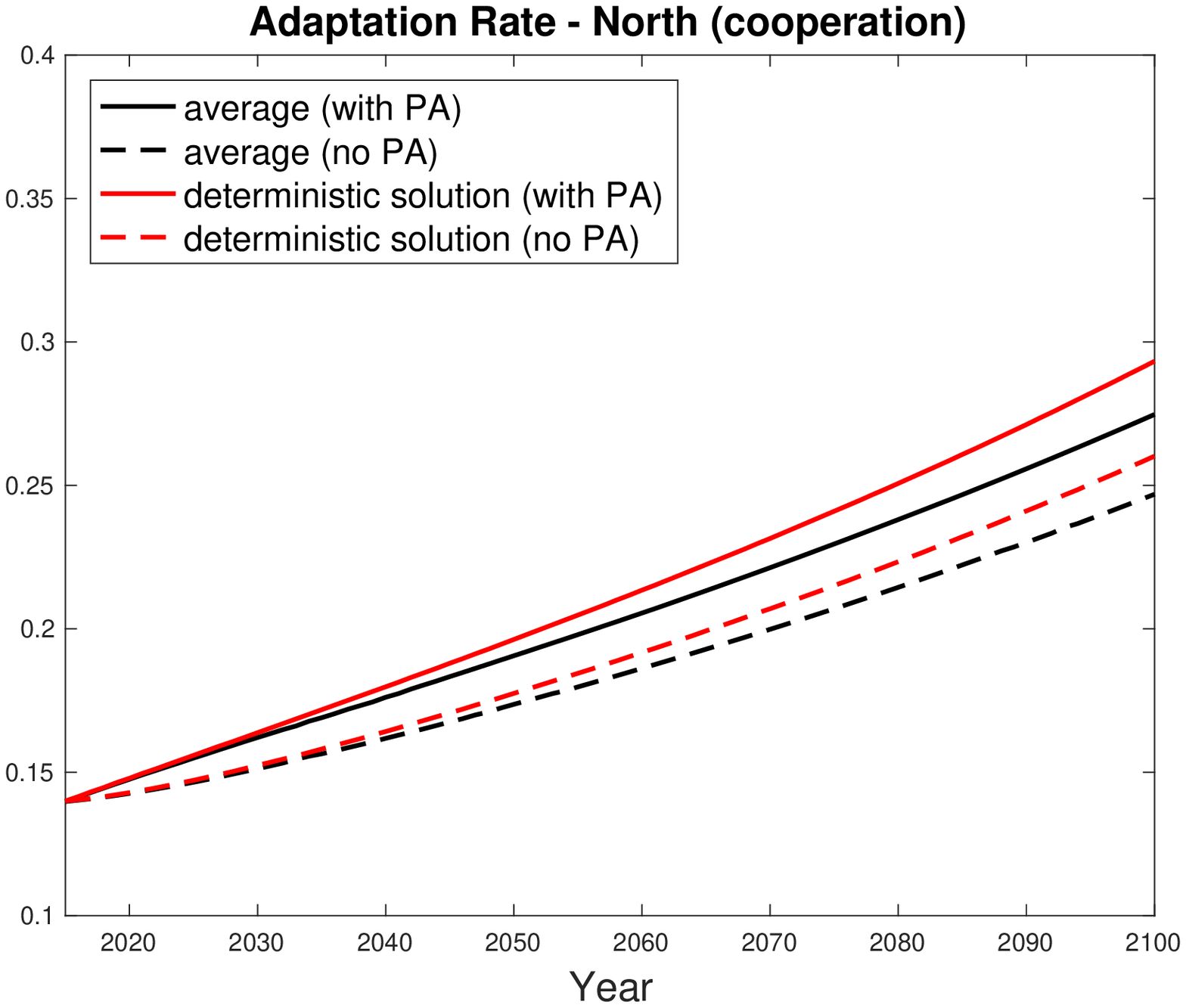} & \includegraphics[width=0.45\textwidth,height=0.21\textheight]{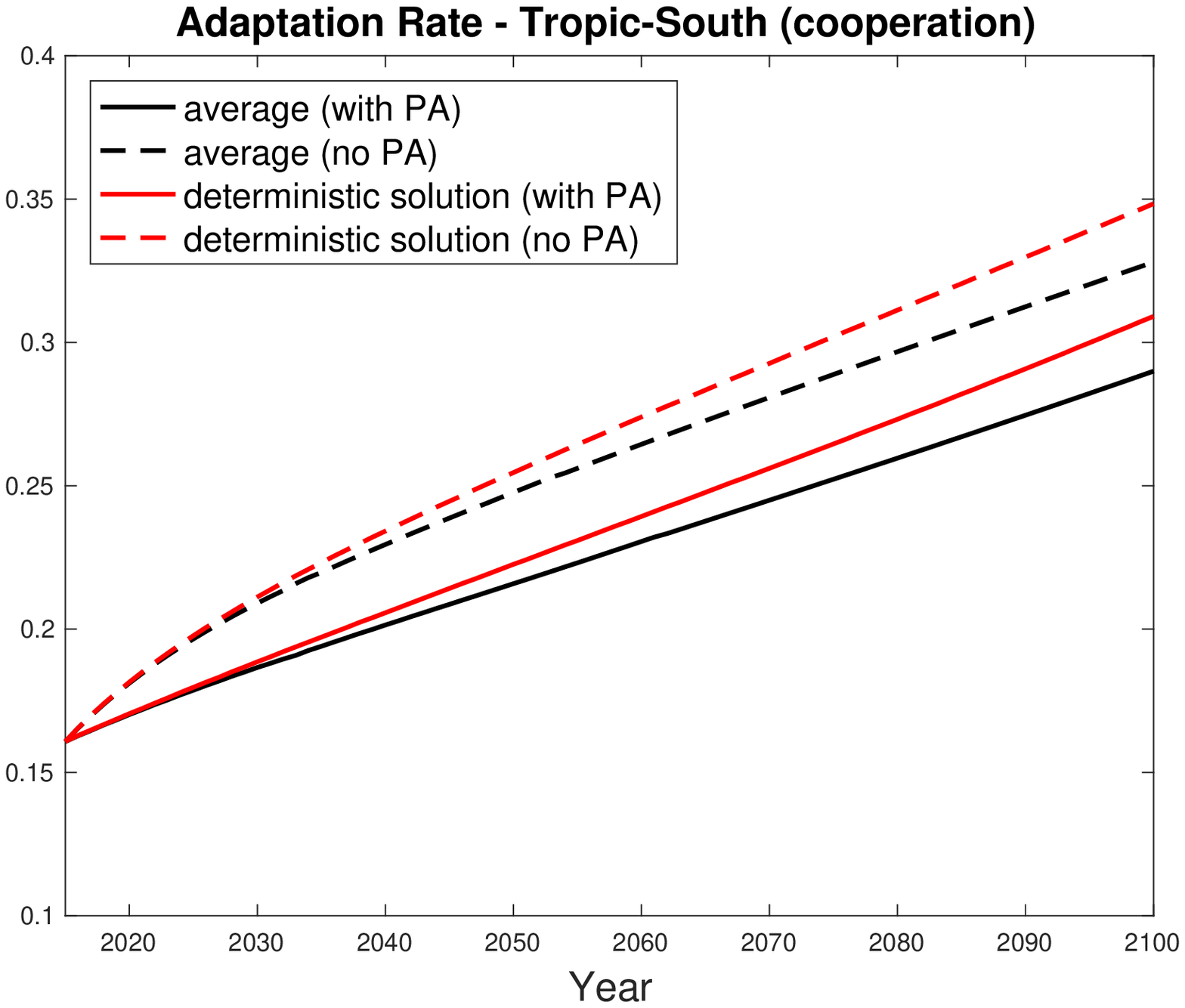}\tabularnewline
\end{tabular}
\par\end{centering}
\caption{Bias of Temperature and Adaptation from Ignoring Heat Transport and
PA \label{fig:TA-adapt-bias-noPA}}
\end{figure}

Figure \ref{fig:TA-adapt-bias-noPA} shows that in the social planner's
model ignoring heat transport and PA significantly underestimates
the atmospheric temperature anomaly and adaptation rates in the North,
and significantly overestimates them in the Tropic-South. The FBNE
model has similar pattern so here we omit its figure. 

\subsection{Bias from Ignoring SLR, Adaptation, and Capital Transfer }

Table \ref{tab:Initial-SCC-ignore} lists the initial regional carbon
tax if we ignore SLR, adaptation, or transfer of capital between regions.
Since the FBNE has no capital transfer between regions, Table \ref{tab:Initial-SCC-ignore}
lists only the cooperative solution. We see that ignoring SLR significantly
underestimate the regional carbon taxes. For example, for the social
planner's stochastic model without SLR, the initial regional carbon
tax is \$294/tC and \$207/tC for the North and the Tropic-South respectively,
only about 65\% of the corresponding ones with SLR. This is because
SLR imposes more damage, so it implies higher regional carbon tax.
Ignoring adaptation significantly overestimate the regional carbon
taxes. For example, for the social planner's stochastic model without
adaptation, the initial regional carbon tax is \$855/tC and \$601/tC
for the North and the Tropic-South respectively, about 88\% higher
than the corresponding ones with adaptation. This is because adaptation
can reduce damages, so it can reduce regional carbon taxes. Ignoring
capital transfer (i.e., the closed economy) overestimates the regional
carbon taxes for the North and underestimates the regional carbon
taxes for the Tropic-South. For example, for the social planner's
stochastic model without capital transfer between regions, the initial
regional carbon tax is \$540/tC for the North, about 19\% higher than
the corresponding ones with capital transfer; and in the Tropic-South,
the initial regional carbon tax is \$275/tC without the capital transfer,
about 14\% less than the one with the capital transfer. This happens
because the capital transfer makes the North to have less capital,
then less output and less damages, so the North has less regional
carbon taxes; oppositely, the capital transfer makes the Tropic-South
to have more capital and then more output and more damages, so the
Tropic-South has higher regional carbon taxes. In addition, we also
examine the impact of ignoring permafrost thaw and find that it has
little impact. The FBNE has the same pattern with the social planner's
solutions for ignoring the elements. Table \ref{tab:Initial-SCC-ignore}
also shows that the initial regional carbon taxes in the FBNE are
always significantly lower than the ones in the cooperative world;
the North always has higher regional carbon taxes than the Tropic-South
except the deterministic FBNE case without SLR; and the stochastic
cases always have higher regional carbon taxes than the deterministic
cases, if we ignore SLR, adaptation, or transfer of capital between
regions.

\begin{table}
\caption{Initial regional carbon tax from Ignoring Model Elements \label{tab:Initial-SCC-ignore}}

\centering{}%
\begin{tabular}{l|c|c|c|c|c}
\hline 
Ignored Element & Model & \multicolumn{2}{c|}{Deterministic} & \multicolumn{2}{c}{Stochastic}\tabularnewline
\cline{3-6} 
 &  & North & Tropic-South & \multicolumn{1}{c|}{North} & \multicolumn{1}{c}{Tropic-South}\tabularnewline
\hline 
SLR & Coop. & 84 & 58 & 294 & 207\tabularnewline
 & FBNE & 32 & 33 & 116 & 109\tabularnewline
\hline 
Adaptation & Coop. & 553 & 384 & 855 & 601\tabularnewline
 & FBNE & 355 & 214 & 400 & 299\tabularnewline
\hline 
Capital Transfer & Coop. & 236 & 118 & 540 & 275\tabularnewline
\hline 
\end{tabular}
\end{table}

\subsection{Sensitivity on IES and Risk Aversion}

Different values of IES and risk aversion are used in the literature.
Table \ref{tab:Initial-SCC-IES-RA} lists the initial regional carbon
tax for various IESs $(\psi\in\{0.69,1.5\})$ and risk aversion $(\gamma\in\{3.066,10\})$.
Table \ref{tab:Initial-SCC-IES-RA} shows that a higher IES leads
to a higher regional carbon tax in both regions; that a higher risk
aversion leads to a higher regional carbon tax in both regions; and
that stochastic cases have a higher regional carbon tax than deterministic
cases for the same IES. This finding is consistent with \citet{CaiJuddLontzek_DSICE}.
Table \ref{tab:Initial-SCC-IES-RA} also shows that for all cases
the North has a higher regional carbon tax than the Tropic-South,
and the regional carbon taxes in the FBNE are always significantly
smaller than the ones in the social planner's solution. 

\begin{table}
\caption{Initial regional carbon tax under various IESs ($\psi$) and risk
aversion ($\gamma$)\label{tab:Initial-SCC-IES-RA}}

\centering{}%
\begin{tabular}{c|c|c|c|cc|cc}
\hline 
IES & Model & \multicolumn{2}{c|}{Deterministic} & \multicolumn{4}{c}{Stochastic}\tabularnewline
\cline{3-8} 
($\psi$) &  & North & Tropic & \multicolumn{2}{c|}{North} & \multicolumn{2}{c}{Tropic-South}\tabularnewline
\cline{5-8} 
 &  &  & -South & $\gamma=3.066$ & $\gamma=10$ & $\gamma=3.066$ & $\gamma=10$\tabularnewline
\hline 
0.69 & Coop. & 58 & 35 & 114 & 132 & 69 & 80\tabularnewline
 & FBNE & 29 & 17 & 55 & 63 & 32 & 38\tabularnewline
\hline 
1.5 & Coop. & 198 & 137 & 454 & 519 & 318 & 363\tabularnewline
 & FBNE & 90 & 67 & 185 & 208 & 152 & 174\tabularnewline
\hline 
\end{tabular}
\end{table}

\section{Conclusion\label{sec:Conclusion}}

This paper has taken a first step toward adding heretofore neglected
elements \textendash{} heat and moisture transport from the lower
latitudes toward the Poles, sea level rise, adaptation \textendash{}
into computational IAMs used in policy-relevant climate economics.
When our model is calibrated to data, we find substantial biases in
key quantities such as temperature anomalies, the regional carbon
tax and damages, relative to the case where a social planner neglects
poleward heat and moisture transport, sea level rise, and adaptation.
We also show how potential arrival times of tipping elements in the
high latitudes are affected by the transport phenomena and, most importantly,
how these tipping elements can affect the regional carbon tax between
the two regions.

In our DIRESCU model we have combined the neglected elements with
stochastic arrival of tipping points, cooperation and noncooperation
between the North and Tropic-South, and recursive preferences which
distinguish between risk \textcolor{black}{aversion} and the IES.
We have shown that these aspects of our model have significant impact
on optimal climate policies, so they constitute an important step
forward in calibrated IAMs at the \textquotedblleft coarse grained\textquotedblright{}
level of aggregation. We have also shown that feedback Nash equilibrium
with recursive preferences can be computed in the complex high-dimensional
dynamic stochastic model with the advances of computational algorithms
and hardware resource. 

\textcolor{black}{We expect that future research and extensions in
the context of our model \textendash{} which would involve a finer
regional disaggregation, nonlinear surface albedo feedbacks, and better
approximations of damages from sources such as SLR or tipping points
\textendash{} could provide additional and improved insights.}

\bibliographystyle{ecta}
\bibliography{CBXJ}

\newpage{}
\begin{center}
{\Large{}Online Appendix for ``Climate Policy under Spatial Heat
Transport: Cooperative and Noncooperative Regional Outcomes'' }{\Large\par}
\par\end{center}

\begin{center}
Yongyang Cai \quad{}William Brock\quad{} Anastasios Xepapadeas\quad{}Kenneth
Judd
\par\end{center}

\begin{doublespace}
\global\long\def\thefigure{A.\arabic{figure}}
 \setcounter{figure}{0} \global\long\def\thesection{A.\arabic{section}}
 \setcounter{section}{0} \global\long\def\thetable{A.\arabic{table}}
 \setcounter{table}{0} \global\long\def\theequation{A.\arabic{equation}}
 \setcounter{equation}{0}\global\long\def\thepage{A.\arabic{page}}
\setcounter{page}{1}
\end{doublespace}

\section{Definition of Parameters and Exogenous Paths\label{sec:Definition-of-Parameters}}

In DIRESCU, we approximate the exogenous paths of DICE-2016 in five-year
time steps by their annual analogs. The land carbon emissions $E_{t}^{\mathrm{Land}}$
and exogenous radiative forcing $F_{t}^{\mathrm{EX}}$ are defined
below:

\begin{singlespace}
\begin{equation}
E_{t}^{\mathrm{Land}}=0.95e^{-0.115t}\label{eq:E-Land-CJL}
\end{equation}

\begin{equation}
F_{t}^{\mathrm{EX}}=\begin{cases}
0.5+0.00588t, & \mathrm{if}\;t\leq85\\
1, & \mathrm{otherwise}.
\end{cases}\label{eq:exogenous-radiativ}
\end{equation}

Tables \ref{tab:Param-climate}-\ref{tab:Param-sto} list the values
and/or definition of all parameters, variables and symbols.
\end{singlespace}

\begin{table}[H]
\caption{\textcolor{black}{\small{}Parameters, Variables and Symbols in the
Deterministic Climate System\label{tab:Param-climate}}}

\centering{}\textcolor{black}{\small{}}%
\begin{tabular}{|l|>{\raggedright}p{85mm}|}
\hline 
\textcolor{black}{\small{}$t$} & \textcolor{black}{\small{}time in years ($t=0$ represents year $2015$)}\tabularnewline
{\small{}$i$$\in\{1,2\}$} & {\small{}region $i$ (North or Tropic-South)}\tabularnewline
\textcolor{black}{\small{}$M_{t}^{\mathrm{AT}}$} & \textcolor{black}{\small{}carbon concentration in the atmosphere (billion
tons); $M_{0}^{\mathrm{AT}}=851$}\tabularnewline
\textcolor{black}{\small{}$M_{t}^{\mathrm{UO}}$} & \textcolor{black}{\small{}carbon concentration in upper ocean (billion
tons); $M_{0}^{\mathrm{UO}}=460$}\tabularnewline
\textcolor{black}{\small{}$M_{t}^{\mathrm{DO}}$} & \textcolor{black}{\small{}carbon concentration in deep ocean (billion
tons); $M_{0}^{\mathrm{DO}}=1740$}\tabularnewline
\textcolor{black}{\small{}$\mathbf{M}_{t}=\left(M_{t}^{\mathrm{AT}},M_{t}^{\mathrm{UO}},M_{t}^{\mathrm{DO}}\right)^{\top}$} & \textcolor{black}{\small{}carbon concentration vector }\tabularnewline
\textcolor{black}{\small{}$T_{t,i}^{\mathrm{AT}}$} & \textcolor{black}{\small{}average surface temperature (Celsius); $T_{0,1}^{\mathrm{AT}}=1.36$,
$T_{0,2}^{\mathrm{AT}}=0.765$}\tabularnewline
\textcolor{black}{\small{}$T_{t}^{\mathrm{OC}}$} & \textcolor{black}{\small{}average ocean temperature (Celsius); $T_{0}^{\mathrm{OC}}=0.0068$}\tabularnewline
\textcolor{black}{\small{}$\mathbf{T}_{t}=\left(T_{t,1}^{\mathrm{AT}},T_{t,2}^{\mathrm{AT}},T_{t}^{\mathrm{OC}}\right)^{\top}$} & \textcolor{black}{\small{}temperature vector }\tabularnewline
{\small{}$S_{t}$} & {\small{}sea level rise (SLR); $S_{0}=0.14$}\tabularnewline
\textcolor{black}{\small{}$F_{t}$} & \textcolor{black}{\small{}global radiative forcing }\tabularnewline
\textcolor{black}{\small{}$F_{t}^{\mathrm{EX}}$} & \textcolor{black}{\small{}exogenous radiative forcing}\tabularnewline
\textcolor{black}{\small{}$\eta=3.68$ } & \textcolor{black}{\small{}radiative forcing parameter }\tabularnewline
\textcolor{black}{\small{}$\mathbf{\Phi}_{\mathrm{M}}$} & \textcolor{black}{\small{}transition matrix of carbon cycle}\tabularnewline
\textcolor{black}{\small{}$\mathbf{\Phi}_{\mathrm{T}}$} & \textcolor{black}{\small{}transition matrix of temperature system}\tabularnewline
{\small{}$\phi_{1,2}=0.0237$, $\phi_{2,1}=0.0388$} & {\small{}parameters in }\textcolor{black}{\small{}transition matrix
of carbon cycle}\tabularnewline
{\small{}$\phi_{2,3}=0.00136$, $\phi_{3,2}=0.00284$} & {\small{}parameters in }\textcolor{black}{\small{}transition matrix
of carbon cycle}\tabularnewline
{\small{}$\xi_{1}=0.0526$, $\xi_{2}=0.08987$} & {\small{}parameters in }\textcolor{black}{\small{}transition matrix
of temperature system}\tabularnewline
{\small{}$\xi_{3}=0.0022$, $\xi_{4}=0.6557$} & {\small{}parameters in }\textcolor{black}{\small{}transition matrix
of temperature system}\tabularnewline
{\small{}$\xi_{5}=0.5565$, $\xi_{6}=0.0$} & {\small{}parameters in }\textcolor{black}{\small{}transition matrix
of temperature system}\tabularnewline
{\small{}$\zeta_{1}^{\mathrm{SLR}}=0.00073$, $\zeta_{2}^{\mathrm{SLR}}=1.4$} & {\small{}parameters in SLR by warming}\tabularnewline
{\small{}$\zeta_{3}^{\mathrm{SLR}}=0.007$} & {\small{}parameters in SLR by warming}\tabularnewline
{\small{}$\zeta_{1}^{\mathrm{Perm}}=1.951$, $\zeta_{2}^{\mathrm{Perm}}=-0.0858$} & {\small{}parameters in emission from permafrost thawing by warming}\tabularnewline
{\small{}$\zeta_{3}^{\mathrm{Perm}}=0.2257$} & {\small{}parameters in emission from permafrost thawing by warming}\tabularnewline
\textcolor{black}{\small{}$M_{*}^{\mathrm{AT}}=588$ } & \textcolor{black}{\small{}equilibrium atmospheric carbon concentration}\tabularnewline
\hline 
\end{tabular}{\small\par}
\end{table}

\begin{table}[H]
\caption{\textcolor{black}{\small{}Parameters, Variables and Symbols in the
Economic System\label{tab:Param-econ}}}

\centering{}\textcolor{black}{\small{}}%
\begin{tabular}{|l|>{\raggedright}p{100mm}|}
\hline 
{\small{}$\mathcal{Y}_{t,i}$} & \textcolor{black}{\small{}gross output}\tabularnewline
{\small{}$Y_{t,i}$} & \textcolor{black}{\small{}output net of damage}\tabularnewline
{\small{}$\widehat{Y}_{t,i}$} & \textcolor{black}{\small{}output net of damage, abatement and adaptation
cost}\tabularnewline
{\small{}$A_{t,i}$} & \textcolor{black}{\small{}total productivity factor (TFP); $A_{0,1}=7.331$,
$A_{0,2}=3.582$}\tabularnewline
{\small{}$\alpha_{1}^{\mathrm{TFP}}=0.013$, $\alpha_{2}^{\mathrm{TFP}}=0.0184$} & {\small{}initial growth of }\textcolor{black}{\small{}TFP}\tabularnewline
{\small{}$d_{1}^{\mathrm{TFP}}=0.0053$, $d_{2}^{\mathrm{TFP}}=0.0061$} & {\small{}change rate of growth of }\textcolor{black}{\small{}TFP }\tabularnewline
{\small{}$L_{t,i}$} & \textcolor{black}{\small{}population (in billions)}\tabularnewline
{\small{}$K_{t,i}$} & \textcolor{black}{\small{}capital (in \$ trillions); $K_{0,1}=146$,
$K_{0,2}=77$}\tabularnewline
\textcolor{black}{\small{}$\alpha=0.3$ } & \textcolor{black}{\small{}output elasticity of capital}\tabularnewline
{\small{}$D_{t,i}^{\mathrm{S}}$} & {\small{}damage (in fraction of output) from sea level rise}\tabularnewline
\textcolor{black}{\small{}$\pi_{1,1}=0.00447$, $\pi_{1,2}=0.00408$ } & \textcolor{black}{\small{}SLR damage parameter}\tabularnewline
\textcolor{black}{\small{}$\pi_{2,1}=0.01146$, $\pi_{2,2}=0.00646$ } & \textcolor{black}{\small{}SLR damage parameter}\tabularnewline
{\small{}$D_{t,i}^{\mathrm{T}}$} & {\small{}damage (in fraction of output) from surface temperature anomaly}\tabularnewline
\textcolor{black}{\small{}$\pi_{3,1}=0.00094$, $\pi_{3,2}=0.00322$ } & \textcolor{black}{\small{}non-SLR damage parameter}\tabularnewline
\textcolor{black}{\small{}$\pi_{4,1}=0.0002$, $\pi_{2,2}=0.00074$ } & \textcolor{black}{\small{}non-SLR damage parameter}\tabularnewline
{\small{}$\Psi_{t,i}$} & {\small{}mitigation expenditure}\tabularnewline
{\small{}$\Upsilon_{t,i}$} & {\small{}adaptation expenditure}\tabularnewline
\textcolor{black}{\small{}$\mu_{t,i}$} & \textcolor{black}{\small{}emission control rate}\tabularnewline
\textcolor{black}{\small{}$E_{t}$, $E_{t,i}^{\mathrm{Ind}}$, $E_{t}^{\mathrm{Land}}$} & \textcolor{black}{\small{}global emission; regional industrial emission;
land emission}\tabularnewline
\textcolor{black}{\small{}$P_{t,i}$} & \textcolor{black}{\small{}adaptation rate}\tabularnewline
\textcolor{black}{\small{}$\sigma_{t,i}$ } & \textcolor{black}{\small{}carbon intensity; $\sigma_{0,1}=0.094$,
$\sigma_{0,2}=0.104$}\tabularnewline
{\small{}$\alpha_{1}^{\sigma}=0.0156$, $\alpha_{2}^{\sigma}=0.0181$} & {\small{}initial declining rate of }\textcolor{black}{\small{}carbon
intensity }\tabularnewline
{\small{}$d_{1}^{\sigma}=0.0063$, $d_{2}^{\sigma}=0.007$} & {\small{}change rate of growth of }\textcolor{black}{\small{}carbon
intensity }\tabularnewline
\textcolor{black}{\small{}$\theta_{2}=2.6$ } & \textcolor{black}{\small{}mitigation cost parameter}\tabularnewline
\textcolor{black}{\small{}$\theta_{1,t,i}$ } & \textcolor{black}{\small{}adjusted cost for backstop}\tabularnewline
\textcolor{black}{\small{}$b_{0,1}=1.71$, $b_{0,2}=2.19$ } & \textcolor{black}{\small{}initial backstop price}\tabularnewline
\textcolor{black}{\small{}$\alpha_{1}^{b}=\alpha_{2}^{b}=0.005$} & \textcolor{black}{\small{}declining rate of backstop price}\tabularnewline
{\small{}$\eta_{1}=0.115$, $\eta_{2}=3.6$} & {\small{}parameters for adaptation cost}\tabularnewline
\textcolor{black}{\small{}$\delta=0.1$ } & \textcolor{black}{\small{}annual depreciation rate }\tabularnewline
\textcolor{black}{\small{}$c_{t,i}$} & \textcolor{black}{\small{}per capita consumption}\tabularnewline
{\small{}$I_{t,i}$} & \textcolor{black}{\small{}investment}\tabularnewline
{\small{}$\Gamma_{t,i}$} & {\small{}adjustment cost for economic interaction between regions}\tabularnewline
{\small{}$B=1$} & {\small{}parameter for economic interaction cost}\tabularnewline
{\small{}$\psi\in\{0.69,1.5\}$} & {\small{}IES}\tabularnewline
\textcolor{black}{\small{}$u$} & \textcolor{black}{\small{}per capita utility function}\tabularnewline
\textcolor{black}{\small{}$\beta=0.985$ } & \textcolor{black}{\small{}discount factor}\tabularnewline
\hline 
\end{tabular}{\small\par}
\end{table}

\begin{table}[H]
\caption{\textcolor{black}{\small{}Additional Parameters, Variables and Symbols
in the Stochastic Model\label{tab:Param-sto}}}

\centering{}\textcolor{black}{\small{}}%
\begin{tabular}{|l|>{\raggedright}p{85mm}|}
\hline 
\textcolor{black}{\small{}$D=50$} & \textcolor{black}{\small{}duration of tipping process}\tabularnewline
{\small{}$J_{t}$} & {\small{}damage level; $J_{0}=0$}\tabularnewline
{\small{}$\overline{J}=0.15$} & {\small{}final damage level}\tabularnewline
{\small{}$\Delta$} & {\small{}annual increment of damage level}\tabularnewline
{\small{}$\chi_{t}\in\{0,1\}$} & {\small{}indicator for whether tipping has happened; $\chi_{0}=0$}\tabularnewline
{\small{}$p_{t}$} & {\small{}tipping probability}\tabularnewline
{\small{}$\varrho=0.00063$} & {\small{}hazard rate for tipping}\tabularnewline
\textcolor{black}{\small{}$\gamma\in\{3.066,10\}$ } & \textcolor{black}{\small{}risk aversion parameter }\tabularnewline
{\small{}$\mathbf{x}_{t}$} & {\small{}vector of state variables}\tabularnewline
{\small{}$V_{t}$} & {\small{}value function at time $t$}\tabularnewline
{\small{}$V_{t}^{\mathrm{SP}}$} & {\small{}value function for the social planner's problem}\tabularnewline
{\small{}$V_{t}^{\mathrm{FBNE}}$} & {\small{}value function for the feedback Nash equilibrium problem}\tabularnewline
\hline 
\end{tabular}{\small\par}
\end{table}

\section{Calibration of the Climate System\label{sec:Calibration-Climate}}

\begin{figure}[h]
\begin{centering}
\includegraphics[width=0.5\textwidth,height=0.25\textheight]{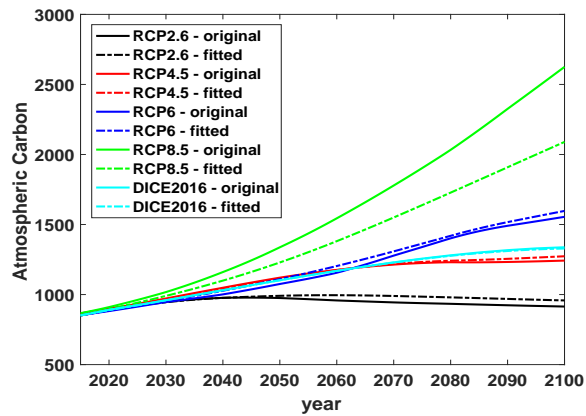}
\par\end{centering}
\caption{Fitting Atmospheric Carbon Concentration\label{fig:Atmospheric-Carbon-Concentration}}
\end{figure}

Figure \ref{fig:Atmospheric-Carbon-Concentration} shows that our
calibrated carbon cycle can approximate well for all scenarios except
RCP8.5. Since RCP8.5 is the business-as-usual scenario, and our model
is solving problems with optimal mitigation policy, the deviation
of approximation for RCP8.5 has little impact on our solutions.

\begin{figure}
\begin{centering}
\begin{tabular}{cc}
\includegraphics[width=0.45\textwidth,height=0.21\textheight]{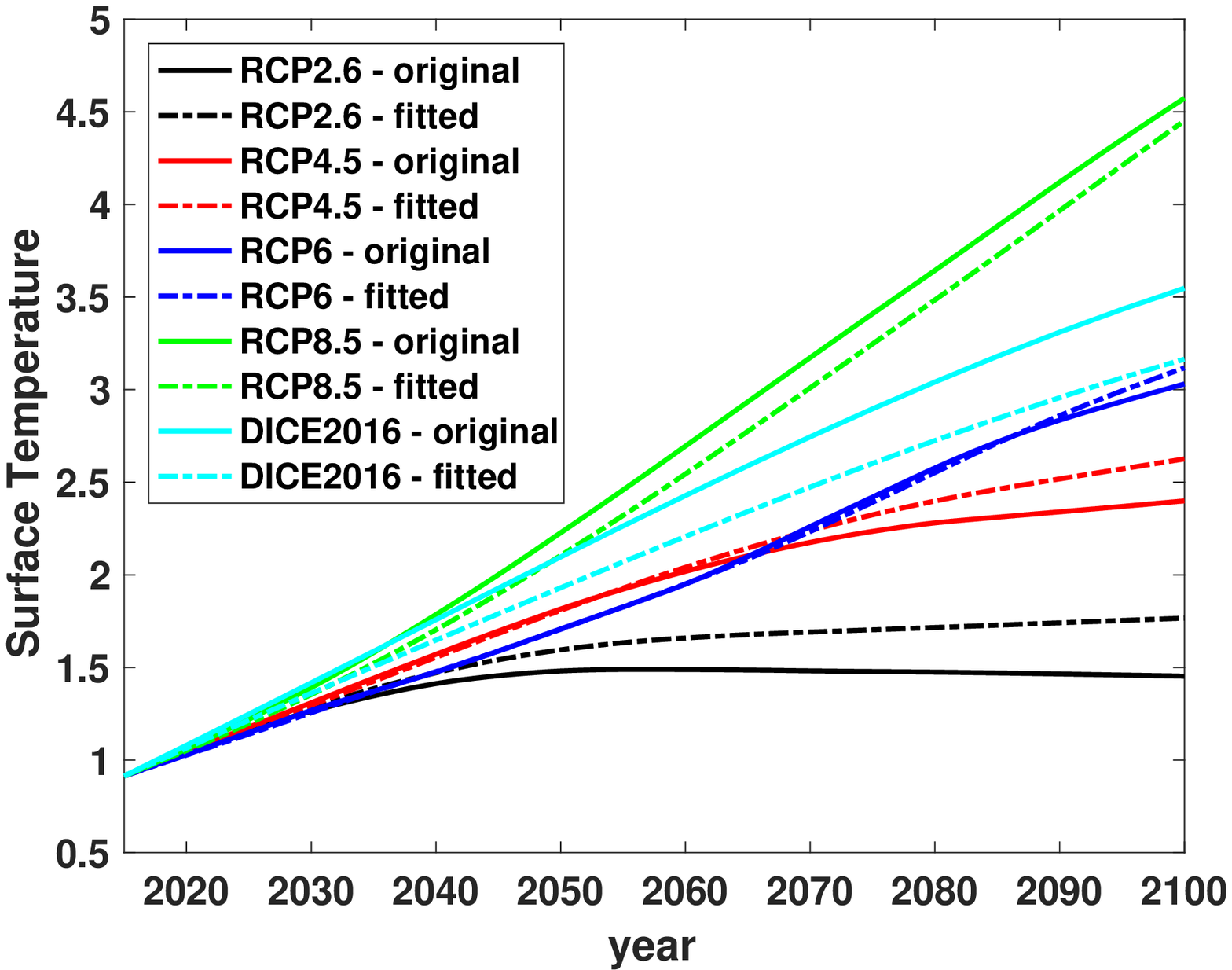} & \includegraphics[width=0.45\textwidth,height=0.21\textheight]{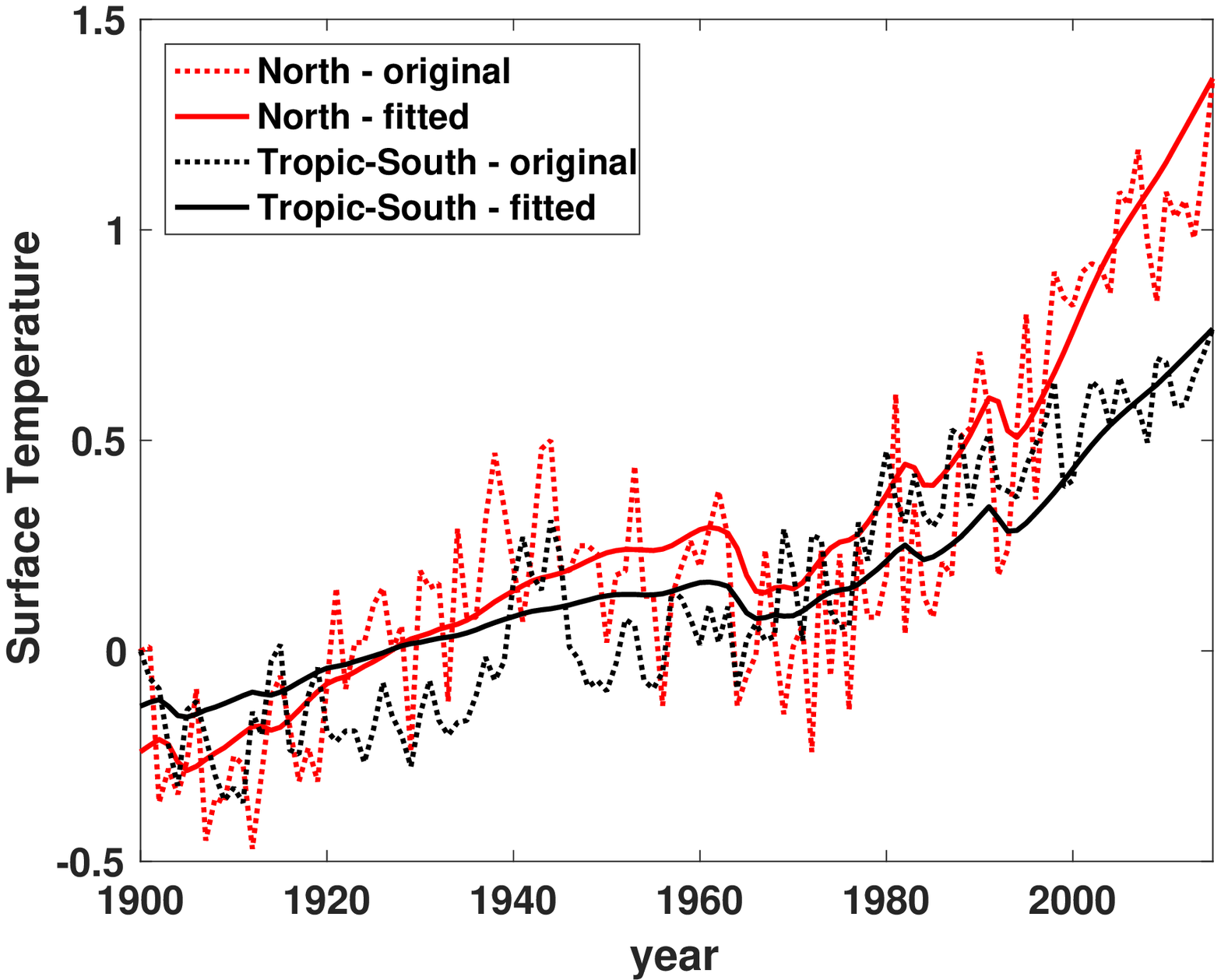}\tabularnewline
\end{tabular}
\par\end{centering}
\caption{Fitting Surface Temperature \label{fig:Fitting-Surface-Temperature}}
\end{figure}

\begin{figure}
\begin{centering}
\includegraphics[width=1\textwidth,height=0.7\textheight]{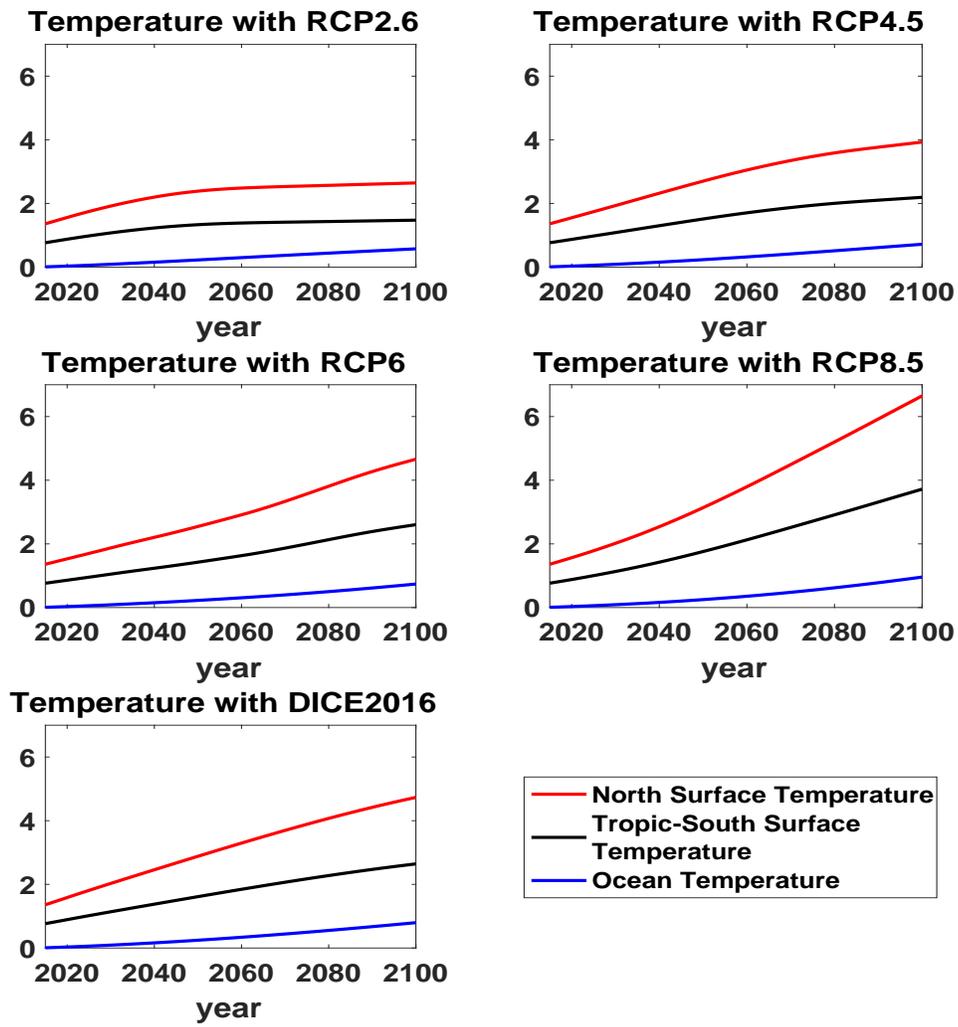}
\par\end{centering}
\caption{Spatial Temperature Using RCP or DICE Radiative Forcing Scenarios\label{fig:Spatial-temperature-using}}
\end{figure}

Figure \ref{fig:Fitting-Surface-Temperature} shows that our calibrated
temperature system can approximate well for all these scenarios. Figure
\ref{fig:Spatial-temperature-using} displays the corresponding spatial
surface temperature and ocean temperature pathways from the calibrated
temperature system, for the RCP scenarios and the DICE-2016 optimal
scenario. It also shows that the spatial surface temperatures in 2081-2100
are close to the ones given in IPCC (2013).

\begin{figure}
\begin{centering}
\includegraphics[width=0.5\textwidth,height=0.25\textheight]{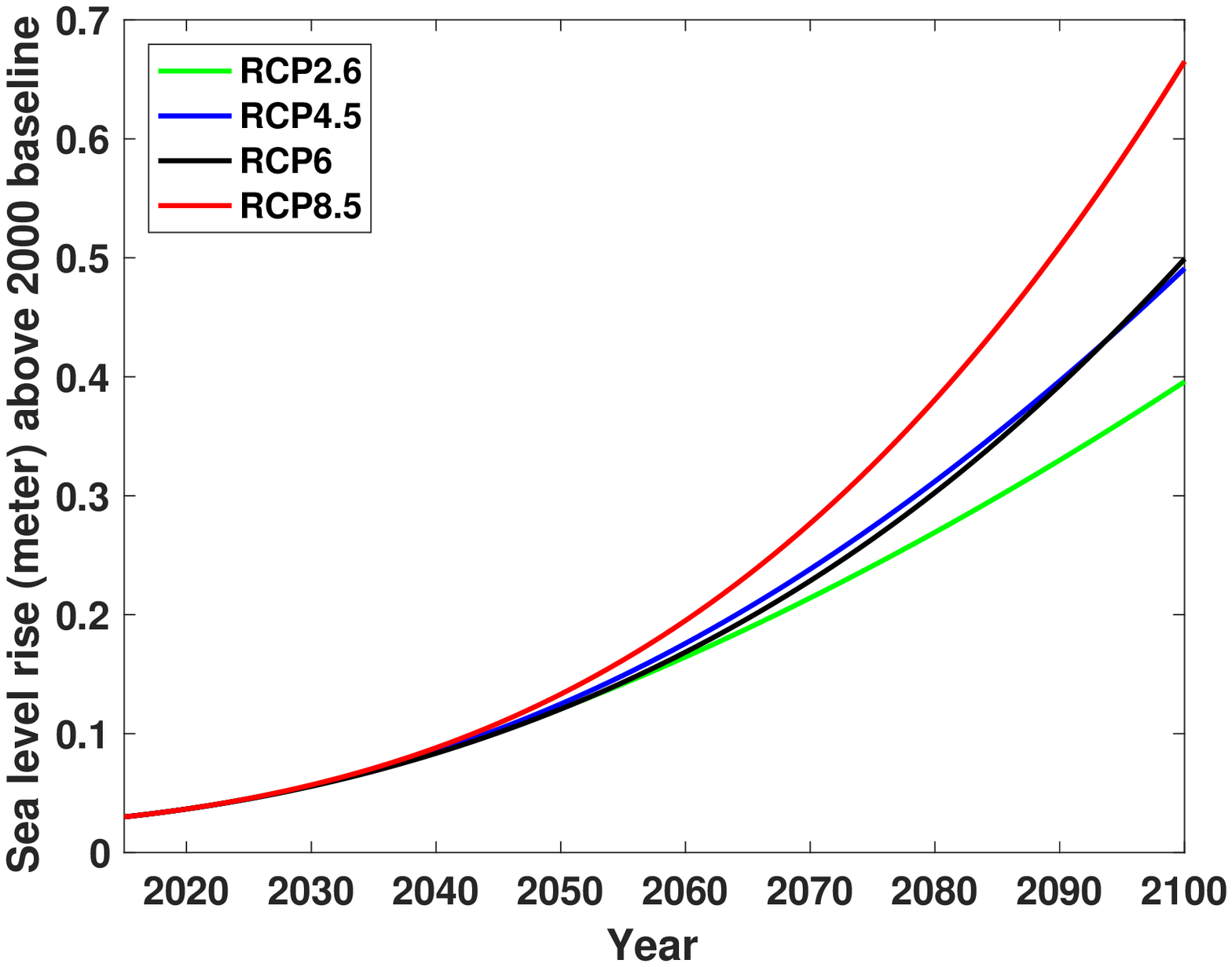}
\par\end{centering}
\caption{Fitting SLR\label{fig:Fitting-SLR-TA}}
\end{figure}

Figure \ref{fig:Fitting-SLR-TA} shows that our fitted paths of SLR
(above the level in 2000) for RCP2.6, RCP4.5 and RCP6 are quite close
to the mean projections in \citet{IPCC2013} and \citet{Kopp2014}. 

\begin{figure}
\begin{centering}
\begin{tabular}{cc}
\includegraphics[width=0.45\textwidth,height=0.21\textheight]{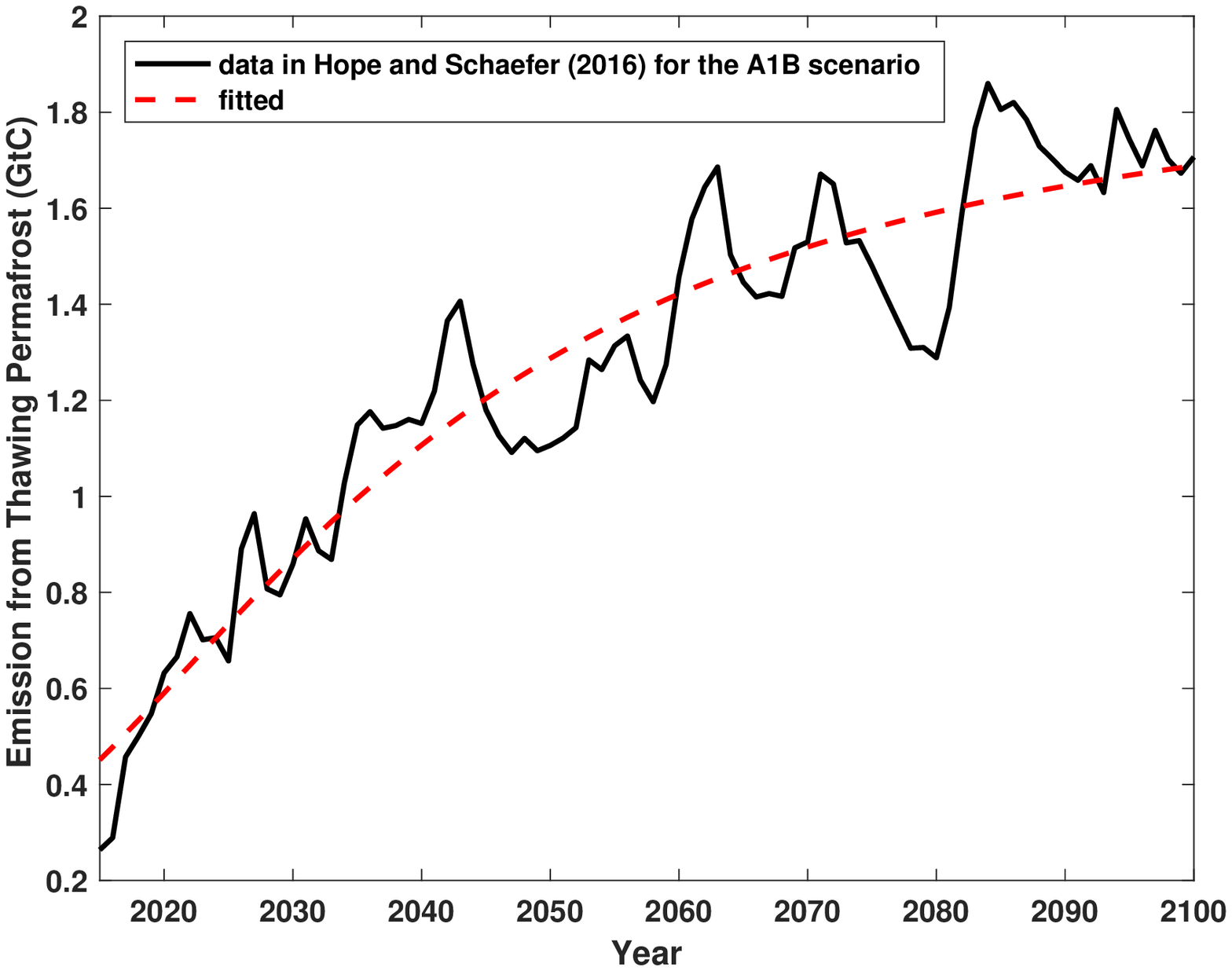} & \includegraphics[width=0.45\textwidth,height=0.21\textheight]{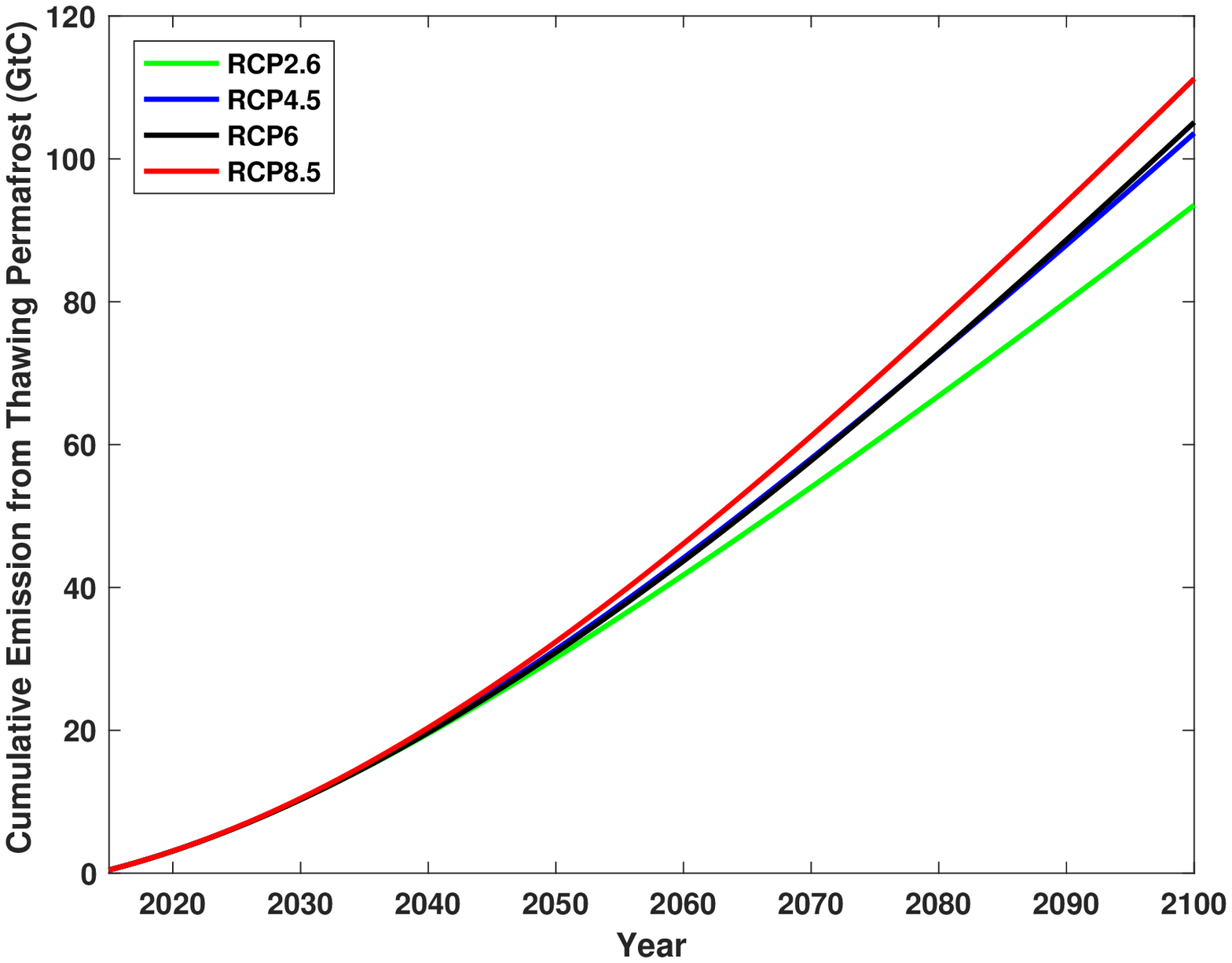}\tabularnewline
\end{tabular}
\par\end{centering}
\caption{Fitting the Carbon Emission Data from Thawing Permafrost and Projecting
Future Cumulative Emission Paths from Thawing Permafrost since 2010\label{fig:Fitting-permafrost}}
\end{figure}

The left panel of Figure \ref{fig:Fitting-permafrost} shows that
our function (\ref{eq:E-perm}) for estimating emissions from thawing
permafrost fits data well, and the right panel of Figure \ref{fig:Fitting-permafrost}
shows projected future cumulative emission paths from thawing permafrost
since 2010 for four RCP scenarios. We see that the cumulative carbon
emission under RCP8.5 is also inside the likely range given in \citet{schuur_2015}.\footnote{Since the amount of GHGs in permafrost is finite, we can have an upper
bound constraint on cumulative emissions from permafrost. But since
our model solution never hits the upper bound, numerically this constraint
does not matter.} 

\section{Calibration of the Economic System\label{sec:Calibration-Econ}}

The left panel of Figure \ref{fig:Fitting-Total-Factor} shows that
our calibrated TFP paths match RICE projections over both the North
and the Tropic-South. The right panel of Figure \ref{fig:Fitting-Total-Factor}
shows that our carbon intensity estimate (\ref{eq:carbon-intensity})
closely approximates the corresponding RICE projections.
\begin{center}
\begin{figure}
\begin{centering}
\begin{tabular}{cc}
\includegraphics[width=0.5\textwidth,height=0.25\textheight]{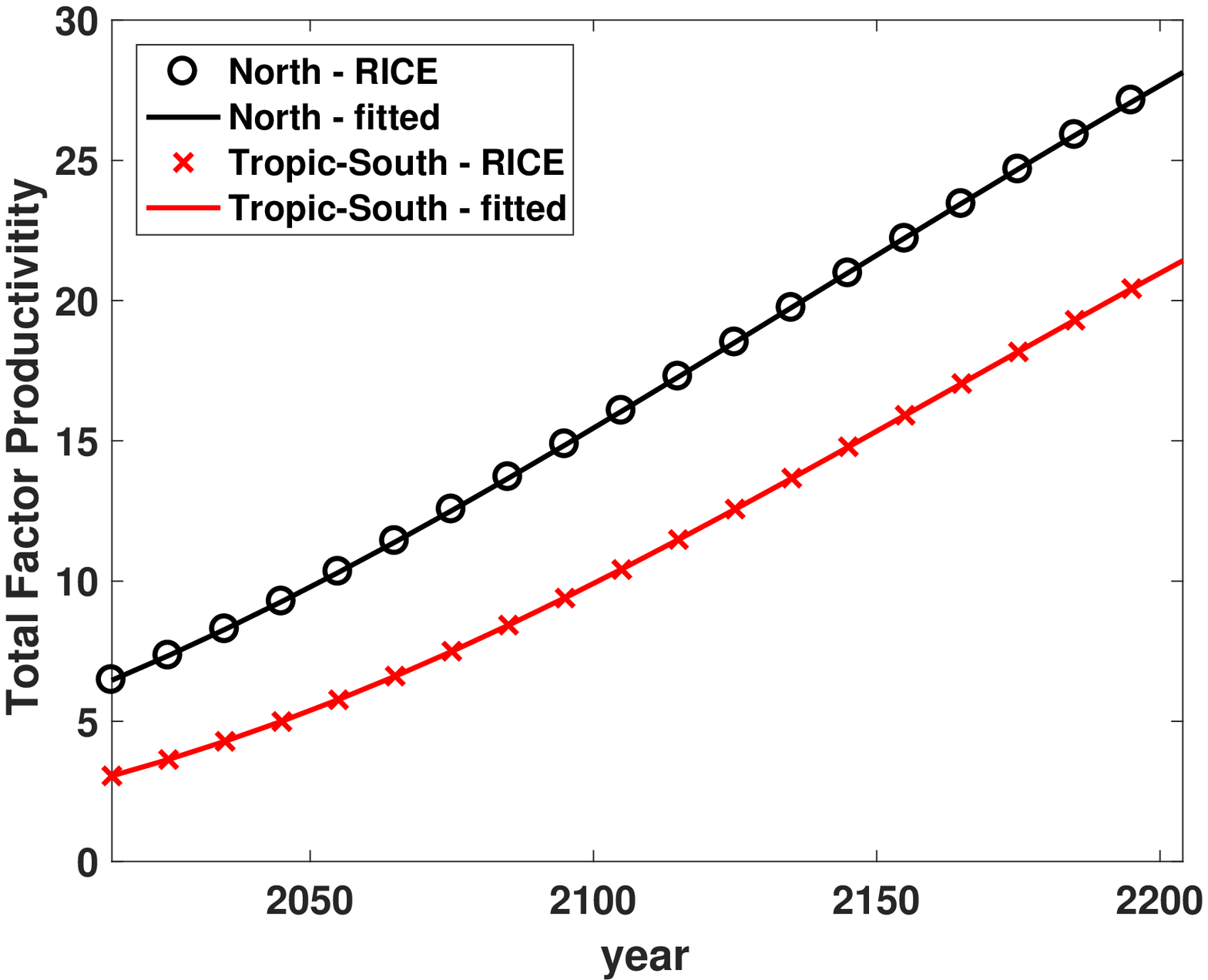} & \includegraphics[width=0.5\textwidth,height=0.25\textheight]{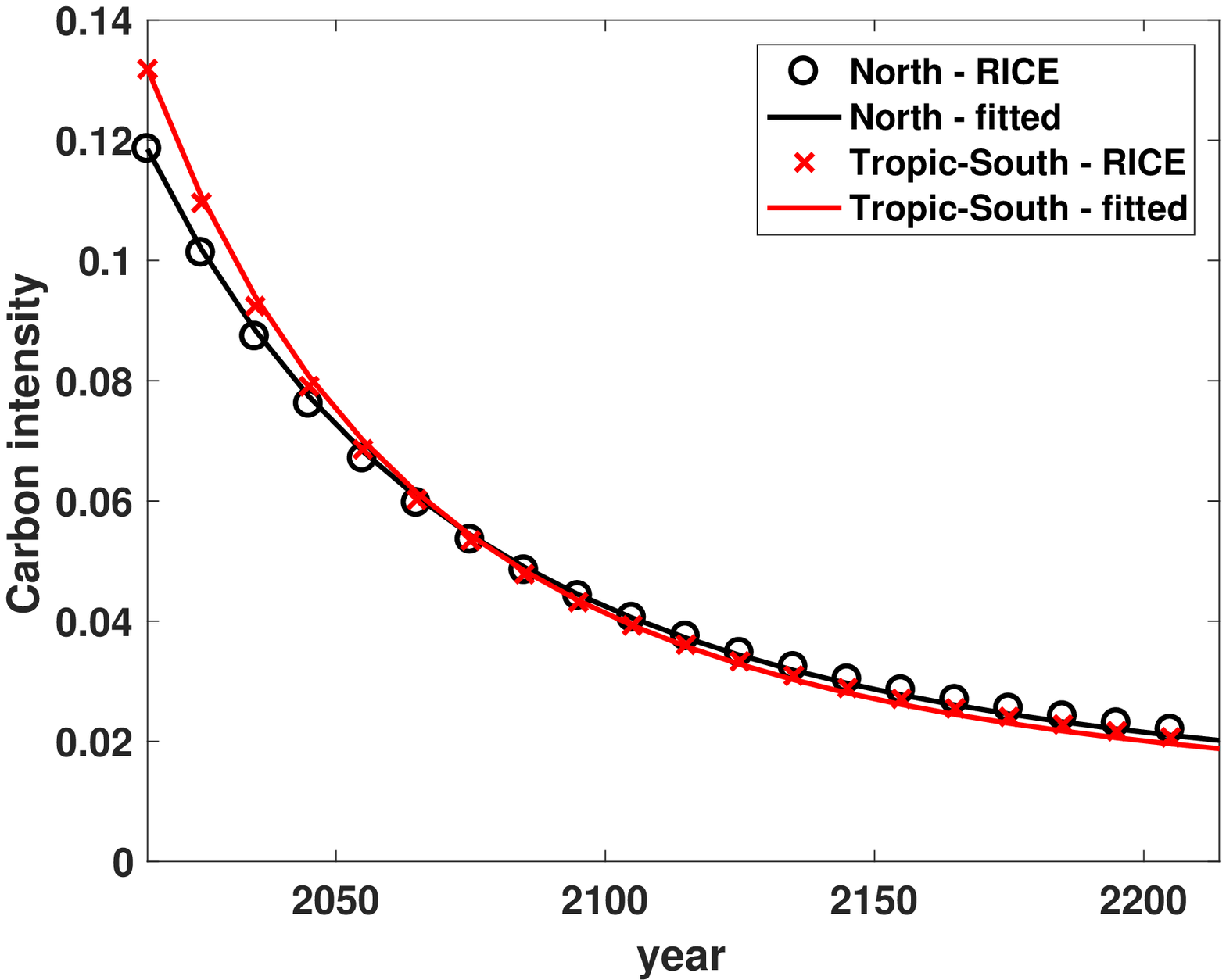}\tabularnewline
\end{tabular}
\par\end{centering}
\caption{Fitting Total Factor Productivity (the left panel) and Carbon Intensity
(the right panel)\label{fig:Fitting-Total-Factor}}
\end{figure}
\par\end{center}

\begin{figure}
\begin{centering}
\begin{tabular}{cc}
\includegraphics[width=0.5\textwidth,height=0.25\textheight]{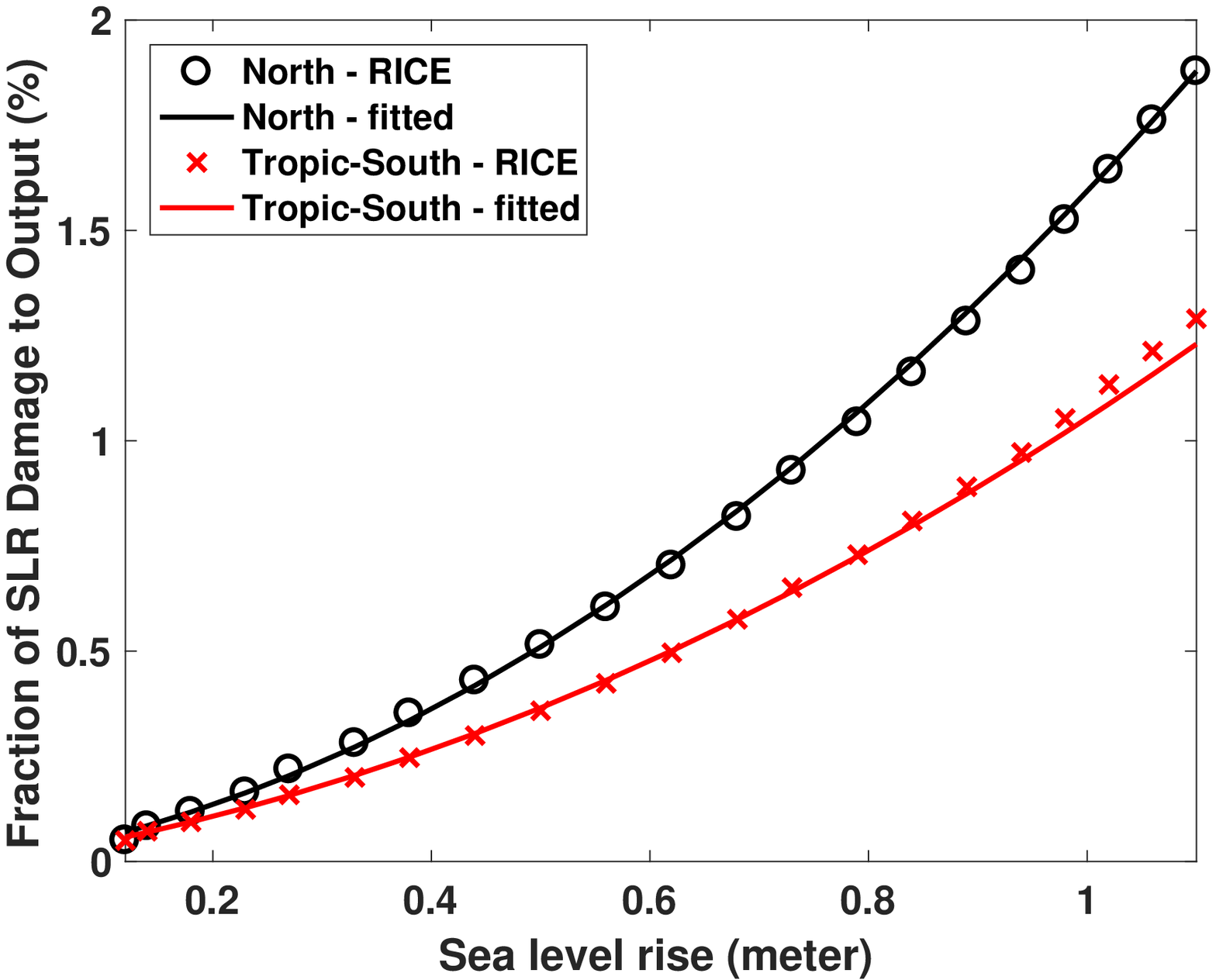} & \includegraphics[width=0.5\textwidth,height=0.25\textheight]{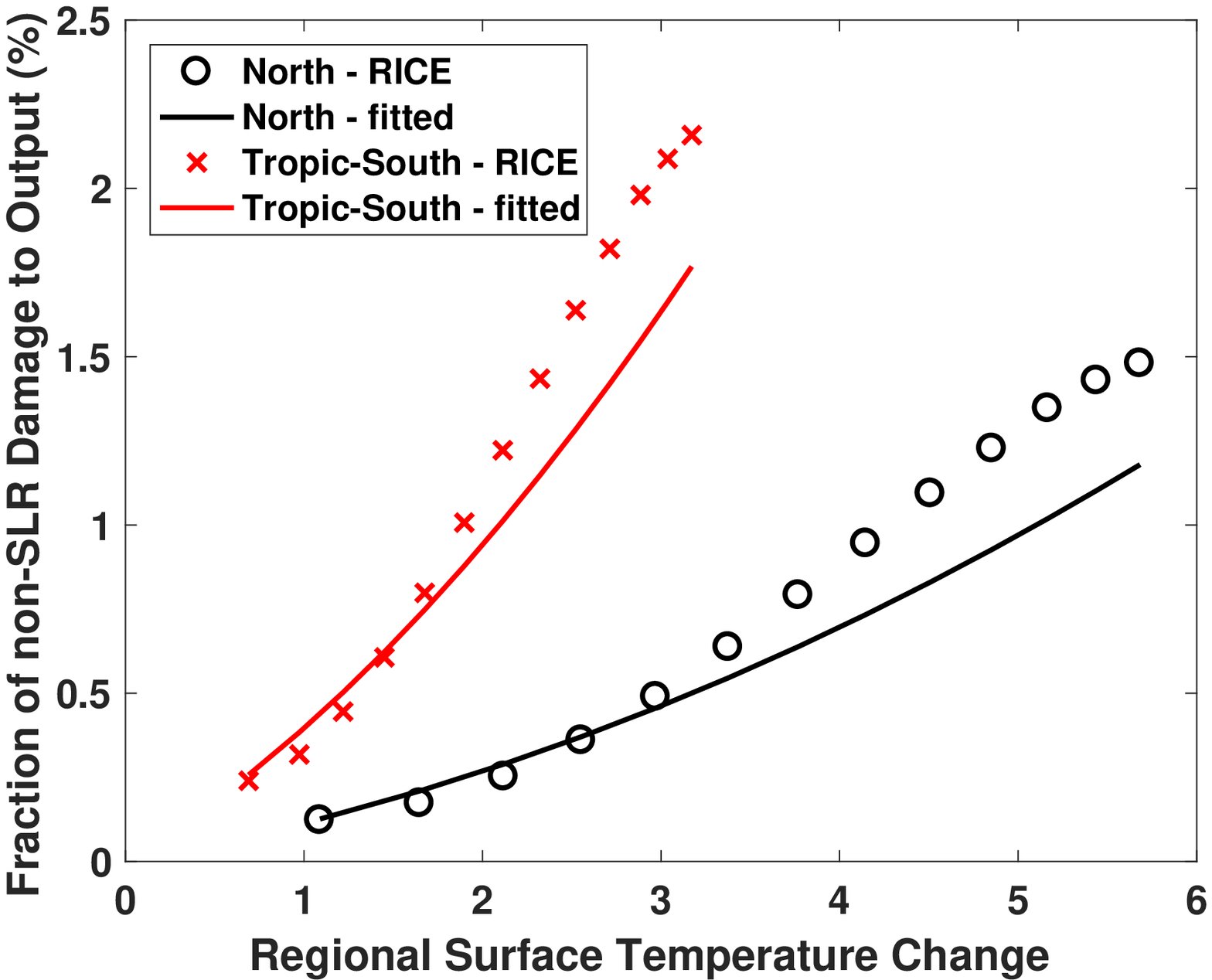}\tabularnewline
\end{tabular}
\par\end{centering}
\caption{Fitting SLR Damage (the left panel) and non-SLR Damage (the right
panel) to Output \label{fig:Fitting-dam}}
\end{figure}

The left panel of Figure \ref{fig:Fitting-dam} shows that our estimate
of SLR damage function (\ref{eq:SLR-dam}) is close to the RICE projection
of SLR damage for both the North and the Tropic-South. The right panel
of Figure \ref{fig:Fitting-dam} shows that our non-SLR damage functions
(\ref{eq:Dam_Temp_Quad}) fit the RICE projection of non-SLR damage
in fraction of output. 

\section{First Order Conditions for FBNE\label{sec:First-Order-Conditions}}

The system of first-order conditions (\ref{eq:FOC}) represents the
following equations:

\begin{eqnarray}
0 & = & u'(c_{t,i})-\beta\frac{\partial\mathcal{G}_{t,i}(\mathbf{x}_{t+1})}{\partial K_{t+1,i}},\ i=1,2,\label{eq:FOC_c}\\
0 & = & \frac{\partial\widehat{Y}_{t,i}}{\partial P_{t,i}},\ i=1,2,\label{eq:FOC_P}\\
0 & = & \frac{\partial\mathcal{G}_{t,i}(\mathbf{x}_{t+1})}{\partial K_{t+1,i}}\frac{\partial\widehat{Y}_{t,i}}{\partial\mu_{t,i}}+\frac{\partial\mathcal{G}_{t,i}(\mathbf{x}_{t+1})}{\partial M_{t+1}^{\mathrm{AT}}}\frac{\partial E_{t,i}^{\mathrm{Ind}}}{\partial\mu_{t,i}},\ i=1,2,\label{eq:FOC_miu}
\end{eqnarray}
where we use $\frac{\partial K_{t+1,i}}{\partial c_{t,i}}=-L_{t,i}$
in deriving (\ref{eq:FOC_c}), and $\frac{\partial K_{t+1,i}}{\partial\widehat{Y}_{t,i}}=1$
and $\frac{\partial M_{t+1}^{\mathrm{AT}}}{\partial E_{t,i}^{\mathrm{Ind}}}=1$
in deriving (\ref{eq:FOC_miu}). 

Here, 
\begin{eqnarray}
\frac{\partial\mathcal{G}_{t,i}(\mathbf{x}_{t+1})}{\partial K_{t+1,i}} & = & \left[\mathbb{E}_{t}\left(\left(V_{t+1,i}^{\mathrm{FBNE}}(\mathbf{x}_{t+1})\right)^{\varTheta}\right)\right]^{1/\varTheta-1}\mathbb{E}_{t}\left(\left(V_{t+1,i}^{\mathrm{FBNE}}(\mathbf{x}_{t+1})\right)^{\varTheta-1}\frac{\partial V_{t+1,i}^{\mathrm{FBNE}}(\mathbf{x}_{t+1})}{\partial K_{t+1,i}}\right)\nonumber \\
\frac{\partial\mathcal{G}_{t,i}(\mathbf{x}_{t+1})}{\partial M_{t+1}^{\mathrm{AT}}} & = & \left[\mathbb{E}_{t}\left(\left(V_{t+1,i}^{\mathrm{FBNE}}(\mathbf{x}_{t+1})\right)^{\varTheta}\right)\right]^{1/\varTheta-1}\mathbb{E}_{t}\left(\left(V_{t+1,i}^{\mathrm{FBNE}}(\mathbf{x}_{t+1})\right)^{\varTheta-1}\frac{\partial V_{t+1,i}^{\mathrm{FBNE}}(\mathbf{x}_{t+1})}{\partial M_{t+1}^{\mathrm{AT}}}\right)\nonumber \\
\frac{\partial\widehat{Y}_{t,i}}{\partial P_{t,i}} & = & Y_{t,i}\left[\left(1-\theta_{1,t,i}\mu_{t,i}^{\theta_{2}}-\eta_{1}P_{t,i}^{\eta_{2}}\right)\Omega_{t,i}(D_{t,i}^{\mathrm{SLR}}+D_{t,i}^{\mathrm{T}})-\eta_{1}\eta_{2}P_{t,i}^{\eta_{2}-1}\right]\label{eq:dY_dP}\\
\frac{\partial\widehat{Y}_{t,i}}{\partial\mu_{t,i}} & = & -\theta_{1,t,i}\theta_{2}\mu_{t,i}^{\theta_{2}-1}Y_{t,i}\nonumber \\
\frac{\partial E_{t,i}^{\mathrm{Ind}}}{\partial\mu_{t,i}} & = & -\sigma_{t,i}\mathcal{Y}_{t,i}\nonumber 
\end{eqnarray}
by assuming $\psi>1$ for convenience. When $\psi<1$, we just need
to replace $\beta$ by $-\beta$, and $V_{t+1,i}^{\mathrm{FBNE}}$
by $-V_{t+1,i}^{\mathrm{FBNE}}$ in the equations.

\section{No Transfer of Capital in the FBNE \label{sec:No-Transfer-of}}

In the FBNE, region $i$ chooses $I_{t,i},c_{t,i},\mu_{t,i},P_{t,i}$,
and migrated capital from the region $i$ to the other. Let $\Delta_{t,1}\geq0$
be migrated capital from region 1 o 2, and $\Delta_{t,2}\geq0$ from
region 2 o 1. Assume the IES $\psi<1$ for convenience. The FBNE solves

\begin{equation}
V_{t,i}(\mathbf{x}_{t})=\max_{I_{t,i},c_{t,i},\mu_{t,i},P_{t,i},\Delta_{t,i}\geq0}\left\{ u(c_{t,i})L_{t,i}+\beta\mathcal{G}_{t,i}(\mathbf{x}_{t+1})\right\} ,\label{eq:BellmanCE-2}
\end{equation}
for $i=1,2$, subject to the transition laws (\ref{eq:M-law}), (\ref{eq:T-law}),
(\ref{eq:SLR-law}), (\ref{eq:k-law}), (\ref{eq:law-damLevel}),
(\ref{eq:tipIndicator}), and the following region-specific market
clearing constraint 
\begin{eqnarray}
I_{t,1}+c_{t,1}L_{t,1}+\Gamma_{t,1}-\Delta_{t,1}+\Delta_{t,2} & = & \widehat{Y}_{t,1}\label{eq:budget1}\\
I_{t,2}+c_{t,2}L_{t,2}+\Gamma_{t,2}+\Delta_{t,1}-\Delta_{t,2} & = & \widehat{Y}_{t,2}\label{eq:budget2}
\end{eqnarray}
respectively, where 
\[
\Gamma_{t,i}=\frac{B}{2}\widehat{Y}_{t,i}\left(\frac{\Delta_{t,1}+\Delta_{t,2}}{\widehat{Y}_{t,i}}\right)^{2}
\]

Let $\lambda_{t,i}\geq0$ be shadow prices for (\ref{eq:budget1})
and (\ref{eq:budget2}), and $\tau_{t,i}\geq0$ for $\Delta_{t,i}\geq0$,
for $i=1,2$. Then from the KKT conditions of (\ref{eq:BellmanCE-2})
for region $i$, we have 
\begin{eqnarray*}
u'(c_{t,i}) & = & \lambda_{t,i}\\
\lambda_{t,i}\left(1-\frac{\partial\Gamma_{t,i}}{\partial\Delta_{t,i}}\right) & = & \tau_{t,i}\\
\tau_{t,i}\Delta_{t,i} & = & 0
\end{eqnarray*}
Thus, since $\lambda_{t,i}=u'(c_{t,i})>0$ and $\frac{\partial\Gamma_{t,i}}{\partial\Delta_{t,i}}=B\frac{\Delta_{t,1}+\Delta_{t,2}}{\widehat{Y}_{t,i}}<1$
(as $B=1$ and it is impossible to have $\Delta_{t,1}+\Delta_{t,2}\geq Y_{t,i}$
due to our regional specification), we have $\tau_{t,i}>0$ and then
$\Delta_{t,i}=0$. That is, in the FBNE, there is no transfer of capital
between the regions. 

\section{$L^{1}$ fitting\label{sec:L1-fitting}}

It is typical to choose the $\mathcal{L}^{1}$ norm in the minimization
step of Algorithm 1. In this case, we can avoid the kinks by transforming
the minimization model (\ref{eq:minFeas-1}) to
\begin{eqnarray}
\min &  & \sum_{i,k}(\epsilon'_{i,1,k}+\epsilon'_{i,2,k}),\label{eq:minFeas}\\
\mathrm{s.t.} &  & \nabla_{\mathbf{a}}\left(u_{t,i}(\mathbf{x},\mathbf{a})+\beta\mathcal{G}_{t,i}\left(\mathbf{H}_{t}(\mathbf{x},\mathbf{a},\omega)\right)\right)=\mathbf{\epsilon'}_{i,1}-\mathbf{\epsilon}'_{i,2},\ i\in\mathbb{I},\nonumber \\
 &  & \mathbf{a}\in\mathcal{D}(\mathbf{x},t),\nonumber \\
 &  & 0\leq\epsilon'_{i,j,k}\leq\bar{\epsilon}',\ i\in\mathbb{I};j=1,2;k=1,...,n,\nonumber 
\end{eqnarray}
where $\epsilon'_{i,1},\epsilon'_{i,2}$ are length-$n$ nonnegative
vectors, $\epsilon'_{i,1,k},\epsilon'_{i,2,k}$ are their elements,
and $\bar{\epsilon}'$ is a small positive number. To reduce computational
time, we may instead solve the following problem
\begin{eqnarray}
\min &  & \sum_{i,k}\epsilon''_{i,k},\label{eq:minFeas-2}\\
\mathrm{s.t.} &  & \nabla_{\mathbf{a}}\left(u_{t,i}(\mathbf{x},\mathbf{a})+\beta\mathcal{G}_{t,i}\left(\mathbf{H}_{t}(\mathbf{x},\mathbf{a},\omega)\right)\right)=\mathbf{\epsilon}''_{i},\ i\in\mathbb{I},\nonumber \\
 &  & \mathbf{a}\in\mathcal{D}(\mathbf{x},t),\nonumber \\
 &  & 0\leq\epsilon''_{i,k}\leq\bar{\epsilon}'',\ i\in\mathbb{I};k=1,...,n\nonumber 
\end{eqnarray}
where $\epsilon''_{i}$ is a length-$n$ nonnegative vector, $\epsilon''_{i,k}$
are its elements, and $\bar{\epsilon}''$ is a small positive number.

\section{Terminal Value Function\label{sec:Terminal-Value-Function}}

For the social planner's problem, the terminal value function at the
terminal time $T=300$ (i.e., year 2315) is computed as
\[
V_{T}^{\mathrm{SP}}(\mathbf{x}_{T})=\sum_{i=1}^{2}\sum_{t=300}^{600}\beta^{t-300}u(c_{t,i})L_{T,i},
\]
where we assume that for all $t>T$, all exogenous paths stop changing
and fix their values at terminal time $T$. Emission control rates
are always 1 (i.e., $\mu_{t,i}=1$), adaptation rates are fixed at
a constant level, there are no economic interaction costs (i.e., $\Gamma_{t,i}=0$),
and both $c_{t,i}L_{t,i}/Y_{t,i}$ and $I_{t,i}/Y_{t,i}$ are given
constant values. Moreover, we assume that if the tipping event has
not happened before $T$, then it never happens; or if the tipping
event has happened, then its damage will unfold until its maximum
level. 

For the FBNE, the terminal value functions at $T=300$ are computed
as
\[
V_{T,i}^{\mathrm{FBNE}}(\mathbf{x}_{T})=\sum_{t=300}^{600}\beta^{t-300}u(c_{t,i})L_{T,i}
\]
for $i=1,2$, under the same assumptions of the social planner's problem
for periods after $T=300$. \textcolor{black}{In both cooperative
and noncooperative cases, consideration of alternatives showed that
changes in the terminal value functions at year 2315 had no significant
impact on any results for the twenty-first century.}
\end{document}